\documentclass[11pt]{article}

\usepackage{geometry}
\usepackage[hypertexnames=false]{hyperref}
\usepackage{color}
\usepackage{latexsym}
\usepackage{dsfont}
\usepackage{amssymb}
\usepackage{graphicx}
\usepackage{amsmath, amsfonts,amssymb,theorem,euscript,array,enumerate,amsfonts,mathrsfs}
\usepackage{subfigure}
\usepackage[normalem]{ulem}
\usepackage{apacite}
\usepackage{natbib}

\newtheorem{Theorem}{Theorem}[part]

\newtheorem{Proposition}{Proposition}[part]

\newtheorem{Lemma}{Lemma}[part]

\newtheorem{Remark}{Remark}[part]

\usepackage[T1]{fontenc}
\usepackage[latin1]{inputenc}
\usepackage{hhline}
\usepackage{graphicx}

\usepackage{xcolor}
\usepackage{booktabs,caption}    
\usepackage[flushleft]{threeparttable}

\usepackage{eurosym}

\usepackage{dsfont}
\usepackage{ulem}

\theorembodyfont{\upshape}

\newcommand{\nc}{\newcommand}
\nc{\ind}{\mathds{1}}

\def \trans{^{\scriptscriptstyle{\intercal}}}

\newcommand{\R}{\mathbb{R}}

\newcommand{\E}{\mathcal{E}}
\newcommand{\F}{\mathcal{F}}

\newcommand{\btheta}{\boldsymbol{\theta}}

\DeclareMathOperator{\esssup}{esssup}

\def\esssup_#1{\underset{#1}{\mathrm{ess\,sup\, }}}
\def\essinf_#1{\underset{#1}{\mathrm{ess\,inf\, }}}
\def\argmax_#1{\underset{#1}{\mathrm{arg\,max\, }}}
\def\argmin_#1{\underset{#1}{\mathrm{arg\,min\, }}}

\usepackage{diagbox}
\def\reff#1{{\rm(\ref{#1})}}

\def \ep{\hbox{ }\hfill$\Box$}

\def \Sum{\displaystyle\sum}
\def \Prod{\displaystyle\prod}

\def \Frac{\displaystyle\frac}
\def \Inf{\displaystyle\inf}
\def \Sup{\displaystyle\sup}
\def \Lim{\displaystyle\lim}

\def \Max{\displaystyle\max}
\def \Min{\displaystyle\min}

\def\b1{\bf 1}

\def\Dt#1{\Frac{\partial #1}{\partial t}}

\def \C{\mathbb{C}}

\def \N{\mathbb{N}}
\def \R{\mathbb{R}}

\def \E{\mathbb{E}}
\def \F{\mathbb{F}}

\def \P{\mathbb{P}}

\def \S{\mathbb{S}}

\def \Ac{{\cal A}}
\def \Bc{{\cal B}}

\def \Fc{{\cal F}}

\def \Pc{{\cal P}}

 \def \Rc{{\cal R}}
\def \Sc{{\cal S}}

\def \Vc{{\cal V}}

\def \Vc{{\cal V}}

\def\btheta{ \boldsymbol{\theta}}
\def\bTheta{{\bf \Theta}}
\def\bb{\boldsymbol{b}}
\def\bR{\boldsymbol{R}}
\def\brho{\boldsymbol{\rho}}

\def \ep{\hbox{ }\hfill$\Box$}
\def \epR{\hbox{ }\hfill$\lozenge$}

\def \trans{^{\scriptscriptstyle{\intercal}}}

\def\reff#1{{\rm(\ref{#1})}}

\def\beqs{\begin{eqnarray*}}
\def\enqs{\end{eqnarray*}}
\def\beq{\begin{eqnarray}}
\def\enq{\end{eqnarray}}

\usepackage[pagewise]{lineno} 
\setpagewiselinenumbers

\begin{document}

\title{
Portfolio diversification and model uncertainty: \\  a robust dynamic mean-variance approach
\thanks{The authors benefited from the research grant for the project ``Model uncertainty and stochastic control in financial risk management" under the aegis of the Alliance National University of Singapore (NUS)-Sorbonne Paris Cit\'e (USPC). This work started when the first two authors were visiting the National University of Singapore (NUS), whose hospitality is kindly acknow\-ledged. Part of this work was carried out while the third author was invited to the University Paris Diderot, whose hospitality is kindly appreciated.  
}
}

\author{
Huy\^en PHAM\footnote{LPSM,  Universit\'e de  Paris and CREST-ENSAE, {\sf pham at lpsm.paris}.
The work of this author is supported by the ANR project CAESARS (ANR-15-CE05-0024), and also by FiME
and the ''Finance and Sustainable Development'' EDF - CACIB Chair.}
\quad\quad
Xiaoli WEI\footnote{IEOR,  UC Berkeley, {\sf tyswxl at gmail.com}. }
\quad\quad
Chao ZHOU\footnote{Department of Mathematics, National University of Singapore,  {\sf matzc at nus.edu.sg}. The work of this author is supported by Singapore MOE AcRF Grants R-146-000-271-112, R-146-000-255-114, R-146-000-219-112 and NSFC Grant 11871364.}
}

\date{}

\maketitle

\begin{abstract}
This paper focuses on a dynamic multi-asset mean-variance portfolio selection pro\-blem under model uncertainty. We develop a continuous time framework for taking into account ambiguity aversion about both expected return rates and correlation matrix of the assets, and for studying the join effects on portfolio diversification.
The dynamic setting allows us to consider time varying ambiguity sets, which include the cases where the drift and correlation are estimated on a rolling window of historical data or when the investor takes into account learning on the ambiguity.  In this context, we prove a general separation principle for the associated robust control problem, which allows us to reduce the determination of the optimal dynamic strategy to the parametric computation of the minimal risk premium function.
Our results provide a justification for under-diversification, as documented in empirical studies and in the static models  \cite{GarUppWan06}, \cite{LZ2017}. Furthermore,
we explicitly quantify the degree of under-diversification in terms of correlation bounds  and Sharpe ratios proximities,
and emphasize the different features induced by drift and correlation ambiguity.
In particular, we show that an investor with a poor confidence in the expected return estimation does not hold any risky asset, and on the other hand, trades only one risky asset when the level of ambiguity on correlation matrix is large.  We also provide a complete picture of the diversification for the optimal robust portfolio in the three-asset case.
\end{abstract}

\vspace{5mm}


\vspace{5mm}


\vspace{5mm}


\newpage

\section{Introduction}

\setcounter{equation}{0} \setcounter{Assumption}{0}
\setcounter{Theorem}{0} \setcounter{Proposition}{0}
\setcounter{Corollary}{0} \setcounter{Lemma}{0}
\setcounter{Definition}{0} \setcounter{Remark}{0}

There are many studies on under\--diversification of portfolio in the Finance and Economics literature, where investors hold only a small part of risky assets among a large number of available risky assets.  In the extreme case  the anti-diversification effect means that investors hold only a single stock (or not even any risky asset) and exclude many others.
Empirical studies reported in numerous papers, see  \cite{FrenchandPoteraba91}, \cite{CooperandKaplanis94}, \cite{MitVor07}, \cite{CalCamSod07}, \cite{GuiLiu16}, have shown the evidence
of  portfolio under-diversification in practice.  For example, in \cite{FrenchandPoteraba91}, \cite{CooperandKaplanis94}, it is observed that there exists a concentration on (bias towards)  domestic assets compared to foreign assets in investors' international equity portfolios. These results are in contrast
with the well-diversified portfolio suggested by the classical mean-variance portfolio theory initiated  in a single period model in \cite{Markowitz52}, later in \cite{LiNg2000} for a multi-period model,  and in \cite{ZhouLi2000} for a continuous-time model.  A possible explanation to under-diversification is provided in the Finance and Economics literature  by model uncertainty, often also called ambiguity or Knightian uncertainty.

\medskip

In classical portfolio theory, the model and parameters are assumed to be perfectly known. However, in reality, due to  statistical estimation issues, there is always uncertainty (ambiguity) about the model or parameters. In this case, a robust approach, see e.g. \cite{bental},  can be used to compute the optimal portfolio,
i.e., the investor makes portfolio decisions under the worst case that corresponds to the least favorable scenario implied by a set of ambiguous parameters or by a set of distributions on the
price process, which is usually refereed in operations research literature to  distributionally robust optimization.



\medskip

Abundant research has been conducted to tackle different types of model uncertainty.  Robustness to uncertainty over a set of distributions on market factors
in portfolio optimization has been analyzed in, e.g., \cite{GhaOksOus03}, \cite{NatPacSim08, NatSimUic10}, \cite{DelYe10},  \cite{GohSim10},
\cite{Wiesetal14}, \cite{HanKuh18}, \cite{JiaGua18}, primarily in single-period formulations, except \cite{glaxu13} in a multi-period setting.  Primary market factor is  the price process,
and in this case, relevant sets of uncertain distributions  correspond to ambiguity on the drift (i.e. the expected rate of return), the volatilities and the correlations when there are multiple assets to be traded.  Indeed, drift estimation  is known to be notoriously difficult: it is typically computed by maximum likelihood estimator (MLE) from historical time series of assets  log-return, and we refer e.g. to \cite{Cont2001} for general issues in statistical estimations of log-return,  \cite{ChaVic2003} for generalized method of moments to estimate drift, and the recent paper \cite{biechecia17}  for the recursive construction of drift confidence region.  Moreover, 
the estimation of correlation between assets may be extremely inaccurate, due to the asynchronous data and lead-lag effect, especially when the number of assets is large, see \cite{CizPotBou2001, LieLieMul2004}, \cite{JagMa2003}, \cite{LedWol14}.  Related works on robust portfolio optimization include \cite{Schi07}, \cite{JinZhou15} for uncertainty solely about drift, \cite{DenKer2013}, \cite{MatPosZho15} for ambiguity about volatility (in a probabilistic setup) with a fa\-mily of nondominated probability measures, \cite{LinRie14}, \cite{BiaPin17} for combined uncertainty about both drift and volatility , \cite{NeuNut18} for joint ambiguity about drift, volatility and jumps.  In this existing literature, the common types of drift uncertainty sets are represented by polyhedral set or  ellipsoidal set in \cite{BiaPin17}, and unified by general ellipsoidal set in \cite{GarUppWan06}.  This general ellipsoidal representation for the drift ambiguity
indicates in particular that drift estimation is affected by correlation estimation. Compared to drift ambiguity and volatility ambiguity, there are rather few results dealing  with  correlation ambiguity, let us mention however  \cite{FouPunWon16},  \cite{JiaTia16},  \cite{HuaZhaZhu17}, \cite{IsmailPham17}, and \cite{LZ2017}.

\begin{table}[!htbp]
\tiny
\centering
\begin{threeparttable}
\caption{Some literature on the impact of model uncertainty on portfolio diversification}
\begin{tabular}{c|c|c|c|c}
\toprule
\diagbox{Ambiguity}{Objective} & static MV & static utility & dynamic MV & dynamic utility \\
\hline
factors & \shortstack{\cite{GolDonGar03}\\ (d, N)} & & & \shortstack{\cite{glaxu13}\\(d, N)}\\
\hline
\shortstack{drift \\ \; \\ \; \\ \;} & \shortstack{\cite{GarUppWan06} \\ (d, Y)\\ \cite{BGUW12}\\(d, Y)} & \shortstack{\cite{DowSer92}\\ (1, Y) \\ \; } & & \shortstack{\cite{UW03}\\(d, Y)}\\
\hline
covariance (cov) & & & \shortstack{\cite{IsmailPham17}\\(d, N)} & \shortstack{\cite{MatPosZho15}\\(d, N)} \\
\hline
\shortstack{ drift \& cov \\ \; \\ \; \\ \;} &  & & &  \shortstack{\cite{BiaPin17}\\(d, N)\\ \cite{LinRie14}\\(d, N)} \\
\hline
drift \& cov \& jumps & & & & \shortstack{ \cite{NeuNut18}\\(d, N)}\\
\hline
\shortstack{correlation(corr) \\ \; \\ \; \\ \;} & \shortstack{\cite{LZ2017}\\(d, Y)\\ \;}  & \shortstack{\cite{HuaZhaZhu17}\\(2, Y)\\  \cite{JiaTia16}\\(d, Y)}& \shortstack{\cite{IsmailPham17}\\(2, N)}& \shortstack{\cite{FouPunWon16}\\(2, N) \\ \; }\\
\hline
drift \& corr & & & \multicolumn{2}{c}{This paper(d, Y)} \\
\bottomrule
\end{tabular}
\label{table1}
\begin{tablenotes}
\small
\item  In the above entries $(.,.)$ the first element $1$, $2$ or $d$ refers to the number of risky assets considered in the paper [.], while the second element Y or N indicates whether the portfolio under-diversification is  studied or not.
\end{tablenotes}
\end{threeparttable}
\end{table}

\medskip

Our purpose is to explore the joint effects of ambiguity about drift and correlation on portfolio selection and diversification with mean-variance  (MV) criterion in continuous time. Notice that in the above cited papers, portfolio selection problems are mainly based on expected utility criterion and the effect on portfolio diversification under continuous-time framework is not really studied. Table \ref{table1} summarizes some papers on the model uncertainty and its impact on the portfolio diversification,
which are  related to the study of our paper. The list of related papers is not exhaustive. We distinguish between usual covariance ambiguity and correlation ambiguity in Table \ref{table1}:
Covariance ambiguity refers to the case when  the covariance matrix lies between two given bounds $\underline \Sigma$, $\overline\Sigma$ in the space of positive definite symmetric matrix,
see e.g. \cite{LinRie14}, \cite{MatPosZho15}, \cite{NeuNut18}, \cite{YLZ2018}, or lies in a proper cone, see e.g. \cite{Wiesetal14}. This is not easily interpretable in terms of correlation. Correlation ambiguity means that 
ambiguity is directly formulated on the correlations between the different assets. It turns out that, in contrast with covariance ambiguity, correlation ambiguity is a relevant indicator for
generating under-diversification, as shown in a static model in \cite{HuaZhaZhu17}, \cite{JiaTia16}  and \cite{LZ2017}.
As pointed out in Table \ref{table1},  the existing literature about under-diversification usually focuses on static mean variance criteria or expected utility criteria and on one type of model uncertainty.

\medskip

Our paper considers  both robust mean-variance and utility framework in continuous time to investigate the impact of combined drift and correlation ambiguity on portfolio diversification. Due to the  nonlinear dependence on the wealth expectation, the mean-variance criterion is a non standard control problem. To circumvent this issue, the authors in \cite{JinZhou15} reformulate the mean-variance problem under drift uncertainty into portfolio Sharpe ratio of the terminal wealth. Robust dynamic mean-variance problem under covariance matrix uncertainty, in particular, correlation ambiguity, has been considered in \cite{IsmailPham17} by a McKean-Vlasov dynamic programming approach, but  the authors neither tackle the drift uncertainty nor  study the portfolio diversification in detail, and mainly focus on the two\--asset case $d$ $=$ $2$. One key assumption in \cite{IsmailPham17} is that one can aggregate a family of processes, a condition, which does not hold true anymore in  the case of drift uncertainty. An additional feature of our framework, compared to one-period models, is the consideration of  learning on the ambiguity about the drift and correlation of risky assets: for instance, the investor  typically  gets more and more information about history of asset prices over time, and thus estimation errors about model parameters are reduced.
Moreover,  compared to models focusing only on one type of parameter ambiguity,  our framework provides a unified setting  for the joint effects of drift and correlation ambiguity on portfolio diversification. In particular, we are able to consider correlation structure of the assets in the drift uncertainty modelling.

\medskip


Regarding portfolio diversification, the authors  in \cite{BGUW12}, \cite{GarUppWan06}, \cite{UW03} considered  ambiguity  about the assets' returns. Their frameworks include  uncertainty about the joint distribution of returns for all assets and also for different levels of uncertainty of the marginal distribution of returns for any subsets of these assets.
They showed that the different levels of uncertainty on different asset subclasses could result in significant under-diversification. They also applied their theoretical results to real data and found consistent results with the empirical studies in \cite{CooperandKaplanis94}, \cite{FrenchandPoteraba91} among others, showing that  international equity portfolios are strongly biased towards domestic stocks, and in \cite{Huberman2001} and \cite{Schultz1996},  where a similar lack of diversification  is revealed on domestic portfolios.
The model in \cite{BGUW12}, \cite{GarUppWan06}, \cite{UW03}  offers a partial explanation for the observed under-diversification and bias towards familiar securities.
More recently, the authors in \cite{LZ2017}  considered the uncertainty  about the correlation of the assets. With a static mean-variance investment, they found that
under-diversification of  robust optimal portfolio depends on the level of correlation  ambiguity. They also provided results with market data and showed, using their uncertain (ambiguous) correlation model, that the investor only holds less than 20 (17 stocks on average) among 100 stocks randomly selected from about 500 stocks in S\&P500. In the two-asset case, they also found that the degree of diversification depends on the comparison between the ratio of assets' Sharpe ratios and  the correlation ambiguity parameters.
 A further explanation for under-diversification is that investors can reduce the uncertainty on the model or the parameters through learning. In \cite{VNV10}, the authors built a framework to solve jointly for investment and information choices, with general preferences and information cost functions. They showed that, for some special preferences and information acquisition technologies, investors tend to learn more about the assets with which they are more familiar  (typically, the domestic assets rather than the foreign ones), and become even more familiar with those assets after learning. As a consequence of this learning procedure, the investors select those assets they have learnt at the expense of others for which they have less
information. Their results are consistent with the empirical studies on portfolios of international investors.

\medskip

We emphasize that the consideration of joint ambiguity on the drift and correlation is relevant regarding portfolio diversification for several reasons:
\begin{itemize}
\item As  the portfolio allocation is determined by both the drift and the correlation of the assets, it is consistent that under-diversification should be governed by the ambiguity levels on both
these parameters: actually, the ambiguity level on the drift mainly determines absolute position of each asset (invest or not  in a asset if the drift ambiguity is small or not), while the ambiguity level on the correlation mainly determines relative position of the assets (not diversify if the correlation is very ambiguous and diversify if not, directional trading if the correlation is small comparing to Sharpe ratio proximity and spread trading if large).
\item  In \cite{UW03}, the authors consider both global ambiguity for the returns of all risky assets and different levels of ambiguity for any subset of these risky assets. In the case of equal ambiguity for all risky assets, ambiguity about the return would not bias the portfolio towards a particular asset, thus not explaining under-diversification. Instead, in the case of different levels of ambiguity for subsets of these risky assets, the ambiguity difference would bias the portfolio towards assets with smaller ambiguity, thus explaining under-diversification.
In our setting, even in the case of equal ambiguity for the returns of all risky assets, we can still explain under-diversification through the ambiguity for the correlations of all assets.
Moreover, in the case of different levels of ambiguity for the returns for subsets of these risky assets, we can explain under-diversification through the combination of the ambiguity difference for the returns and the ambiguity for correlations. In particular, we still obtain the optimal strategy in an explicit form, which allows us to understand the different effects of correlation ambiguity and expected return ambiguity.
\item In \cite{BGUW12}, the familiarity and unfamiliarity of assets are modelled by different levels of ambiguity on expected rates of return  in single period setting.
The main feature of their model is that it allows investors to distinguish their ambiguity about one asset class relative to others.  Moreover,  similarly as in  \cite{UW03}, the different levels of ambiguity can explain bias to familiar assets and under-diversification. The authors showed that the correlation coefficient has an important effect on the portfolio weights, notably, an increase in correlation from $50\%$ to $70\%$ roughly doubles the holding of the familiar asset.  Since correlation coefficients are very difficult to estimate with a good accuracy, it is important to take into account correlation ambiguity when building the optimal  portfolio which is less sensitive to estimation inaccuracy. This is our main motivation to consider ambiguity on both expected return rates and correlation and study the effects of ambiguity on portfolio strategy, in particular, under-diversification.
\item  Correlation ambiguity and expected return ambiguity have different features. For example, technically, in a continuous-time setting,  a set of absolute continuous probability measures can be used to model the expected return ambiguity. In contrast, a set of mutually singular probability measures is needed for correlation ambiguity.
This explains why in the existing literature, such as \cite{IsmailPham17} and \cite{MatPosZho15}, only one type of uncertainty is considered.
Economically, if the expected return ambiguity level is large, no risky asset is held (see e.g. \cite{BGUW12}). 
In contrast, if correlation ambiguity is large, one and only one asset is held (see e.g. \cite{LZ2017}). However,  by considering only correlation ambiguity as in \cite{LZ2017}, one could not  explain nonparticipation in which the investor does not make  any risky investment at all.
\end{itemize}




To sum up, the contributions of our paper are fourfold: 

(1) First, we develop a robust model that  takes into account  uncertainty on both  drift and correlation of multiple risky assets for $d$ $\geq$ $2$,  in a dynamic, continuous time mean-variance portfolio setting. The dynamic setting allows us to consider time varying ambiguity sets, which include the cases where the drift and correlation are estimated on a rolling window of historical data or when the investor takes into account learning on the ambiguity.

(2) Secondly, we state a separation principle for the associated robust control problem formulated as a mean-field type differential game, which allows us to reduce the original min-max problem to the parametric computation of minimal risk premium. We derive the separation principle in a general setting with time varying ambiguity sets and for general preference criteria, including expected utility.  In particular, the separation principle also holds true in  single period and multi-period models. We can then generalize results in static models as in   \cite{BGUW12} and  \cite{GarUppWan06} by incorporating  correlation ambiguity, and  study the implications of both expected return and correlation ambiguity.
The main methodology for the separation principle is based on a weak version of the martingale optimality principle. This extends  the classical martingale optimality principle that can not be directly applied  in the presence of drift uncertainty, see detailed comments in Remark \ref{remdis}.

(3) Furthermore, we illustrate our results in rectangular  and ellipsoidal uncertainty set and quantify explicitly the diversification effects on the optimal robust portfolio in terms of the ambiguity level. As in \cite{UW03} and \cite{GarUppWan06}, the uncertainty set is flexible enough to allow for joint uncertainty for all assets or different levels of uncertainty for different subsets of the assets. Both drift uncertainty and correlation uncertainty can lead to  under-diversification. In particular, we find that the robust investor does not trade in assets with large expected return ambiguity and trades only one risky asset in presence of  high level of ambiguity about correlation.
We also obtain closed-form expressions for the robust optimal portfolio, and we provide notably a complete picture of the diversification for the optimal robust portfolio in the case with three risky assets, which is new to the best of our knowledge. 
In the paper \cite{LZ2017} dealing with a rectangular ambiguity set in a static model, the authors proposed an implicit condition for not investing in one asset, i.e.,  under-diversification, in terms of correlation ambiguity. We provide in our dynamic setting an explicit condition in terms of ambiguity set for under-diversification,  and obtain
the optimal strategies in explicit form.
%
For our future studies, we 
may introduce the information acquisition procedure as in \cite{VNV10} or the multi-agent with heterogeneous beliefs setting as in \cite{JiaTia16} in our framework.

(4) Finally, our results suggest that the diversification effect is mainly determined by the relation between correlations of risky assets and Sharpe ratio "proximities" (the ratio of the Sharpe ratios). To the best of our knowledge, it is the first time that such an explicit relation is pointed out  in a general setting. For simplicity, let us illustrate the idea and explain why classical portfolio theory does not lead to under-diversification in the two-asset case. Indeed, the Sharpe ratio proximity provides a standard to measure if the correlation is large or small. When the correlation is a constant, the investor chooses  her  portfolio allocation according to the relationship between the correlation and the Sharpe ratio proximity of the two assets. When the correlation is larger than the Sharpe ratio proximity, the optimal strategy is to make a spread trading. When the correlation is smaller than the Sharpe ratio proximity, the optimal strategy is to make a directional  trading. Then the only case the investor invests in one asset is when the correlation is equal to the Sharpe ratio proximity. However, in practice, it is almost impossible to have this equality with parameters estimated from market data. However, in the case when correlation uncertainty set is an interval, the investor invests in one asset if the Sharpe ratio proximity lies in the correlation uncertainty interval, which is more likely to occur when the uncertainty interval is large due to lack of market data.

\medskip

The rest of paper is organized as follows. Part I is devoted to the theoretical developments regarding the robust optimization problem, and contains two sections.  Section 2 presents the formulation of the model uncertainty setting and the robust mean-variance problem. In Section 3, we derive the separation principle and explicit robust solution. 
Part II concerns the financial applications with two sections. Section 4 provides several examples arising from the separation principle, and the implications for the optimal robust portfolio strategy and the portfolio diversification. Section 5 illustrates through two numerical examples the effects of drift and correlation estimation error, thus ambiguity level, on portfolio Sharpe ratio. 
Finally, Appendix collects proofs of several mathematical results including the separation
principle for general expected utility criteria.

\section*{Part I: Theoretical developments}

\section{Problem formulation}

\setcounter{equation}{0}
\setcounter{Assumption}{0}
\setcounter{Theorem}{0} \setcounter{Proposition}{0}
\setcounter{Corollary}{0} \setcounter{Lemma}{0}
\setcounter{Definition}{0} \setcounter{Remark}{0}

\subsection{Model uncertainty setting} \label{subsecmodel}

We consider a financial market with one risk-free asset, assumed to be constant equal to one, and $d$ risky assets on a finite investment horizon $[0, T]$. Model uncertainty is formulated by using a probabilistic setup as in \cite{NeuNut18}.  We define the canonical state space by
$\Omega=\{\omega=(\omega(t))_{t \in [0, T]} \in C([0, T], \R^d): \omega(0) = 0\}$ representing the continuous paths driving the risky assets. We equip $\Omega$ with the uniform norm and the corresponding Borel $\sigma$-field $\Fc$. We denote by $B=(B_t)_{t \in [0, T]}$ the canonical process, i.e.,  $B_t(\omega)=\omega(t)$,  and by $\F$ $=$ $(\Fc_t)_{0 \leq t \leq T}$ the canonical filtration, i.e. the natural (raw) filtration generated by $B$.

We assume that the investor knows the marginal volatilities $\sigma_i$ $>$ $0$  of each asset $i$ $=$ $1,\ldots,d$, typically through a quadratic variation estimation of the assets,
and we denote by $\mathfrak S$ the known constant \footnote{we could consider deterministic  marginal volatilities, but this does not impact our results, and for simplicity of presentation, we take them constant.} diagonal matrix with  $i$-th diagonal term equal to $\sigma_i$, $i$ $=$ $1,\ldots,d$.
However, there is uncertainty about the drift (expected rate of return) and the correlations of the multi-assets, which are parameters notoriously difficult to estimate in practice.

The ambiguity  about drift and correlation matrix is parametrized by a  family ${\bf \Theta}$ $=$ $\{ \Theta(t): t \in [0,T]\}$  of nonempty sets with
\beqs
\Theta(t) & \subset &  \R^d \times \C^d_{>+},  \;\;\; t \in [0,T],
\enqs
where $\C^d_{>+}$ is the subset of all elements  $\rho$ $=$ $(\rho_{ij})_{1\leq i \neq j\leq d}$ $\in$ $[-1,1]^{d(d-1)}$ with $\rho_{ij}$ $=$ $\rho_{ji}$  s.t. the symmetric  matrix
 $C(\rho)$ with diagonal terms $1$ and  off-diagonal terms $\rho_{ij}$
 \beqs
 C(\rho) & = &
\left(
 \begin{array}{cccc}
 1 & \rho_{12} & \ldots & \rho_{1d} \\
 \rho_{12} & 1 & \ldots & .\\
 \vdots &  \vdots & \ddots & \vdots  \\
 \rho_{1d} & .  & \ldots & 1
 \end{array}
 \right)
 \enqs
 lies in $\S_{>+}^d$, the set of positive definite symmetric matrices in $\R^{d\times d}$. Notice that $\C^d_{>+}$ is an open convex set of
 $[-1,1]^{d(d-1)}$.
 The first component set of $\Theta(t)$ represents the values taken by the (possibly random) drift of the assets at time $t$, while
 the matrices $C(\rho)$, when $\rho$ runs in the second component set of $\Theta(t)$,  represent the correlation matrices of the multi-assets at time $t$.  The introduction of a family of sets $\Theta(t)$, $0\leq t\leq T$, allows us to take into account learning  on  the ambiguity about  the mean return and correlations. 
 The covariance matrices of the assets are given by
\beqs
\Sigma(\rho) &:=& \mathfrak S C(\rho) \mathfrak S \; = \;  \left(
 \begin{array}{cccc}
 \sigma_1^2 & \sigma_1\sigma_2\rho_{12} & \ldots & \sigma_1\sigma_d \rho_{1d} \\
 \sigma_1\sigma_2 \rho_{12} & \sigma_2^2 & \ldots & . \\
 \vdots &  \vdots & \ddots & \vdots  \\
 \sigma_1\sigma_d \rho_{1d} & .  & \ldots & \sigma_d^2
 \end{array}
 \right),
\enqs
and we denote by $\sigma(\rho)$ $=$ $\Sigma^{\frac{1}{2}}(\rho)$ $=$ $\big(\Sigma(\rho)\big)^{\frac{1}{2}}$  the square-root matrix, called volatility matrix.

An element $\btheta$ $=$ $(\theta(t))_{t\in [0,T]}$ $\in$ $\bTheta$ $=$ $\{ \Theta(t): t \in [0,T]\}$ can be viewed as a function on $[0,T]$ s.t. $\theta(t)$ $\in$ $\Theta(t)$ for all $t$ $\in$ $[0,T]$, and we write $\btheta$ $=$ $(\bb,\brho)$  to distinguish the first and second component of this function, with $\bb$ $=$ $(b(t))_{t\in [0,T]}$ and $\brho$ $=$
$(\rho(t))_{t\in [0,T]}$.

Let us now  introduce the family of (squared) risk premium $\bR(\btheta)$ $=$ $\{R(\theta(t)): t \in [0,T]\}$, for $\btheta$ $=$ $(\theta(t))_{t\in [0,T]}$ $\in$ $\bTheta$, by
\beq \label{defpremium}
R(\theta) &=& b\trans \Sigma(\rho)^{-1} b \;=\; \|\sigma(\rho)^{-1}b\|_{_2}^2, \;\;  \mbox{ for }  \theta = (b,\rho) \in \Theta(t), \; t \in [0,T],
\enq
where $\Sigma(\rho)^{-1}$ $=$ $\big(\Sigma(\rho)\big)^{-1}$.
Hereafter, $\trans$ denotes the transpose of matrix and $\|\cdot\|_{_2}$ denotes the Euclidean norm in $\R^d$.
\begin{Remark} \label{remcorrel}
{\rm There exist  different conditions for characterizing the positive definiteness of the correlation matrix $C(\rho)$. For example, Sylvester's criterion states that $C(\rho)$ is positive definite if and only if all  the leading principal minors are positive, e.g., in dimension $d$ $=$ $2$, $\rho$ $\in$ $(-1, 1)$; in dimension $d$ $=$ $3$, $\rho_{ij}$ $\in$ $(-1, 1)$ $1$ $\leq$ $i$ $<$ $j$ $\leq$ $3$ and
$\rho_{12}^2 + \rho_{13}^2 + \rho_{23}^2 -1 -2 \rho_{12}\rho_{13}\rho_{23}$ $<$ $0$. Alternatively, one can  characterize the positive definiteness of $C(\rho)$ using angular coordinates as in \cite{RapiBriMer07}.  Instead of working directly with $C(\rho) \subset \S_{>+}^d$ in the form of matrix, we characterize $C(\rho) \in \Sc_{>+}^d$ in a more explicit way in terms of parameter $\rho \in \C_{>+}^d$ via Sylvester's criterion. 
}
\epR
\end{Remark}

The ambiguity sets $\bTheta$  $=$ $\{ \Theta(t): t \in [0,T]\}$ for the drift and correlation are assumed to satisfy
\beqs
\hspace{-0.7cm} {\bf (H\bTheta)} \hspace{1cm}  t \mapsto \Theta(t) \; \text{ is measurable} \footnotemark,  \;\text{and}\;\; \Theta(t) \mbox{ is a bounded convex set of }  \R^d \times \C^d_{>+}, \;\;\; t \in [0,T].
\enqs

\addtocounter{footnote}{0}
\footnotetext{The set-valued map $\Theta: [0, T] \to 2^{\R^d \times \C_{>+}^d}$ is said to be measurable if and only if for every open set $U \subset \R^d \times \C_{>+}^d$, $\{t \in [0, T]: \Theta(t) \cap  U \neq \varnothing \} \in \Bc([0, T])$. Here $2^{\R^d \times \C_{>+}^d}$ represents the family of all subsets in $\R^d \times \C_{>+}^d$ and $\Bc([0, T])$ is Borel $\sigma$-algebra on $[0, T]$. See e.g., \cite{Frysz2004} for more details on the measurability of set-valued maps.}

A relevant class for practical applications of ambiguity sets $\bTheta$ satisfying ${\bf (H\bTheta)}$  is the following.
Let $\{J_1, \ldots, J_l, \ldots, J_p\}$, $1$ $\leq$ $p$ $\leq$ $d$, be a partition  of $\{1, \ldots, d\}$, and denote by $ |J_l|$ the cardinality of $J_l$, $l$ $=$ $1$, $\ldots$, $p$.
We consider ambiguity set $\Theta(t)$ in the form
\beq \label{exaTheta}
\Theta(t) \;=\; \{(b, \rho) \in \R^d \times \Gamma(t): \|\sigma_{J_{l}}(\rho)^{-1}(b_{J_l} -\hat b_{J_l}(t))\|_{_2} \leq \delta_{l}(t),\;\; l =1, \ldots,p\},
\enq
for some convex set
$\Gamma(t)$ of $\C^d_{>+}$,  where $\Sigma_{J_l}(\rho)$ is  the $|J_l|$ $\times$ $|J_l|$ variance-covariance matrix of assets in subclass $J_l$ and its square root $\sigma_{J_l}(\rho)$ $=$
$(\Sigma_{J_l}(\rho))^{\frac{1}{2}}$. Here $\hat b_{J_l}(t)$ is a known vector, representing  an estimate of mean return vector $b_{J_l}$ of assets in $J_l$ at time $t$,  and
$\delta_l(t)$ $\geq$ $0$ represents a level of ambiguity around $\hat b_{J_l}(t)$ due to estimation error as well as her level of uncertainty aversion. 

\begin{Remark}
{\rm
In the particular case when the number of subclasses  is equal to the number of risky assets, i.e., $p$ $=$ $d$, $\Theta(t)$ is a rectangular set in the form
$\prod_{i=1}^d [\hat b_i(t) -\sigma_i\delta_i(t), \hat b_i(t) +\sigma_i\delta_i(t)]$ $\times$ $\Gamma(t)$, interpreted as a product set of  {{uncertainty regions}} with size determined by the  level $\delta_i(t)$ for each asset
$i$ $=$ $1,\ldots,d$. Instead of setting  {{uncertainty regions}} for the assets individually, one can do it jointly for all assets by considering the case when $p$ $=$ $1$, which corresponds to an  ellipsoidal set in the form $\{(b, \rho) \in \R^d \times \Gamma(t): \|\sigma(\rho)^{-1}(b -\hat b(t))\|_{_2} \leq \delta(t)\}$. An extension of the two above sub-cases, allowing for separate estimation and   {{uncertainty regions}} for different subclasses of assets (due e.g. to different  available histories across the assets) is to consider ambiguity sets as in \reff{exaTheta}. This is an extension of expected rate of return uncertainty considered in \cite{bental}, \cite{GarUppWan06} for a single period model by allowing an additional ambiguity on correlation and learning on estimation error through the
deterministic level $\delta_l(t)$.

Theoretically, $\Gamma(t)$ can be in the rectangular form $\{\rho \in (-1, 1)^d: \rho_{ij} \in [\hat\rho_{ij}(t)- \epsilon_{ij}(t), \hat\rho_{ij}(t)- \epsilon_{ij}(t)]\}$, where $\hat\rho(t) = (\hat\rho_{ij}(t))_{1 \leq i, j \leq d}$ is estimator of correlation, and $\epsilon_{ij}(t)$ represents uncertainty level around $\hat\rho_{ij}(t)$. This rectangular correlation set falls in $\C_{>+}^d$ with suitable choice of $\epsilon_{ij}(t)$. When $\Gamma(t)$ $=$ $\C^d_{>+}$, this means that  the investor has at time $t$  a full ambiguity about correlation on the $d$-risky assets.  In the opposite case, when
$\Gamma(t)$ is a singleton, this means that the investor knows (or is fully confident about) the value of the correlation at time $t$. Similarly, the case $\delta_l(t)$ $=$ $0$ means that the
mean return vector $b_{J_l}$ for assets in the subclass $J_l$ is known or the investor is fully confident about her estimate at time $t$.
}
\epR
\end{Remark}

\begin{Remark}
{\rm
An interesting extension of our model uncertainty setting would be to consider ambiguity sets that may evolve randomly in time $\Theta(t, \omega)$, e.g., through a factor process or price, and in this case, this would add an additional state variable in the value function and the optimal strategy. For example, in \cite{biaetal17} the threshold  $\delta$ may depend on the current and past asset prices,  which corresponds to an adaptive estimation error from the information flow of the observed asset prices. However, it is not immediate how to extend in this case the weak martingale optimality principle for characterizing the optimal strategy, and whether the separation theorem still holds. 
}
\epR
\end{Remark}

 We denote by $\Vc_\bTheta$ the set of $\F$-progressively measurable processes $\theta_.$ $=$ $(\theta_t)_t$ $=$ $(b_t,\rho_t)_t$ $=$ $(b_.,\rho_.)$ valued in $\bTheta$,
in the sense that  
$\theta_t$ is valued in $\Theta(t)$, $0\leq t\leq T$,  and introduce the set of probability measures $\Pc^\bTheta$:
 \beqs
 \Pc^\bTheta & = & \{ \P^{\theta_.} : \theta_. \in \Vc_\bTheta\},
 \enqs
 where $\P^{\theta_.}$ is the probability measure on $(\Omega,\Fc)$ s.t. $B$ is a semimartingale on $(\Omega,\Fc,\P^{\theta_.})$ with absolutely continuous characteristics (w.r.t. the Lebesgue
 measure $dt$) $(b_.,\Sigma(\rho_.))$.  The probabilities $\P^{\theta_.}$ are in general non-equivalent, and actually mutually singular, and we say that a property holds
 $\Pc^\bTheta$-quasi surely
 ($\Pc^\bTheta$-q.s. in short) if it holds $\P^{\theta_.}$-a.s. for all $\theta_.$ $\in$ $\Vc_\bTheta$.

 The price process $S$ $=$ $(S^1,\ldots,S^d)$  of the $d$ risky assets valued in $(0,\infty)^d$  is given by the dynamics
 \beqs
 dS_t &=& {\rm diag}(S_t) dB_t, \;\;\; 0 \leq t\leq T, \; \Pc^\bTheta-q.s.  \\
 &=&  {\rm diag}(S_t)\big( b_t dt + \sigma(\rho_t) dW_t^\theta), \;\;\; \P^{\theta_.}-a.s., \;\; \mbox{ for } \theta_. \; = \; (b_.,\rho_.) \in \Vc_\bTheta,
 \enqs
 where $W^\theta$ is a $d$-dimensional Brownian motion under $\P^{\theta_.}$.  Here ${\rm diag}(S_t)$ is the diagonal matrix with $i$-th element equal to $S_t^i$.
 Notice that in this uncertainty modeling, we allow the unknown drift and correlation to be a priori random process, valued in $\bTheta$.

\subsection{Robust mean-variance problem}

An admissible portfolio strategy $\alpha=(\alpha_t)_{0 \leq t \leq T}$ representing the amount invested in the $d$ risky assets, is a $\R^d$-valued
$\F$-progressively measurable process, satisfying the integrability condition
\beq \label{alphauniform}
\sup_{\P^{\theta_.} \in \Pc^\bTheta} \E_{\theta_.}\Big[ \int_0^T|\alpha_t\trans b_t | dt \;
+ \; \int_0^T \alpha_t\trans \Sigma(\rho_t) \alpha_t dt \Big] \; < \;  \infty,
\enq
and denoted by $\alpha \in \Ac$. Hereafter,  $\E_{\theta_.}$ denotes the expectation under $\P^{\theta_.}$.
This integrability condition \reff{alphauniform} ensures that ${\rm diag}(S)^{-1}\alpha$ is $S$-integrable under any $\P$ $\in$ $\Pc^\bTheta$.
For  a portfolio strategy $\alpha \in \Ac$, and an initial capital $x_0 \in \R$, the dynamics of the self-financed wealth process is driven by
\beq
dX_t^\alpha & = & \alpha_t\trans {\rm diag}(S_t)^{-1} dS_t \; = \;  \alpha_t\trans dB_t, \;\;\; 0 \leq t\leq T, \; X_0^\alpha = x_0, \; \Pc^\bTheta-q.s. \nonumber \\
&=&   \alpha_t\trans \big(  b_t dt +   \sigma(\rho_t) dW_t^\theta\big), \;\; 0 \leq t \leq T, \; X_0^\alpha =x_0 \in \R, \;\; \P^{\theta_.}-a.s. \label{Xalphadyna}
\enq
for all $\theta_.$ $=$ $(b_.,\rho_.)$ $\in$ $\Vc_\bTheta$.

Given a risk aversion parameter $\lambda$ $>$ $0$, the worst-case mean-variance functional under ambiguous drift and correlation is
\beqs
J_{wc}(\alpha) &=& \inf_{\P^{\theta_.}\in\Pc^\bTheta} \Big( \E_{\theta_.} [X_T^\alpha]  - \lambda {\rm Var}_{\theta_.}(X_T^\alpha) \Big) \; < \; \infty, \; \alpha \in \Ac,
\enqs
where ${\rm Var}_{\theta_.}(.)$ denotes the variance under $\P^{\theta_.}$, and the robust mean-variance portfolio selection is formulated as
\beq\label{robustMV}
\left\{
\begin{array}{rcl}
V_0 & := & \Sup_{\alpha\in\Ac} J_{wc}(\alpha)  \; = \;  \Sup_{\alpha\in\Ac} \inf_{\theta_.\in\Vc_\bTheta} J(\alpha,\theta_.) \\
\end{array}
\right.
\enq

Notice that problem \reff{robustMV} is a non standard stochastic differential game due to the pre\-sence of the variance term in the criterion, which prevents the use of classical control method by dynamic programming or maximum principle.  We complete this section by recalling  the solution to the mean-variance problem when there is no ambiguity on the model parameters, and which will serve later as benchmark for comparison when studying the  uncertainty case.

\subsection{Case of no model uncertainty} \label{nouncertainty}

When $\Theta(t)$ $=$ $\{\theta^o(t) = (b^o(t),\rho^o(t)) \}$ is a singleton for any $t$ $\in$ $[0,T]$, we are reduced to the Black-Scholes model with time varying deterministic drift $b^o(t)$,
deterministic co\-variance matrix $\Sigma^o(t)$ $:=$ $\Sigma(\rho^o(t))$, volatility $\sigma^o(t)$ $:=$ $\sigma(\rho^o(t))$,
and deterministic risk premium $R^o(t)$ $:=$ $R(\theta^o(t))$.  In this case, two notions of strategy are adopted, see e.g. \cite{ZhouLi2000,FisLiv2016, PhaWei2017} for pre-committed strategy, which we consider here, and e.g. \cite{HuJinZhou2012, BjoKhaMur2017} for equilibrium.
It is known that the optimal mean-variance strategy is given in feedback form by
\beqs
\alpha_t^* &=& \Lambda^o(X_t^*) (\Sigma^o(t))^{-1}b^o(t) ,\;\;\; 0 \leq t \leq T,
\enqs
where $X^*$ is the wealth process associated to $\alpha^*$,  and $\Lambda^o(X_t^*)> 0$ with 
\beqs
\Lambda^o(x): =  x_0 + \frac{e^{\int_0^T R^o(t) dt}}{2\lambda} -  x, \;\;\; x \in \R,
\enqs
while the optimal performance value is
\beqs
V_0 &=& x_0 + \frac{1}{4\lambda}\big[ e^{\int_0^T R^o(t) dt} - 1 \big].
\enqs
The vector $(\Sigma^o(t))^{-1} b^o(t)$, which depends only on the model parameters of the risky assets, determines the allocation in the risky assets.  The above expression of
$\alpha^*$ shows that, once we know the exact values of the rate of return and covariance matrix, one diversifies  her portfolio among all the assets according to the components of the vector $(\Sigma^o(t))^{-1} b^o(t)$,
and this is weighted by the scalar term $\Lambda^o(X_t^*)$, which depends on the risk aversion of the investor via the parameter $\lambda$, on  the current wealth but also on the initial capital $x_0$ (which is sometimes refereed to as the pre-commitment of the mean-variance criterion). Notice that
$\Lambda^o(X_t^*)$ is positive. Indeed, observe that
\beqs
d\Lambda^o(X_t^*) \; = \; - dX_t^* &=& -(\alpha_t^*)\trans(b^o(t) dt + \sigma^o(t) dW_t^o) \\
&=& - \Lambda^o(X_t^*) (R^o(t) dt + \big((\sigma^o(t))^{-1}b^o(t)\big)\trans dW_t^o), \;\;\; 0 \leq t \leq T,
\enqs
with $\Lambda^o(X_0^*)$ $=$ $\frac{1}{2\lambda}e^{\int_0^T R^o(t) dt}$ $>$ $0$, which shows clearly that $\Lambda^o(X_t^*)$ $>$ $0$, $0\leq t\leq T$, and
decreases with $\lambda$.

Let us discuss in particular  the allocation in the two-asset case. Notice that the vector $(\Sigma^o(t))^{-1} b^o(t)$ of allocation is then given by
\beqs
(\Sigma^o(t))^{-1} b^o(t) &=& \frac{1}{1- |\rho^o(t)|^2}
\left( \begin{array}{c}
\frac{\beta_1^o(t) - \rho^o(t) \beta_2^o(t)}{\sigma_1} \\
\frac{\beta_2^o(t) - \rho^o(t) \beta_1^o(t)}{\sigma_2}
\end{array}
\right) \; =: \; \left( \begin{array}{c} \kappa_1^o(t) \\  \kappa_2^o(t) \end{array} \right),
\enqs
where $\beta^o_i(t)$ $=$ $b^o_i(t)/{\sigma_i}$ is the 
Sharpe ratio of the $i$-th asset, $i$ $=$ $1,2$, at time $t$. To fix the idea, assume that
$\beta_1^o(t)$ $>$ $\beta_2^o(t)$ $>$ $0$. We then see that $\kappa_1^o(t)$  $>$ $0$, while $\kappa_2^o(t)$ $\geq$ $0$ if and only if
$\frac{\beta_2^o(t)}{\beta_1^o(t)}$ $\geq$ $\rho^o(t)$. The interpretation is the following: the ratio $\frac{\beta_2^o(t)}{\beta_1^o(t)}$ $\in$ $(0,1)$ measures the
``proximity" in terms of 
Sharpe ratio between the two assets,
and has to be compared with the correlation $\rho^o(t)$ between these assets in order to determine whether it is optimal to invest according to a
directional trading,  i.e.,  $\kappa_1^o(t)\kappa_2^o(t)$ $>$ $0$ (thus here long in both assets) or according to a spread trading, i.e., $\kappa_1^o(t)\kappa_2^o(t)$ $<$ $0$ (long in the first asset and short in the second one) or according to under-diversification, i.e., $\kappa_1^o(t)\kappa_2^o(t)$ $=$ $0$ (only long in the first asset). Notice that  under-diversification only occurs when $\rho^o(t)$ $=$ $\frac{\beta_2^o(t)}{\beta_1^o(t)}$, a condition ``rarely"   satisfied in practice.
For example, when both assets have close 
Sharpe ratios, and their correlation is not too high, then one optimally invests in both assets with a directional trading.
In contrast, when one asset has a much larger 
Sharpe ratio than the other one, or when the correlation between the assets is high, then  one optimally invests in both assets with a spread trading.

\vspace{3mm}

In the sequel, we  study the quantitative impact of the drift and correlation uncertainty 
on the optimal robust mean-variance strategy, in particular regarding the portfolio diversification.

\medskip

\section{Separation principle and robust solution}

\setcounter{Theorem}{0} \setcounter{Proposition}{0}
\setcounter{Corollary}{0} \setcounter{Lemma}{0}
\setcounter{Definition}{0} \setcounter{Remark}{0}

The main result of this section is to state a separation principle for solving the robust dynamic mean-variance problem.

\begin{Theorem}
[Separation Principle]
\label{robustoptimal}
Let us consider a parametric set $\bTheta$ for model uncertainty as in {\bf (H$\bTheta$)}.
Suppose that  there exists a pair $\btheta^*= (\theta^*(t))_t = (\bb^*,\brho^*) = (b^*(t),\rho^*(t))_t$ $\in$  $\bTheta$ solution to
$\arg\Min_{\btheta \in \bTheta} \bR(\btheta)$, i.e., $\theta^*(t)$ $\in$ $\arg\Min_{\theta \in \Theta(t)} R(\theta)$, for all $t$ $\in$ $[0,T]$.
Then the robust mean-variance problem \reff{robustMV} admits an optimal portfolio strategy given  in feedback form  by
\beq
\alpha_t^*
&  =  &  \Lambda^*(X_t^*)\Sigma(\rho^*(t))^{-1}b^*(t), \;\;\; 0 \leq t \leq T, \;\;  \Pc^\bTheta-q.s., \label{robustalpha}
\enq
where $X^*$ is the state process associated to $\alpha_t^*$, and $\Lambda^*(X_t^*)$ $>$ $0$ with
\beq \label{Lambdatheta*}
\Lambda^*(x) \; := \; x_0 +\frac{e^{\int_0^TR(\theta^*(s))ds}}{2\lambda}  - x,  \;\; x \in \R.
\enq
Moreover, the corresponding initial value function is
\beqs
V_0 &=& x_0 +  \frac{1}{4\lambda}\big[ e^{\int_0^T R(\theta^*(s))ds}-1\big].
\enqs
\end{Theorem}

\noindent {\bf Interpretation.}  Theorem \ref{robustoptimal} means that the robust mean-variance problem \reff{robustMV}  can be solved in two steps according to a separation principle.
(i) First, at each time $t$ $\in$ $[0,T]$, we search for the infimum of the risk premium function $\theta$ $\in$ $\Theta(t)$ $\mapsto$ $R(\theta)$ as defined in \reff{defpremium}, which depends only on the inputs of the uncertainty model.  Existence and explicit determination of an element $\btheta^*$ $=$ $(\bb^*,\brho^*)$  $\in$ $\bTheta$
attaining this infimum will be discussed and illustrated all along the paper through several examples.
(ii) The solution to  \reff{robustMV} is then given by the solution to the mean-variance problem in the Black-Scholes model with time varying deterministic drift $b^*(t)$ and correlation $\rho^*(t)$, see Section \ref{nouncertainty}, and the worst-case scenario of the robust dynamic mean-variance problem is simply given by the family of deterministic parameters
$\btheta^*$ $=$ $(\bb^*,\brho^*)$. Some interesting features show up, especially regarding portfolio diversification, as detailed in the next section.
\epR

\begin{Remark}
{\rm The existence of the solution $\btheta^*$ to $\arg\Min_{\btheta \in \bTheta} \bR(\btheta)$  is guaranteed under \reff{exaTheta} 
whenever the ambiguity sets
$\Gamma(t)$, $t$ $\in$ $[0,T]$, on correlation are compact as the risk premium function $R$ is continuous. Since we also want to consider the case of full ambiguity on correlation, i.e.,  when  $\Gamma(t)$ $=$ $\C^d_{>+}$, which is an open set, we do not impose such compactness condition.
}
\epR
\end{Remark}

\begin{Remark}[Relation with static model]
{\rm
Actually, the worst-case scenario of static robust mean-variance problem is also determined by the minimal risk premium. By analogue with model uncertainty described as in Section \ref{subsecmodel}, we characterize ambiguity about the rate of return $b$ and correlation $\rho$ as a bounded convex set, i.e. $\theta$ $=$ $(b, \rho)$ $\in$ $\Theta$, and formulate the single-period robust mean-variance problem under ambiguity 
as
\beqs
\Sup_{\alpha \in \R^d}\Inf_{\theta \in \Theta} \Big(\alpha\trans b -  \lambda \alpha\trans \Sigma(\rho)\alpha\Big) \;=:\; \Sup_{\alpha \in \R^d}\Inf_{\theta\in \Theta} J_{static}(\alpha, \theta).
\enqs
where $\alpha$ is the portfolio held in  $d$ risky assets.
Assume that there exists $\theta^*$ $=$ $(b^*, \rho^*)$ $\in$ $\arg\Min_{\theta \in \Theta} R(\theta)$, we then obtain from Lemma \ref{lemsaddle2} in Appendix that $(\alpha^*, \theta^*)$ with $\alpha^*$ $:=$ $\frac{1}{2\lambda} \Sigma(\rho^*)^{-1}b^*$ is a saddle point of $J(\alpha, \theta)$, and
\beqs
\Sup_{\alpha \in \R^d}\Inf_{\theta \in \Theta} J_{static}(\alpha, \theta) &=& \Inf_{\theta \in \Theta} \Sup_{\alpha \in \R^d}J_{static}(\alpha, \theta) \;=\; J_{static} (\alpha^*, \theta^* ) \;=\;\frac{1}{4\lambda}R(\theta^*).
\enqs
Therefore, the key point in all single period, multi-period and continuous-time models is the computation of infimum of risk premium, which will be discussed in the next section.
}
\epR
\end{Remark}

\vspace{3mm}

The rest of this section is devoted to the proof of Theorem \ref{robustoptimal}, and the methodology is based on the following weak version of the martingale optimality principle.
The usual martingale optimality principle is introduced in \cite{Elkar1981} for optimal stopping problems, and later applied in \cite{RouElK2000, HuImkMul05} for utility maximization, in \cite{MatPosZho15} for robust utility maximization under the uncertain volatility  model, and in \cite{YLZ2018} for robust portfolio-consumption strategies with uncertainty on both drift and volatility, to name a few.
A comparison of the usual martingale optimality principle and our weak version is given in Remark \ref{remmarU} below.

\begin{Lemma}[Weak optimality principle]\label{MOP}
Let $\{V_t^{\alpha, \theta_.}, t \in [0, T], \alpha \in \Ac, \theta_. \in \Vc_\bTheta\}$ be a family of real-valued processes in the form
\beqs
V_t^{\alpha, \theta_.}: &=& v_t(X_t^\alpha, \E_{\theta_.}[X_t^\alpha]),
\enqs
for some measurable functions $v_t$ on $\R \times \R$, $t$ $\in$ $[0,T]$, such that :
\begin{itemize}
\item [(i)] $v_T(x, \bar x)$ $=$ $x-\lambda(x -\bar x)^2$, for all $x, \bar x$ $\in$ $\R$,
\item  [(ii)] the function $t$ $\in$ $[0,T]$ $\mapsto$ $\E_{\theta_.^*}[V_t^{\alpha, \theta_.^*}]$
is nonincreasing for all $\alpha$ $\in$ $\Ac$ and some $\theta_.^*$ $\in$ $\Vc_\bTheta$,
\item [(iii)]  $\E_{\theta_.}[V_T^{\alpha^*, \theta_.}-V_0^{\alpha^*, \theta_.}]$ $\geq$ $0$,  for some $\alpha^*$ $\in$ $\Ac$ and all $\theta_.$ $\in$ $\Vc_\bTheta$.
\end{itemize}
Then, $\alpha^*$ is an optimal portfolio strategy for the robust mean-variance problem \reff{robustMV} with a worst-case scenario $\theta_.^*$, and
\beq
V_0 & = & J_{wc}(\alpha^*)\; = \; \Sup_{\alpha \in \Ac}\Inf_{\theta_. \in \Vc_\bTheta}J(\alpha, \theta_.) \; = \;
\Inf_{\theta_. \in \Vc_\bTheta}\Sup_{\alpha \in \Ac}J(\alpha, \theta_.) \; = \; v_0(x_0, x_0) \label{robustMVV0} \\
&=&   J(\alpha^*, \theta_.^*).  \nonumber
\enq
\end{Lemma}
{\bf Proof.}\; First, observe that $V_0^{\alpha, \theta_.}$ $=$ $v_0(x_0, x_0)$ is a constant that does not depend on $\alpha$, $\theta_.$, and from condition (i)
that $\E_{\theta_.}[V_T^{\alpha, \theta_.}]$ $=$ $J(\alpha, \theta_.)$ for all $\alpha \in \Ac$, $\theta_. \in \Vc_\bTheta$. Then, from condition (ii), we see that
\beqs
v_0(x_0, x_0) \; = \; \E_{\theta_.^*}[V_0^{\alpha, \theta_.^*}] &\geq&  \E_{\theta_.^*}[V_T^{\alpha, \theta_.^*}] \; = \; J(\alpha, \theta_.^*),
\enqs
for all $\alpha \in \Ac$, and thus $v_0(x_0, x_0)$ $\geq$ $\Sup_{\alpha \in \Ac}J(\alpha, \theta_.^*)$ $\geq$ $\Inf_{\theta_. \in \Vc_\bTheta}\Sup_{\alpha \in \Ac}J(\alpha, \theta_.)$. Similarly,  from condition (iii), we have $v_0(x_0, x_0)$ $\leq$ $J(\alpha^*, \theta_.)$ for all $\theta_.$ $\in$ $\Vc_\bTheta$, and thus $v_0(x_0, x_0)$ $\leq$
$\Inf_{\theta_. \in \Vc_\bTheta} J(\alpha^*, \theta_.)$ $=$ $J_{wc}(\alpha^*)$ $\leq$ $\Sup_{\alpha \in \Ac}\Inf_{\theta_. \in \Vc_\bTheta} J(\alpha, \theta_.)$. Recalling that we always have $\Sup_{\alpha \in \Ac}\Inf_{\theta_. \in \Vc_\bTheta} J(\alpha, \theta_.)$ $\leq$ $\Inf_{\theta_. \in \Vc_\bTheta} \Sup_{\alpha \in \Ac} J(\alpha, \theta_.)$, we obtained the required equality in \reff{robustMVV0}. Then, finally,  from (ii) with $\alpha^*$ and (iii) with
$\theta_.^*$, we  obtain that $v_0(x_0, x_0)$ $=$ $J(\alpha^*, \theta_.^*)$.
\ep

\begin{Remark} \label{remmarU}
{\rm The usual martingale optimality principle  for stochastic differential games as in robust portfolio selection problem, and with classical expected utility criterion for some nondecreasing and concave utility function $U$ on $\R$, e.g.,  $U(x)$ $=$ $-e^{-\eta x}$, $\eta$ $>$ $0$:
\beqs
\sup_{\alpha\in\Ac}\inf_{\theta_.\in\Vc_\bTheta}\E_{\theta_.}[U(X_T^\alpha)],
\enqs
would consist of finding a family of processes $V_t^{\alpha,\theta_.}$ in the form $v_t(X_t^\alpha)$ for some measurable  functions $v_t$ on $\R$
s.t. (i) $v_T(x)$ $=$ $U(x)$, (ii') the process  $(V_t^{\alpha,\theta_.^*})_t$ is a supermartingale under $\P_{\theta_.^*}$ for all $\alpha$ and some $\theta_.^*$, and (iii') the process
$(V_t^{\alpha^*,\theta_.})_t$ is a submartingale under $\P_{\theta_.}$ for some $\alpha^*$ and all $\theta_.$.  Due to the nonlinear dependence on the law of the state wealth process via the variance term in the mean-variance criterion, making the problem a {\it priori} time inconsistent, we have to adopt a weaker version of the optimality principle.  First, the functions $v_t$ depend not only on the state process $X_t^\alpha$ but also on its mean $\E_{\theta_.}[X_t^\alpha]$. Second, we replace condition
(ii') by the weaker condition (ii) on the mean in Lemma \ref{MOP}. Third, condition (iii') is substituted by the weaker condition (iii), which is even weaker than (iii'') $t$ $\mapsto$
$\E_{\theta_.}[V_t^{\alpha^*, \theta_.}]$  is nondecreasing for  some $\alpha^*$ and all $\theta_.$.
This asymmetry of condition between (ii) and (iii) is explained in more detail in  Remark \ref{remdis}.
}
\epR
\end{Remark}

\vspace{2mm}

We shall also use the following property  on the infimum of the risk premium function.


\begin{Lemma} \label{lemH}
Given $\bTheta$ as in {\bf (H$\Theta$)}, and assuming that there exists $\btheta^*$ $=$ $(\bb^*, \brho^*)$ $\in$ $\arg\Min_{\btheta \in \bTheta} \bR(\btheta)$, let us define the function $H_t$ on $\Theta(t)$, $t$ $\in$ $[0, T]$, by
\beq \label{defH}
H_t(\theta) &:= & b\trans \Sigma(\rho^*(t))^{-1} \Sigma(\rho) \Sigma(\rho^*(t))^{-1}b^*(t),\;\;\; \mbox{ for } \theta = (b, \rho) \in \Theta(t).
\enq
Then, we have for all $(b, \rho)$ $\in$ $\Theta(t)$:
\beq \label{inegH}
R(\theta^*(t)) - 2H_t(b, \rho^*(t)) + H_t(b^*(t), \rho) & \leq & 0.
\enq
\end{Lemma}
{\bf Proof.}
See Section \ref{appenlemH} in Appendix.

\vspace{3mm}

In the following, we provide details of the proof for Theorem \ref{robustoptimal} using Lemma \ref{MOP} and Lemma \ref{lemH}.

\noindent {\bf Proof of Theorem \ref{robustoptimal}}. We aim to construct a family of processes $\{V_t^{\alpha, \theta_.}, t \in [0, T], \alpha \in \Ac, \theta_. \in \Vc_\bTheta\}$ as in Lemma \ref{MOP}, and given the linear-quadratic structure of our optimization problem, we look for measurable functions $v_t$ in the form:
\beq\label{linquav0}
v_t(x, \bar x) &= & K_t(x -\bar x)^2 +  Y_t x + \chi_t, \;\;\; t \in [0, T], (x, \bar x) \in \R^2,
\enq
for some deterministic processes $(K_t, Y_t, \chi_t)_t$ to be determined. Condition (i) in Lemma \ref{MOP} fixes  the terminal condition
\beq \label{termK}
K_T \; = \; - \lambda, \;\; Y_T \; = \; 1, \;\; \chi_T \;= \; 0.
\enq

We now consider $\btheta^*$ $\in$ $\bTheta$ as in Theorem \ref{robustoptimal}, hence defining in particular a deterministic process $\btheta^*$ $=$ $(\theta^*(t))_t$
$\in$ $\Vc_\bTheta$,
and $\alpha^*$  given by \reff{robustalpha}. Let us first check that $\alpha^*$ $\in$ $\Ac$.
The corresponding wealth process $X^*$ satisfies under any $\P^{\theta_.}$, $\theta_.$ $=$ $(b_.,\rho_.)$  $\in$ $\Vc_\bTheta$, a linear stochastic differential equation with bounded random coefficients (notice that $b_.$ and $\sigma(\rho_.)$  are bounded processes), and thus by standard estimates: $\E_{\theta_.}\big[ \sup_{0\leq t\leq T} |X_t^*|^2]$ $\leq$
$C(1+|x_0|^2)$ for some constant $C$ independent of $\theta_.$ $\in$ $\Vc_\bTheta$.  It follows immediately that $\alpha^*$ satisfies the integrability condition in \reff{alphauniform}, i.e., $\alpha^*$ $\in$ $\Ac$.

The main issue now is to show that such a pair $(\alpha^*,\theta_.^*)$ satisfies conditions (ii)-(iii) of Lemma \ref{MOP}.

\vspace{2mm}

\noindent $\bullet$ {\it Step 1. condition (ii) of Lemma \ref{MOP}.}

\noindent For any $\alpha$ $\in$ $\Ac$, with associated wealth process $X$ $:=$ $X^\alpha$,
let us compute the derivative of the deterministic function $t$ $\mapsto$ $\E_{\btheta^*}[V_t^{\alpha,\btheta^*}]$ $=$
$\E_{\btheta^*}[v_t(X_t^{},\E_{\btheta^*}[X_t])]$ with $v_t$ as in \reff{linquav0}.
From the dynamics of $X$ $=$ $X_t^\alpha$ in \reff{Xalphadyna} under $\P^{\btheta^*}$ and by applying It\^o's formula, we obtain
\beqs
\label{expX}
\frac{d \E_{\btheta^*}[X_t]}{dt}  &=&\E_{\btheta^*}[\alpha_t\trans  b^*(t)],  \\
\frac{d{\rm Var}_{\btheta^*}(X_t)}{dt} &=& 2{\rm Cov}_{\btheta^*}(X_t, \alpha_t\trans b^*(t)) + \E_{\btheta^*}[\alpha_t\trans\Sigma(\rho^*(t))\alpha_t].
\enqs
From the quadratic form of $v_t$ in \reff{linquav0}, with $(K,Y,\chi)$ differentiable in time, we then have
\beq \label{dVt}
\frac{d\E_{\btheta^*}[V_t^{\alpha, \btheta^*}]}{dt}
&=& \frac{d\E_{\btheta^*} [v_t(X_t, \E_{\btheta^*}[X_t])]}{dt} \nonumber \\
&=& \dot K_t {\rm Var}_{\btheta^*}(X_t) + K_t\frac{d{\rm Var}_{\btheta^*}(X_t^{})}{dt}
+  \dot Y_t \E_{\btheta^*}[X_t] + Y_t\frac{d\E_{\btheta^*}[X_t]}{dt} +\dot \chi_t \nonumber\\
&=& \dot K_t {\rm Var}_{\btheta^*}(X_t)  + \dot Y_t \E_{\btheta^*}[X_t] + \dot \chi_t + \E_{\btheta^*}[G_t(\alpha)],
\enq
where  $\dot K_t$, $\dot Y_t$ and $\dot \chi_t$ represent the time derivatives of $K_t$, $Y_t$ and $\chi_t$ respectively, and
\beqs
G_t(\alpha) &:=& \alpha_t\trans Q_t \alpha_t + \alpha_t\trans \big[2U_t(X_t -\E_{\btheta^*}[X_t]) + O_t\big],
\enqs
with the deterministic coefficients
\beqs
Q_t \; = \; K_t \Sigma(\rho^*(t)), \;\; \; U_t \; = \; K_t b^*(t), \;\;\; O_t \;=\; Y_t b^*(t).
\enqs
By square completion, we rewrite $G_t(\alpha)$ as
\beqs
G_t(\alpha) &=&  \big(\alpha_t -\hat a_t(X_t,\E_{\btheta^*}[X_t])\big)\trans Q_t \big(\alpha_t - \hat a_t(X_t,\E_{\btheta^*}[X_t]) \big)-\zeta_t,
\enqs
where for $t$ $\in$ $[0,T]$, $x,\bar x$ $\in$ $\R^2$,
\beqs
\hat a_t(x,\bar x) &:= &  -Q_t^{-1}U_t(x  - \bar x ) -  \frac{1}{2} Q_t^{-1}O_t,
\enqs
and
\beqs
\zeta_t &:=&
U_t\trans Q_t^{-1} U_t {\rm Var}_{\btheta^*}(X_t) + \frac{1}{4}O_t\trans Q_t^{-1}O_t
\; = \;  K_t R(\theta^*(t)) {\rm Var}_{\btheta^*}(X_t) + \frac{Y_t^2}{4 K_t} R(\theta^*(t)).
\enqs
The expression in \reff{dVt} is then rewritten as
\beq
\frac{d\E_{\btheta^*}[V_t^{\alpha, \btheta^*}]}{dt} &=&
(\dot K_t -K_t R(\theta^*(t))){\rm Var}_{\btheta^*}(X_t) +  \dot Y_t  \E_{\btheta^*}[X_t]  + \;\dot \chi_t -  \frac{Y_t^2}{4 K_t} R(\theta^*(t))  \label{derivEV} \\
& & \;\;\;  + \;  K_t  \E_{\btheta^*}\big[\big(\alpha_t -\hat a_t(X_t,\E_{\btheta^*}[X_t])\big)\trans \Sigma(\rho^*(t))  \big(\alpha_t -\hat a_t(X_t,\E_{\btheta^*}[X_t]) \big) \big]. \nonumber
\enq
Therefore, whenever
\begin{equation} \label{sysK}
\left\{
\begin{array}{rcl}
\dot K_t -K_t R(\theta^*(t)) & = & 0, \\
\dot Y_t  &=& 0, \\
\dot \chi_t - \frac{Y_t^2}{4 K_t} R(\theta^*(t))  &=& 0,
\end{array}
\right.
\end{equation}
holds for all $t$ $\in$ $[0, T]$, which yields, together with the terminal condition \reff{termK}, the explicit forms:
\beq \label{GamLamYchi}
K_t \; = \; - \lambda e^{\int_t^T R(\theta^*(s))ds} \; < \; 0,  \;\; Y_t \; = \; 1, \;\;  \chi_t \; = \;  \frac{1}{4\lambda}\big[e^{\int_t^T R(\theta^*(s))ds}-1\big],
\enq
we have
\beqs
\frac{d\E_{\btheta^*}[V_t^{\alpha, \btheta^*}]}{dt} &=&
K_t  \E_{\btheta^*}\big[\big(\alpha_t -\hat a_t(X_t,\E_{\btheta^*}[X_t])\big)\trans \Sigma(\rho^*)  \big(\alpha_t -\hat a_t(X_t,\E_{\btheta^*}[X_t]) \big) \big],
\enqs
which is nonpositive for all $\alpha$ $\in$ $\Ac$, i.e., the process $V_t^{\alpha, \btheta^*}$ satisfies the condition (ii) of Lemma \ref{MOP}.  Moreover, notice that in this case,
\beq \label{v0}
V_0^{\alpha,\theta^*} \; = \; v_0(x_0,x_0) &= & x_0 +  \frac{1}{4\lambda}\big[e^{\int_0^T R(\theta^*(t))dt}-1\big],
\enq
and
\beq \label{hataopt}
\hat a_t(x,\bar x) &=& - \Sigma(\rho^*(t))^{-1}b^*(t) \big(x  - \bar x  -  \frac{1}{2\lambda} e^{\int_t^T R(\theta^*(s))ds} \big).
\enq
Notice that in this step, we have not yet used the property that $\btheta^*$ attains the infimum of the risk premium function. This will be used in the next step.

\vspace{2mm}

\noindent $\bullet$ {\it Step 2. condition (iii) of Lemma \ref{MOP}.}

\noindent   Let us now prove that $V_0^{\alpha^*,\theta_.}$ $\leq$ $\E_{\theta_.}[V_T^{\alpha^*,\theta_.}]$, for all $\theta_.$ $\in$ $\Vc_\bTheta$.  A sufficient condition is  the nondecreasing monotonicity of the function $t$ $\mapsto$ $\E_{\theta_.}[V_t^{\alpha^*,\theta_.}]$, by proving that $\frac{d\E_{\theta_.}[V_t^{\alpha^*, \theta_.}]}{dt}$ is nonnegative, for all  $\theta_.$ $\in$ $\Vc_\bTheta$.
However, while this non\-decreasing property is valid when there is no uncertainty on the drift, this does not hold true in the general uncertainty case as shown in Remark \ref{remdis}.
We then proceed by computing directly the difference: $\E_{\theta_.}[V_T^{\alpha^*,\theta_.}]$ $-$ $V_0^{\alpha^*,\theta_.}$. Notice from \reff{robustalpha}, \reff{Xalphadyna}, that the dynamics of $\Lambda^*(X^*)$, with $\Lambda^*(x)$ defined in \reff{Lambdatheta*},  under $\P^{\theta_.}$, $\theta_. \in \Vc_\bTheta$, is given by
\beqs
d \Lambda^*(X_t^*)  &=& - \Lambda^*(X_t^*) (b^*(t))\trans\Sigma(\rho^*(t))^{-1} \big[ b_t dt + \sigma(\rho_t) dW_t^\theta \big],
\enqs
with $\Lambda^*(x_0)$ $=$ $\frac{e^{\int_0^T R(\theta^*(t)) dt}}{2\lambda}$. By setting $N_t^*$ $:=$ $\frac{2\lambda}{e^{\int_0^T R(\theta^*(t)) dt}} \Lambda^*(X_t^*)$, we deduce that
\beqs
N_t^* &=& \exp\Big( -\int_0^t \big( b_s\trans \Sigma(\rho^*(t))^{-1} b^*(t)   + \frac{1}{2}  (b^*(t))\trans \Sigma(\rho^*(t))^{-1}\Sigma(\rho)\Sigma(\rho^*(t))^{-1} b^*(t) \big) ds  \\
& & \hspace{2.5cm}   - \; \int_0^t (b^*(t))\trans \Sigma(\rho^*(t))^{-1}\sigma(\rho_s) dW_s^\theta \Big), \;\;\; 0 \leq t \leq T, \; \P^{\theta_.}-a.s. \\
X_t^* &=& x_0 + \frac{e^{\int_0^T R(\theta^*(t))dt}}{2\lambda} (1 - N_t^*), \;\;\; 0 \leq t \leq T, \; \Pc^\bTheta-q.s.,
\enqs
and  thus
\begin{equation} \label{Xt*expvar}
\left\{
\begin{array}{ccl}
\E_{\theta_.}[X_t^*] & = & x_0 + \frac{e^{\int_0^T R(\theta^*(t))dt}}{2\lambda}(1 -\E_{\theta_.}[N_t^*]), \\
{\rm Var}_{\theta_.}(X_t^*) & = & \frac{e^{2 \int_0^T R(\theta^*(t))dt}}{4\lambda^2} {\rm Var}_{\theta_.}(N_t^*).
\end{array}
\right.
\end{equation}
By using the quadratic form \reff{linquav0} of $v_t$, together with the terminal condition \reff{termK}, \reff{v0}, and  \reff{Xt*expvar}, we then  obtain for all $\theta_.$
$\in$ $\Vc_\bTheta$:
\begin{align}
\E_{\theta_.}[V_T^{\alpha^*,\theta_.}] - V_0^{\alpha^*, \theta_.} &= \E_{\theta_.}\big[ v_T(X_T^*,\E_{\theta_.}[X_T^*]) \big] - v_0(x_0,x_0)  \nonumber \\
&= - \lambda {\rm Var}_{\theta_.}(X_T^*) +  \E_{\theta_.}[X_T^*]  -  x_0 - {1 \over4\lambda}(e^{\int_0^T R(\theta^*(t))dt}-1) \nonumber \\
&= - \frac{e^{2 \int_0^T R(\theta^*(t)) dt}}{4\lambda} {\rm Var}_{\theta_.}(N_T^*) + \frac{e^{\int_0^T R(\theta^*(t))dt}}{2\lambda}(1 -\E_{\theta_.}[N_T^*]) - {1 \over4\lambda}
(e^{\int_0^T R(\theta^*(t))dt}-1)  \nonumber \\
&=   \frac{e^{\int_0^T R(\theta^*(t)) dt}}{4\lambda}  \Big(  1 - e^{\int_0^T R(\theta^*(t)) dt} \E_{\theta_.}[ |N_T^*|^2]  \Big)
+ \frac{1}{4\lambda} \Big( e^{\int_0^T R(\theta^*(t)) dt} \E_{\theta_.}[N_T^*] - 1 \Big)^2  \nonumber \\
& \geq    \frac{e^{\int_0^T R(\theta^*(t)) dt}}{4\lambda}  \Big(  1 - e^{\int_0^T R(\theta^*(t)) T} \E_{\theta_.}[ |N_T^*|^2]  \Big)
\; = : \;  \frac{e^{\int_0^T R(\theta^*(t)) T}}{4\lambda} \Delta_T^*(\theta_.).  \label{Delta}
\end{align}

Noting that $N^*$ is rewritten in terms of $H$ introduced in Lemma \ref{lemH} as
\beqs \label{nut}
N_t^* &=&  \exp\Big( -\int_0^t \big( H_s(b_s,\rho^*(s))   + \frac{1}{2}  H_s(b^*(s), \rho_s) \big) ds
- \; \int_0^t (b^*(s))\trans \Sigma(\rho^*(s))^{-1}\sigma(\rho_s) dW_s^\theta \Big),
\enqs
for $t$ $\in$ $[0,T]$,   $\P^{\theta_.}-a.s.$,  and observing that $|(b^*(s))\trans \Sigma(\rho^*(s))^{-1}\sigma(\rho_s)|^2$ $=$ $H_s(b^*(s),\rho_s)$, we see that
\beqs
|N_t^*|^2 &=& \exp\Big( - \int_0^t \big( 2 H_s(b_s,\rho^*(s))  - H(b^*(s),\rho_s) \big) ds \Big) M_t^*,
\enqs
where
\beqs
M_t^* & := & \exp\Big( - 2 \int_0^t  |(b^*(s))\trans \Sigma(\rho^*(s))^{-1}\sigma(\rho_s)|^2 ds -  2 \int_0^t (b^*(s))\trans \Sigma(\rho^*(s))^{-1}\sigma(\rho_s)
dW_s^\theta \Big),
\enqs
is  an exponential Dol\'eans-Dade local martingale under any $\P^{\theta_.}$, $\theta$ $\in$ $\Vc_\bTheta$. Actually, the Novikov criterion is satisfied. Indeed,
\beqs
& & \E_{\theta_.}\Big[ \exp\Big(  \frac{1}{2} \int_0^T|2 (b^*(t))\trans \Sigma(\rho^*(t))^{-1}\sigma(\rho_t)|^2 dt  \Big)\Big]   =
\E_{\theta_.}\Big[ \exp\Big(  2 \int_0^T H_t(b^*(t),\rho_t)  dt  \Big)\Big] \\
&=& \E_{\theta_.}\Big[\exp\Big(2 \int_0^T \kappa(b^*(t), \rho^*(t))\trans\Sigma(\rho_t)\kappa(b^*(t), \rho^*(t))dt\Big)\Big] \\
&=& \E_{\theta_.}\Big[\exp\Big(2 \int_0^T \Sum_{i=1}^d \kappa^i(b^*(t), \rho^*(t))^2 + 2\Sum_{1 \leq i < j \leq d}\rho_{ij, t}\kappa^i(b^*(t), \rho^*(t))\kappa^j(b^*(t), \rho^*(t))dt\Big)\Big] \\
&\leq&  \E_{\theta_.}\Big[\exp\Big(\int_0^T \Sum_{i=1}^d \kappa^i(b^*(t), \rho^*(t))^2 + 2 \Sum_{1 \leq i < j \leq d} |\kappa^i(b^*(t), \rho^*(t))|\kappa^j(b^*(t), \rho^*(t))|\Big)\Big] \\
&=&  \E_{\theta_.}\Big[\exp\Big(2 \int_0^T (\Sum_{i=1}^d |\kappa^i(b^*(t), \rho^*(t))|)^2 dt \Big)\Big] \;  < \; \infty,
\enqs
where the first  inequality comes from the fact that the process $\rho_{ij, t}$, $1$ $\leq$ $i$ $<$ $j$ $\leq$ $d$, is valued in $(-1, 1)$. Therefore,
$(M_t^*)_{0\leq t\leq T}$ is a  martingale under any $\P^{\theta_.}$, $\theta_.$ $\in$ $\Vc_\bTheta$.  Consequently, we have
\beqs
\Delta_T^*(\theta_.) &=& 1 -  \E_{\theta_.} \Big[  \exp\Big(  \int_0^t \big( R(\theta^*(s)) - 2 H_s(b_s,\rho^*(s))  + H_s(b^*(s),\rho_s) \big) ds \Big)  M_T^* \Big]  \\
& \geq & 1 -  \E_{\theta_.} [  M_T^* ] \; = \; 1 - M_0^* \; = \; 0,
\enqs
where we used \reff{inegH} in the above inequality.  From \reff{Delta}, this proves condition (iii) of Lemma \reff{MOP}, and finally concludes the proof of Theorem \ref{robustoptimal}.
\ep

\begin{Remark}\label{remfeedback}
{\rm The optimal strategy  $\alpha^*$ given in \reff{robustalpha} can be expressed in feedback form as
\beq \label{expressalpha*}
\alpha_t^* &=& \hat a_t(X_t^*,\E_{\btheta^*}[X_t^*]), \;\;\;  0 \leq t \leq T, \; \Pc^\bTheta-q.s.,
\enq
where $\hat a_t$ is defined in \reff{hataopt}. Indeed, denoting by $\hat\alpha$ $\in$ $\Ac$ the process defined by
$\hat\alpha_t$ $=$
$\hat a_t(\hat X_t,\E_{\btheta^*}[\hat X_t])$, $0 \leq t \leq T, \; \Pc^\bTheta-q.s.$, where $\hat X$ is the wealth process associated to $\hat\alpha$, we see from
\reff{Xalphadyna} that
$\hat X$ satisfies the dynamics under $\P^{\btheta^*}$:
\beqs
d\hat X_t &=& - \Big[ \hat X_t - \E_{\btheta^*}[\hat X_t] -  \frac{1}{2\lambda} e^{\int_t^T R(\theta^*(s))ds} \Big]
(b^*(t))\trans \Sigma(\rho^*(t))^{-1}\big[ b^*(t) dt + \sigma(\rho^*(t)) dW_t^{\theta^*}].
\enqs
By taking expectation under $\P^{\btheta^*}$, we get: $d\E_{\btheta^*}[\hat X_t]$ $=$ $ \frac{1}{2\lambda} e^{\int_t^T R(\theta^*(s))ds} R(\theta^*)dt$, and thus
\beqs
\E_{\btheta^*}[\hat X_t] &=& x_0 + \frac{e^{\int_0^T R(\theta^*(t))dt}}{2\lambda} \big[  1 -  e^{- \int_0^t R(\theta^*(s))ds} \big], \\
\hat\alpha_t &=&  \Lambda^*(\hat X_t) \Sigma(\rho^*(t))^{-1} b^*(t), \;\;\; 0 \leq t \leq T, \; \Pc^\bTheta-q.s.
\enqs
This implies that $\hat X$ and $X^*$ satisfy the same linear SDE under $\P^{\theta_.}$, for any $\theta_.$ $\in$ $\Vc_\bTheta$,  and so $\hat X_t$ $=$ $X_t^*$,
$0 \leq t \leq T$,  $\Pc^\bTheta$-q.s. This proves that $\alpha^*$ $=$ $\hat\alpha$, equal to \reff{expressalpha*}.
}
\epR
\end{Remark}

\begin{Remark} \label{remdis}
{\rm By similar  derivation as in \reff{derivEV}, and using \reff{sysK},  \reff{expressalpha*}, we have that for all $\theta_.$ $=$ $(\theta_t)_t$ $=$ $(b_t,\rho_t)_t$ $\in$ $\Vc_\bTheta$, $t$ $\in$ $[0,T]$,
\beq
\frac{d\E_{\theta_.}[V_t^{\alpha^*, \theta_.}]}{dt} &=& K_t \big(R(\theta^*(t)) - R(\theta_t) \big) {\rm Var}_{\theta_.}(X_t^*)
+ \frac{1}{4K_t} \big(R(\theta^*(t)) - R(\theta_t) \big) \label{EVtheta} \\
& & \;\;\; + \;  K_t  \E_{\theta_.}\Big[\big( \hat a_t(X_t^*,\E_{\btheta^*}[X_t^*])  -\hat a_t(X_t^*,\E_{\theta_.}[X_t^*])\big)\trans \Sigma(\rho_t)  \nonumber \\
& & \hspace{1.5cm}  \big(\hat a_t(X_t^*,\E_{\btheta^*}[X_t^*])  -\hat a_t(X_t^*,\E_{\theta_.}[X_t^*]) \big) \Big] \nonumber  \\
& \geq &  K_t  \E_{\theta_.}\Big[\big( \hat a_t(X_t^*,\E_{\btheta^*}[X_t^*])  -\hat a_t(X_t^*,\E_{\theta_.}[X_t^*])\big)\trans \Sigma(\rho_t)  \nonumber \\
& & \hspace{1.5cm}  \big(\hat a_t(X_t^*,\E_{\btheta^*}[X_t^*])  -\hat a_t(X_t^*,\E_{\theta_.}[X_t^*]) \big) \Big]  \label{rhs}
\enq
by definition of $\btheta^*$ $\in$ ${\rm arg}\min_{\btheta \in \bTheta} \bR(\btheta)$, and as $K_t$ $<$ $0$.  In the case when there is no uncertainty on the drift, i.e., for any
$\theta_.$ $=$ $(b_.,\rho_.)$ $\in$ $\Vc_\bTheta$, $b_.$ is a deterministic function equal to $b^o(t)$, $t$ $\in$ $[0, T]$, the dynamics of $X^*$ under any $\P^{\theta_.}$, $\theta$  $\in$ $\Vc_\bTheta$, is given by
\beqs
dX_t^* &=& \big[ x_0 + \frac{e^{\int_0^T R(\theta^*(t))dt}}{2\lambda} - X_t^*\big] (b^o(t))\trans \Sigma(\rho^*(t))^{-1} \big[ b^o(t) dt + \sigma(\rho_t) dW_t^\theta \big],
\enqs
from which, we deduce by taking expectation under $\P^{\theta_.}$:
\beqs
\E_{\theta_.}[X_t^*] &=& x_0 + \frac{e^{\int_0^T R(\theta^*(t))dt}}{2\lambda} \big[  1 -  e^{- \int_0^t R(\theta^*(s))ds} \big].
\enqs
This means that the expectation under $\P^{\theta_.}$ of the optimal wealth process $X^*$ does not depend on $\theta_.$ $\in$ $\Vc_\bTheta$, and
the r.h.s. of \reff{rhs} is then equal to zero. Therefore, the function
$t$ $\mapsto$ $\E_{\theta_.}[V_t^{\alpha^*, \theta_.}]$ is nondecreasing for all $\theta_.$ $\in$ $\Vc_\bTheta$, which implies in particular condition (iii) of Lemma \ref{MOP}.

However, in the case of drift uncertainty, we cannot conclude as above, and actually this nondecreasing property does not always hold true.  Indeed, consider for example the case where there
is  only drift uncertainty in a single asset model $d$ $=$ $1$, with $\Theta(t)$ $=$ $\Theta$ $=$ $\{ \theta \in [\underline{b},\bar b]\}$,
$0$ $\leq$ $\underline{b}$ $<$ $\bar b$, and known variance $\Sigma^o$ normalized to one.  Notice that $R(\theta)$ $=$ $\theta^2$, and
$\theta^*$ $=$ ${\rm arg}\min_{\theta\in \Theta} R(\theta)$ $=$ $\underline{b}$.
For any constant process equal to $\theta$ $\in$ $\Theta$, we can compute explicitly from  \reff{Xt*expvar} the expectation and variance of $X^*$ under $\P^{\theta}$:
\beqs
\E_{\theta}[X_t^*] & = & \frac{1}{2\lambda} e^{R(\theta^*)T}\big[1 -e^{- \theta \theta^*t}\big], \\
{\rm Var}_{\theta}(X_t^*) &= & \frac{1}{4\lambda^2}e^{2R(\theta^*)T}\big[e^{(R(\theta^*)-2 \theta\theta^*)t} -e^{-2 \theta\theta^*t}\big].
\enqs
Plugging into \reff{EVtheta}, and using also the expression of $K$, $\hat a$ in \reff{GamLamYchi}, \reff{hataopt}, we have for all $\theta$ $\in$ $\Theta$, $t \in [0, T]$, after some straightforward rearrangement:
\beqs
\frac{d\E_{\theta}[V_t^{\alpha^*, \theta}]}{dt}
&=& \frac{1}{2\lambda} e^{R(\theta^*)T}\Big[ce^{-2ct}  -  e^{-R(\theta^*)t}(1 -e^{-ct})\Big(\frac{R(\theta^*)}{2}-\big(\frac{R(\theta^*)}{2} +c\big) e^{-ct} \Big)\Big] \\
& = : & f(t,c),
\enqs
where we set $c$ $=$ $(\theta-\theta^*)\theta^*$ $\geq$ $0$.
Now, we easily see that  for all $t$ $\in$ $[0,T]$, $f(t,c)$ converges to $-\frac{R(\theta^*)}{4\lambda} e^{R(\theta^*)(T-t)}$ $<$ $0$, as $c$ goes to infinity.
Then, by continuity of $f$ with respect to $c$,  we deduce that for $\theta$ large enough (hence for $c$ large enough), $\frac{d\E_{\theta}[V_t^{\alpha^*, \theta}]}{dt}$  is negative, which means that the function $t$ $\mapsto$ $\E_{\theta}[V_t^{\alpha^*, \theta}]$ is not nondecreasing for all $\theta$ $\in$ $\Theta$. Actually, we have proved in Theorem  \ref{robustoptimal} the weaker condition (iii) of Lemma \ref{MOP}, that is, $V_0^{\alpha^*,\theta}$ $\leq$ $\E_{\theta}[V_T^{\alpha^*,\theta}]$, for all $\theta_.$ $\in$ $\Vc_\bTheta$.
}
\epR
\end{Remark}

\begin{Remark}
{\rm Assume that there is only 
correlation ambiguity, hence the known drift is a deterministic function denoted by $b^o(t)$, and $\hat\rho(t)$ is a point estimation of correlation lying in 
$\Gamma(t)$. Then the robust portfolio strategy $\alpha_t^*$ in \reff{robustalpha} is less risky than standard mean-variance portfolio strategy denoted by $\alpha_t^{MV, *}$ (to distinguish with robust portfolio strategy) in Section \ref{nouncertainty} in the sense that, for each $t$ $\in$ $[0, T]$
\beq \label{rmklessrisk}
{\rm Var}_{\hat\theta}(\alpha_t^*): =\E_{\hat\theta}[(\alpha_t^*)\trans\Sigma(\hat\rho(t))\alpha_t^*] & \leq & \E_{\hat\theta}[(\alpha_t^{MV, *})\trans \Sigma(\hat\rho(t))\alpha_t^{MV, *}] 
=: {\rm Var}_{\hat\theta}(\alpha_t^{MV, *}),
\enq
where $\E_{\hat\theta}$ is expectation under probability measure $\P^{\hat\theta}$ with 
$\hat\theta:=(b^o(t), \hat\rho(t))$. Indeed, from the expression of $\alpha_t^*$ and $\alpha_t^{MV, *}$, we then have
\beqs
{\rm Var}_{\hat\theta}(\alpha_t^*) &=& \frac{1}{4\lambda^2}(b^o(t))\trans\Sigma(\rho^*(t))^{-1}\Sigma(\hat\rho(t))\Sigma(\rho^*(t))^{-1} b^o(t)e^{\int_t^T R(b^o(s), \rho^*(s))ds}\\
& & \;\;\;\;e^{-\int_0^t \big(R(b^o(s), \rho^*(s))ds- (b^o(s))\trans\Sigma(\rho^*(s))^{-1}\Sigma(\hat\rho(s))^{-1}\Sigma(\rho^*(s))^{-1}b^o(s)\big)ds},\\
{\rm Var}_{\hat\theta}(\alpha_t^{MV, *}) &=& \frac{1}{4\lambda^2}R(b^o(t), \hat\rho(t))e^{\int_t^T R(b^o(s), \hat\rho(s))ds}.
\enqs
From Lemma \ref{lemH} with $b^o(t)$ $=$ $b^*(t)$, together with $\rho^*(t)$ $\in$ $\arg\Min_{\rho \in \Gamma(t)} R(b^o(t), \rho)$, we obtain for $t$ $\in$ $[0, T]$
\beqs
(b^o(t))\trans\Sigma(\rho^*(t))^{-1}\Sigma(\hat\rho(t))\Sigma(\rho^*(t))^{-1} b^o(t) \;\leq\; R(b^o(t), \rho^*(t)) \; \leq \; R(b^o(t), \hat\rho(t)),
\enqs
which implies \reff{rmklessrisk}. This point was  observed  in \cite{LZ2017} under a single period setting, and extended here in a continuous-time setting.
However, notice that  this result does not always hold  in the case of both drift and correlation ambiguity.
}
\epR
\end{Remark}

\paragraph{Conclusion of Part I} We complete  this part by  highlighting  the key mathematical result in this paper about the separation principle and (weak) martingale optimality principle for solving robust portfolio selection problem.  A related methodology has been used in \cite{IsmailPham17}, however  only in the case of ambiguity set for the covariance matrix: it is pointed out in their Remark 4.3 that the employed method relying on a verification theorem for McKean-Vlasov control problem (which itself is derived from an associated optimality principle in the Wasserstein space of probability measures)  cannot tackle the ambiguity about mean return rate.  In  the paper \cite{FouPunWon16}, which considers a special setting with two-asset model and uncertain correlation, a similar separation principle is obtained, but not explicitly written in terms of a risk premium function, and it is not clear how their conditions (see Theorem 2.2 in \cite{FouPunWon16}) can be expressed in a multi-asset case.  An important contribution of  our paper is to state this separation principle  in a more general framework including uncertainty  both on the mean return rates and on the correlations of multi-assets, and with ambiguity sets that may decrease  over time, taking into account, for example, learning about the true parameter to reduce the estimation error.    On the other hand, such a result holds not only for mean-variance problems, but for other popular classes of performance measures like utility criteria, and also in discrete-time setting.
This is detailed and discussed in Appendix A, where we used martingale optimality principle as explained in Remark  \ref{remmarU}. 
It could be also applied to other time-inconsistent robust optimization, like V@R, or CV@R, in the future work. 
Finally, we point out that we are not able to tackle uncertainty on both marginal volatilities and correlation, as we would lose in this case the convexity of covariance matrix on parameter, which is required in the proof of the separation principle.

\section*{Part II: Applications}

\section{Applications and examples}

\setcounter{Theorem}{0} \setcounter{Proposition}{0}
\setcounter{Corollary}{0} \setcounter{Lemma}{0}
\setcounter{Definition}{0} \setcounter{Remark}{0}

We provide in this section several examples for the determination of the minimal risk premium arising from the separation principle in Theorem \ref{robustoptimal}, and the implications for the optimal robust portfolio strategy and the portfolio diversification.
We shall focus in this section on ambiguity sets $\bTheta$ $=$ $\{\Theta(t),t\in [0,T]\}$ as in \reff{exaTheta}, i.e.,
in the ellipsoidal form
\beq \label{Thetabeyond}
\Theta(t) \;=\; \{(b, \rho) \in \R^d \times \Gamma(t): \|\sigma_{J_{l}}(\rho)^{-1}(b_{J_l} -\hat b_{J_l}(t))\|_2 \leq \delta_{l}(t),\;\; l =1, \ldots, p\}.  \label{ThetaJm}
\enq



Given $\Theta(t)$ as in \reff{Thetabeyond}, we denote by $\hat\beta_i(t)$ $:=$ $\frac{\hat b_i(t)}{\sigma_i}$ the instantaneous Sharpe ratio of the $i$-th asset associated with  estimated mean return  $\hat b_i(t)$, and marginal volatility $\sigma_i$ $>$ $0$, $i$ $=$ $1,\ldots,d$. In what follows, we assume that $\Max_{1 \leq j \leq d}|\hat\beta_j(t)|$ $\neq$ $0$ (otherwise $\hat\beta_j(t)$ $=$ $0$ for each $1$ $\leq$ $j$ $\leq$ $d$, i.e., $\hat b(t)$ $=$ $0$, meaning that
the optimal portfolio strategy is to never trade, i.e., $\alpha_t^*$ $=$ $0$).
We define the Sharpe ratio "proximity" between $i$-th asset and $j$-th asset, $1$ $\leq$ $i$ $\neq$ $j$ $\leq$ $d$, by
\beq
\hat\varrho_{ij}(t) \; = \; \hat\varrho_{ji}(t) &:= & \frac{\hat\beta_j(t)}{\hat\beta_i(t)} 1_{\{|\hat\beta_i(t)| > |\hat\beta_j(t)|\}} + \frac{\hat\beta_i(t)}{\hat\beta_j(t)} 1_{\{|\hat\beta_i(t)| \leq |\hat\beta_j(t)|\}}\in [-1, 1],
\label{hatvarrhoij}
\enq
with the convention that $\hat\varrho_{ij}(t)$ $=$ $\hat\varrho_{ji}(t)$ $=$ $0$ when $\hat\beta_i(t)$ $=$ $\hat\beta_j(t)$ $=$ $0$.


\medskip

We first provide the general explicit expression of the robust optimal strategy in the case of ellipsoidal ambiguity set.

\begin{Proposition}\label{alphaellip}
Let $\Theta(t)$ be an ellipsoidal set as in \reff{Thetabeyond} with $p$ $=$ $1$,  and assume that there exists $\rho^*(t)$ $\in$
${\rm arg}\Min_{\rho\in\Gamma(t)}\big\|\sigma(\rho)^{-1} \hat b(t)\big\|_2$. Then, an optimal portfolio strategy for \reff{robustMV} is given by, for  $t$ $\in$ $[0, T]$,
\beq \label{optimalalpha*ellip}
\alpha_t^* &=& \Big[x_0 + \frac{1}{2\lambda} e^{\int_0^T(\|\sigma(\rho^*(s))^{-1}\hat b(s)\|_2 -\delta(s))^2 1_{\{\|\sigma(\rho^*(s))^{-1}\hat b(s)\|_2 > \delta(s)\}} ds}-X_t^*\Big] \nonumber\\
 & & \hspace{0.5cm}\Big(1 -\frac{\delta(t)}{\|\sigma(\rho^*(t))^{-1}\hat b(t)\|_2} \Big)1_{\{\|\sigma(\rho^*(t))^{-1}\hat b(t)\|_2 > \delta(t)\}}\Sigma(\rho^*(t))^{-1}\hat b(t).
\enq
\end{Proposition}
{\bf Proof.} See Section \ref{appenlemellipoid} in Appendix.

\begin{Remark} [Financial interpretation: no risky investment] \label{remnorisky}
{\rm We have seen in the previous section that $\rho^*(t)$ exists when $\Gamma(t)$ is compact (in particular when it is a singleton, i.e., there is no ambiguity on correlation) or when $\Gamma(t)$ $=$ $\C_{>+}^d$, i.e., there is full ambiguity on correlation.  From \reff{optimalalpha*ellip}, we observe notably that whenever $\delta(t)$ $\geq$ $\|\sigma(\rho^*(t))^{-1}\hat b(t)\|_ 2$, $\alpha^*_t$ $=$ $0$.  In other words, when, at time $t$,  the investor is poorly confident about her estimation on the expected rate of return $\hat b(t)$, or when the
level of uncertainty aversion about the expected rate of return is high, then she does not make risky investment at all.
}
\epR
\end{Remark}

\subsection{Full ambiguity correlation and anti-diversification}
In this paragraph, we consider the case of full ambiguity on correlation, i.e., $\Gamma(t)$ $=$ $\C_{>+}^d$, and investigate the impact on optimal robust portfolio strategy.
\begin{Theorem}[Full ambiguity correlation]\label{thmThetafull}
Let $\Theta(t)$ be an ellipsoidal set as in \reff{Thetabeyond}, with $\Gamma(t)$ $=$ $\C_{>+}^d$ for all $t$ $\in$ $[0, T]$, and assume that for each $1$ $\leq$ $l$ $\leq$ $p$, there exists a function $m_l(t)$ $\in$ $J_l$ s.t.
 $|\hat\beta_{m_l(t)}(t)|$ $>$ $\Max_{ j \in J_l, j \neq m_l(t)}|\hat\beta_j(t)|$. Assume further that there exists a function $k(t)$ $\in$ $\{1, \ldots, p\}$ s.t. $|\hat\beta_{m_{k(t)}(t)}(t)|$ $-$ $\delta_{k(t)}(t)$
$>$ $\Max_{1 \leq l \leq p, l \neq k(t)}(|\hat\beta_{m_l}(t)| -\delta_{l}(t))$. Then an optimal portfolio strategy for the robust mean-variance problem \reff{robustMV} is
explicitly given by, for $t$ $\in$ $[0, T]$,
\beqs
\alpha_t^* &=&
\Big[ x_0 + \frac{1}{2\lambda} e^{\int_0^T (|\hat\beta_{m_{k(s)}(s)}(s)| -\delta_{k(s)}(s))^2 1_{\{|\hat\beta_{m_{k(s)}(s)}(s)| > \delta_{k(s)}(s)\}}ds} - X_t^*\Big]\\
& & \hspace{0.6cm} \Big(1-\frac{\delta_{{k(t)}}(t)}{|\hat\beta_{m_{k(t)}(t)}(t)|}\Big)1_{\{|\hat\beta_{m_{k(t)}(t)}(t)| > \delta_{k(t)}(t)\}}(0, \ldots,0, \frac{\hat b_{m_{k(t)}(t)}(t)}{\sigma_{m_{k(t)}(t)}^2}, 0, \ldots, 0)\trans.
\enqs
\end{Theorem}
{\bf Proof.} See Section \ref{appenproThetafull} in Appendix.

\begin{Remark}\label{rmkThetafull}
{\rm Notice that in the particular case when $p$ $=$ $1$, the conditions of Theorem \ref{thmThetafull} simply assume that there exists $m(t)$ $\in$ $\{1, \ldots, d\}$ s.t.
$|\hat\beta_{m(t)}(t)|$ $>$ $\Max_{1 \leq j \leq d, j \neq m(t)} |\hat\beta_j(t)|$. Then an optimal portfolio strategy for the robust mean-variance problem \reff{robustMV} is
explicitly given by, for $t$ $\in$ $[0, T]$,
\beqs
\alpha_t^* &=&
\big[x_0 + \frac{1}{2\lambda} e^{\int_0^T (|\hat\beta_{m(s)}(s)| -\delta(s))^2 1_{\{|\hat\beta_{m(s)}(s)| > \delta(s)\}} ds } - X_t^*\big]\\
& & \hspace{0.6cm} \Big(1 -\frac{\delta(t)}{|\hat\beta_{m(t)}(t)|}\Big) 1_{\{|\hat\beta_{m(t)}(t)| > \delta(t)\}} (0, \ldots, 0, \frac{\hat b_{m(t)}(t)}{\sigma_{m(t)}(t)^2}, 0, \ldots, 0)\trans.
\enqs
}
\epR
\end{Remark}

\begin{Remark}[Financial interpretation: anti-diversification] \label{remanti}
{\rm
Observe that both reference Sharpe ratio and drift ambiguity level play an important role in the portfolio choice of an  investor. 
If this investor is poorly confident on the drift estimate, i.e., whenever all $\delta_l(t)$, $1$ $\leq$ $l$ $\leq$ $p$ are large enough, then she does not make risky investments at all, i.e., $\alpha_t^*$
$=$ $0$.
When the reference Sharpe ratio of an asset is large enough to offset the effect of its own drift ambiguity but not others, she would only invest in this asset, namely the one with the highest lower Sharpe ratio (the reference Sharpe ratio (absolute value) of $i$-th asset minus its drift estimation error), which might change from one subclass to another subclass with time.

This anti-diversification result under full ambiguity about correlation has been also observed in \cite{LZ2017} for a single-period mean-variance problem without drift uncertainty, and is  extended here in a continuous-time framework. Moreover, under this general setting, the single  risky asset that the investor trades may  change over time because of the time varying ambiguity set.

Compared  to Remark \ref{remnorisky}, we see that expected return rates ambiguity and correlation ambiguity have different effects on portfolio selection. When expected return rates ambiguity is large, no risky investment is made, while when correlation  ambiguity is large, one and only one risky asset is traded.}\epR
\end{Remark}
\subsection{Partial diversification}
\noindent $\bullet$ {\bf Two-asset model: $d$ $=$ $2$}\\
\noindent We provide a complete picture of the optimal robust portfolio strategy in a two-asset model with ambiguous drift and correlation.

\begin{Theorem}[Ambiguous drift and correlation in the two-asset case]  \label{theoopt2}
Let $\Theta(t)$\\ $=$
 $\{(b, \rho) \in \R^2 \times [\underline \rho(t), \bar \rho(t)]: \|\sigma(\rho)^{-1}(b-\hat b(t))\|_2 \leq \delta(t)\}$, with
$-1<\underline\rho(t)\leq\bar\rho(t)<1$, $t$ $\in$ $[0, T]$, and assume at each time $t$ $\in$ $[0, T]$,
$\max(|\hat\beta_1(t)|, |\hat\beta_2(t)|)$ $\neq$ $0$.

Then, an optimal portfolio strategy is given by
\beqs
\alpha_t^* &=& \left\{\begin{array}{cl}
\big[x_0 + \frac{1}{2\lambda} e^{\int_0^T R(\theta^*(s))ds} - X_t^*\big](1 -\frac{\delta(t)}{\max(|\hat\beta_1(t)|, |\hat\beta_2(t)|)})\\
\;\; 1_{\{\max(|\hat\beta_1(t)|,|\hat\beta_2(t)|)> \delta(t)\}}\left(\begin{matrix}\frac{\hat b_1(t)}{|\sigma_1|^2}1_{\{|\hat\beta_1(t))| > |\hat\beta_2(t)|\}}\\ \frac{\hat b_2(t)}{|\sigma_2|^2}1_{\{|\hat\beta_2(t)|>|\hat\beta_1(t)|\}}\end{matrix}\right), & \mbox{ if }\;\; \hat\varrho_{12}(t) \in [\underline \rho(t), \bar\rho(t)]\\[1cm]

\big[x_0 +\frac{1}{2\lambda} e^{\int_0^TR(\theta^*(s))ds}-X_t^*\big]
\Big(1-\frac{\delta(t)}{\|\sigma(\bar\rho(t))^{-1}\hat b(t)\|_2} \Big)\\
\hspace{1.5cm} 1_{\{\|\sigma(\bar\rho(t))^{-1}\hat b(t)\|_2 > \delta(t)\}}\Sigma(\bar\rho(t))^{-1}\hat b(t), & \mbox{ if } \;\;\bar\rho(t) < \varrho_{12}(t)\\[0.5cm]

\big[x_0 +\frac{1}{2\lambda} e^{\int_0^T R(\theta^*(s))ds}-X_t^*\big]\Big(1 -\frac{\delta(t)}{\|\sigma(\underline\rho(t))^{-1}\hat b(t)\|_2}\Big)\\
\hspace{1.5cm} 1_{\{\|\sigma(\underline \rho(t))^{-1}\hat b(t)\|_2 > \delta(t)\}}\Sigma(\underline \rho(t))^{-1}\hat b(t), &  \mbox{ if }\;\; \underline\rho(t) > \varrho_{12}(t)\end{array}\right.
\enqs
where for each $s$ $\in$ $[0, T]$
\beq \label{prop2Rintegr}
& &  R(\theta^*(s))\\
& =& (\max(|\hat\beta_1(s)|, |\hat\beta_2(s)|)-\delta(s))^2 1_{\{\max(|\hat\beta_1(s)|, |\hat\beta_2(s)|)> \delta(s)\}} 1_{\{\hat\varrho_{12}(s) \in [\underline \rho(s), \bar\rho(s)]\}} \nonumber\\
& & \; + \; (\|\sigma(\bar\rho(s))^{-1}\hat b(s)\|_2 -\delta(s))^2 1_{\{\|\sigma(\bar\rho(s))^{-1}\hat b(s)\|_2  > \delta(s)\}}1_{\{\bar\rho(s) < \hat\varrho_{12}(s)\}} \nonumber\\
& & \;+\; (\|\sigma(\underline\rho(s))^{-1}\hat b(s))\|_2 -\delta(s))^2 1_{\{\|\sigma(\underline \rho(s))^{-1}\hat b(s)\|_2  > \delta(s)\}} 1_{\{\underline\rho(s) > \varrho_{12}(s)\}} \nonumber.
\enq

\end{Theorem}
{\bf Proof} See Section \ref{appenpro2} in Appendix.
\begin{Remark}
{\rm When
there is only 
ambiguity on correlations and correlation ambiguity set does not vary with time,
we retrieve the results obtained in \cite{IsmailPham17} for the correlation ambiguity between two assets (see their Theorem 4.2). Our Theorem includes in addition the case when there is uncertainty on the expected rate of return and the ambiguity sets vary over time. 
Also, with  time varying ambiguity sets, we see that  the investor may switch between well-diversification and under-diversification over time, moreover, when under-diversification occurs, she may only invest in the first asset or the second asset; when well-diversification occurs, she may switch between directional trading and spread trading. This feature reflects dynamic changes in real markets.}\epR
\end{Remark}
\begin{Remark}[Financial interpretation]
{\rm
At a given time $t$ $\in$ $[0, T]$, we have three possible cases depending on the relation between the bounds of correlations, more precisely,  the value of variance risk ratio on the bounds of correlations, 
and 
Sharpe ratio proximity. 
Notice that $R(\hat b(t), \rho)$ is convex, the first order derivative $\frac{\partial R(\hat b(t), \rho)}{\partial \rho}$ $=$ $-\sigma_1\sigma_2\hat\kappa^1_t(\rho)\hat\kappa^2_t(\rho)$ is increasing, hence $\rho$ $\in$ $[\underline\rho(t), \bar\rho(t)]$ $\mapsto$ $\hat\kappa^1_t(\rho)\hat\kappa^2_t(\rho)$ decreasing. 
Moreover, we have $\Lim_{\rho \to 1} \hat\kappa^1_t(\rho)\hat\kappa^2_t(\rho)$ $<$ $0$, $\Lim_{\rho \to -1} \hat\kappa^1_t(\rho)\hat\kappa^2_t(\rho)$ $>$ $0$ and $\hat\kappa^1_t(\hat\varrho_{12}(t))\hat\kappa^2_t(\hat\varrho_{12}(t))$ $=$ $0$.

{{In the first case when correlation ambiguity interval includes $\hat\varrho_{12}(t)$,  anti-diversification phenomenon occurs, i.e., one and only one asset is invested. Indeed, note that when the range of correlation ambiguity interval is larger than $\max\{1 - \hat\varrho_{12}(t),  1 + \hat\varrho_{12}(t)\}$, Sharpe ratio proximity $\hat\varrho_{12}(t)$ $\in$ $[\underline\rho(t), \bar\rho(t)]$, or in other words, the correlation can be either larger or smaller than the Sharpe ratio proximity. In this case, the optimal strategy under the worst-case scenario is to invest in the asset with highest Sharpe ratio.}}

In the second case when $\bar\rho(t)$ $<$ $\hat\varrho_{12}(t)$, meaning that the correlation taking value in $[\underline{\rho}(t),\bar\rho(t)]$ is small compared to the Sharpe ratio proximity, then it is  optimal to invest in both assets with a directional trading, that is, buying or selling simultaneously.
And  the worst-case correlation refers to the highest corre\-lation $\bar\rho(t)$ 
where the diversification effect is minimal.
This case corresponds to $\hat\kappa^1_t(\bar\rho(t))\hat\kappa^2_t(\bar\rho(t))$ $>$ $0$. Together with the monotonicity of $\hat\kappa_t^1(\rho)\hat\kappa_t^2(\rho)$ , we have $\hat\kappa^1_t(\rho)\hat\kappa^2_t(\rho)$ $>$ $0$ for any $\rho$ $\in$ $[\underline\rho(t), \bar\rho(t)]$, meaning that it is optimal to take directional trading, and the worst-scenario correlation is upper bound of correlation under which the diversification effect is minimal. 


In the third case when $\underline\rho(t)$ $>$ $\hat\varrho_{12}(t)$, meaning that the correlation taking value in $[\underline{\rho}(t),\bar\rho(t)]$ is large  compared to the Sharpe ratio proximity, then it is  optimal to invest in both assets with a spread trading, that is, buying one and selling another. And the worst-case correlation corresponds to the lowest correlation
$\underline{\rho}(t)$ where the  profit from the spread trading is minimal.
This case 
corresponds to $\hat\kappa_t^1(\underline\rho(t))\hat\kappa_t^2(\underline\rho(t))$ $<$ $0$. By analogy with the second case, we have $\hat\kappa_t^1(\rho)\hat\kappa_t^2(\rho)$ $<$ $0$ for any $\rho$ $\in$ $[\underline\rho(t), \bar\rho(t)]$, meaning that it is  optimal to invest with spread trading with lower bound of correlation.

This diversification result with only correlation uncertainty has been also observed in the literature \cite{FouPunWon16} for a continuous-time expected utility problem, \cite{LZ2017} for a single-period mean-variance problem and \cite{IsmailPham17} for a continuous-time mean-variance problem, and is extended here in a continuous time framework with time varying ambiguity set for both drift and correlation uncertainty. One interesting additional finding in continuous time  is that at some time we can be in the first case where under-diversification occurs, and next at a future time,  in case 2 or 3 with directional or spread trading due to a change in the relation between the bounds of correlations and Sharpe ratio proximity.}\epR
\end{Remark}

\noindent $\bullet$ {\bf Three-asset model:   $d=3$}

\noindent We finally provide an explicit description of the optimal robust strategy in a three-asset model under drift uncertainty and ambiguous correlation where ambiguity set does not vary with time for simplicity. We introduce the so-called variance risk ratio $\hat\kappa(\rho)$,
\beqs
\Sigma(\rho)^{-1}\hat b  = : \hat\kappa(\rho)=(\hat\kappa^1(\rho), \hat\kappa^2(\rho), \hat\kappa^3(\rho))\trans,
\enqs
which represents (up to a scalar term) the vector of allocation in the assets when the drift is $\hat b$ and the correlation is $\rho$.


\begin{Theorem}\label{theoopt3}
Let $\Theta$ $=$ $\{(b, \rho)\in \R^3  \times \Gamma: \|\sigma(\rho)^{-1}(b -\hat b)\|_2 \leq \delta\}$, with $\Gamma$ $=$
$[\underline \rho_{12}, \bar\rho_{12}]$ $\times$ $[\underline \rho_{13}, \bar\rho_{13}]$ $\times$
$[\underline \rho_{23}, \bar\rho_{23}]$ $\subset$ $\C_{>+}^3$, and assume w.l.o.g. that $|\hat\beta_1|\geq  |\hat\beta_2| \geq |\hat\beta_3|$ and $\hat\beta_1$ $\neq$ $0$.
Then, we have the following possible exclusive cases:
\begin{itemize}
 \item [{\bf 1.}] (Anti-diversification) If $\hat\varrho_{12} \in [\underline \rho_{12}, \bar\rho_{12}]$, and $\hat\varrho_{13} \in [\underline \rho_{13}, \bar\rho_{13}]$, then an optimal portfolio strategy is explicitly given by
\beqs
\alpha_t^* &=& \big[x_0 +\frac{1}{2\lambda}e^{(|\hat\beta_1|-\delta)^2 T}-X_t^*\big]\Big(1 -\frac{\delta}{|\hat\beta_1|}\Big)1_{\{|\hat\beta_1| >\delta\}}
\left(\begin{matrix}\frac{\hat b_1}{\sigma_1^2}\\ 0\\ 0\end{matrix}\right), \;\; 0 \leq t \leq T, \; \Pc^\bTheta- q.s..
\enqs
\item [{\bf 2.}] (Under-diversification: no investment in the third asset)
 \begin{itemize}
 \item [(i)] If $\bar\rho_{12} < \hat\varrho_{12}$, and $\hat\kappa^3(\bar\rho_{12}, \bar\rho_{13}, \bar\rho_{23})\hat \kappa^3(\bar\rho_{12}, \underline \rho_{13}, \underline \rho_{23}) \leq 0$, then an optimal portfolio stra\-tegy is
 \beqs
\left(\begin{matrix}\alpha_t^{1,*}\\ \alpha_t^{2, *} \end{matrix}\right) &=& \big[x_0 + \frac{1}{2\lambda} e^{(\|\sigma_{-3}(\bar\rho_{12})^{-1}\hat b_{-3}\|_2 -\delta)^2 T} -X_t^*\big]\Big(1 -\frac{\delta}{\|\sigma_{-3}(\bar\rho_{12})^{-1}\hat b_{-3}\|_2}\Big) \\
& & \hspace{4cm} 1_{\{\|\sigma_{-3}(\bar\rho_{12})^{-1}\hat b_{-3}\|_2 > \delta \}}\Sigma_{-3}(\bar\rho_{12})^{-1}\hat b_{-3}\\
\alpha_t^{3, *} &\equiv& 0,
 \enqs
and if $\|\sigma_{-3}(\bar\rho_{12})^{-1}\hat b_{-3}\|_2 > \delta$, then $\alpha_t^{1, *}\alpha_t^{2, *}$ $>$ $0$.
 \item [(ii)] If $\underline \rho_{12} > \hat\varrho_{12}$, and $\hat \kappa^3(\underline \rho_{12}, \underline \rho_{13}, \bar\rho_{23}) \hat \kappa^3(\underline \rho_{12}, \bar\rho_{13}, \underline \rho_{23})  \leq 0$, then an optimal portfolio stra\-tegy is
 \beqs
 \left(\begin{matrix}\alpha_t^{1, *}\\ \alpha_t^{2, *} \end{matrix}\right) &=& \big[x_0 + \frac{1}{2\lambda} e^{(\|\sigma_{-3}(\underline \rho_{12})^{-1}\hat b_{-3}\|_2)-\delta)^2T} -X_t^*\big]
 \Big(1-\frac{\delta}{\|\sigma_{-3}(\underline \rho_{12})^{-1} \hat b_{-3}\|_2}\Big) \\
 & & \hspace{4cm} 1_{\{\|\sigma_{-3}(\underline \rho_{12})^{-1}\hat b_{-3}\|_2 > \delta\}}\Sigma_{-3}(\bar\rho_{12})^{-1}\hat b_{-3}\\
 \alpha_t^{3, *} &\equiv& 0,
 \enqs
 and if $\|\sigma_{-3}(\underline \rho_{12})^{-1}\hat b_{-3}\|_2$ $>$ $\delta$, then $\alpha_t^{1, *}\alpha_t^{2, *}$ $<$ $0$.
 \end{itemize}
 \item [{\bf 3.}] (Under-diversification: no investment in the second asset)
 \begin{itemize}
\item[(i)]  If $\bar\rho_{13} < \hat\varrho_{13}$, and
 $\hat \kappa^2(\bar\rho_{12}, \bar\rho_{13}, \bar\rho_{23})\hat \kappa^2(\underline \rho_{12}, \bar\rho_{13}, \underline \rho_{23})$ $\leq$ $0$, then an optimal portfolio stra\-tegy is
 \beqs
\left(\begin{matrix}\alpha_t^{1, *}\\ \alpha_t^{3, *}\end{matrix}\right) &=& \big[x_0 + \frac{1}{2\lambda} e^{(\|\sigma_{-2}(\bar\rho_{13})^{-1}\hat b_{-2}\|_2 -\delta)^2 T} -X_t^*\big]\Big(1 -\frac{\delta}{\|\sigma_{-2}(\bar\rho_{13})^{-1}\hat b_{-2}\|_2}\Big) \\
&& \hspace{4cm} 1_{\{\|\sigma_{-2}(\bar\rho_{13})^{-1}\hat b_{-2}\|_2 > \delta\}}  \Sigma_{-2}(\bar\rho_{13})^{-1}\hat b_{-2}\\
\alpha_t^{2, *} &\equiv& 0,
 \enqs
 and if $\|\sigma_{-2}(\bar\rho_{13})^{-1}\hat b_{-2}\|_2$ $>$ $\delta$, then $\alpha_t^{1, *}$$\alpha_t^{3, *}$  $>$ $0$.
 \item [(ii)] If $\underline \rho_{13} > \hat\varrho_{13}$, and $\hat\kappa^2( \underline \rho_{12},\underline \rho_{13}, \bar\rho_{23})  \hat\kappa^2(\bar\rho_{12}, \underline \rho_{13}, \underline \rho_{23}) \leq 0$,  then an optimal portfolio strategy is given by
 \beqs
 \left(\begin{matrix}\alpha_t^{1, *}\\ \alpha_t^{3, *}\end{matrix}\right) &=& \big[x_0 + \frac{1}{2\lambda} e^{(\|\sigma_{-2}(\underline\rho_{13})^{-1}\hat b_{-2}\|_2 -\delta)^2 T} -X_t^*\big]\Big(1 -\frac{\delta}{\|\sigma_{-2}(\underline\rho_{13})^{-1}\hat b_{-2}\|_2}\Big) \\
 & & \hspace{4cm}1_{\{\|\sigma_{-2}(\underline\rho_{13})^{-1}\hat b_{-2}\|_2 > \delta\}} \Sigma_{-2}(\underline \rho_{13})^{-1}\hat b_{-2}\\
 \alpha_t^{2, *} &\equiv& 0,
 \enqs
 and if $\|\sigma_{-2}(\underline\rho_{13})^{-1}\hat b_{-2}\|_2 > \delta$, then $\alpha_t^{1, *}$$\alpha_t^{3, *}$ $<$ $0$.
 \end{itemize}
 \item [{\bf 4.}] (Under-diversification: no investment in the first asset)
 \begin{itemize}
 \item[(i)]  If $\bar\rho_{23} < \hat\varrho_{23}$, and
 $\hat\kappa^1(\underline \rho_{12}, \underline \rho_{13}, \bar\rho_{23})\hat\kappa^1(\bar\rho_{12}, \bar\rho_{13}, \bar\rho_{23})$ $\leq$ $0$, then an optimal portfolio strategy is
 \beqs
 \left(\begin{matrix}\alpha_t^{2, *} \\ \alpha_t^{3, *}\end{matrix}\right) &=& \big[x_0 + \frac{1}{2\lambda} e^{(\|\sigma_{-1}( \bar\rho_{23})^{-1}\hat b_{-1}\|_2 -\delta)^2T} -X_t^*\big]\Big(1 -\frac{\delta}{\|\sigma_{-1}(\bar\rho_{23})^{-1}\hat b_{-1}\|_2}\Big) \\
 & & \hspace{4cm}1_{\{\|\sigma_{-1}(\bar\rho_{23})^{-1}\hat b_{-1}\|_2> \delta\}} \Sigma_{-1}(\bar\rho_{23})^{-1}\hat b_{-1}\\
 \alpha_t^{1, *} &\equiv& 0,
 \enqs
 and if $\|\sigma_{-1}(\bar\rho_{23})^{-1}\hat b_{-1}\|_2> \delta$, then $\alpha_t^{2, *}$$\alpha_t^{3, *}$ $>$ $0$.
 \item [(ii)] If $\underline \rho_{23} > \hat\varrho_{23}$, and $\hat\kappa^1(\underline \rho_{12}, \bar\rho_{13}, \underline \rho_{23})\hat\kappa^1(\bar\rho_{12}, \underline \rho_{13}, \underline \rho_{23}) \leq 0$, then an optimal portfolio strategy is
 \beqs
\left(\begin{matrix}\alpha_t^{2, *} \\ \alpha_t^{3, *} \end{matrix}\right)&=& \big[x_0 + \frac{1}{2\lambda} e^{(\|\sigma_{-1}( \underline\rho_{23})^{-1}\hat b_{-1}\|_2 -\delta)^2T} -X_t^*\big]\Big(1 -\frac{\delta}{\|\sigma_{-1}(\underline\rho_{23})^{-1}\hat b_{-1}\|_2}\Big) \\
& & \hspace{4cm} 1_{\{\|\sigma_{-1}(\underline\rho_{23})^{-1}\hat b_{-1}\|_2 > \delta\}} \Sigma_{-1}(\underline\rho_{23})^{-1}\hat b_{-1}\\
\alpha_t^{1, *} &\equiv& 0,
 \enqs
and if $\|\sigma_{-1}(\underline\rho_{23})^{-1}\hat b_{-1}\|_2 > \delta$, then $\alpha_t^{2, *}$ $\alpha_t^{3, *}$ $<$ $0$.
 \end{itemize}
 \item [{\bf 5.}] (Well-diversification)
 \begin{itemize}
 \item [(i)]  If $\hat\kappa^1\hat\kappa^2(\bar\rho_{12}, \bar\rho_{13}, \bar\rho_{23}) >0$, and $\hat\kappa^1\hat\kappa^3(\bar\rho_{12}, \bar\rho_{13}, \bar\rho_{23})>0$,  then an optimal portfolio strategy is given by
 \beqs
 \alpha_t^* &=& \big[x_0 +\frac{1}{2\lambda}e^{(\|\sigma(\bar\rho_{12}, \bar\rho_{13}, \bar\rho_{23})^{-1}\hat b\|_2 -\delta)^2T} -X_t^*\big]\Big(1 -\frac{\delta}{\|\sigma(\bar\rho_{12}, \bar\rho_{13}, \bar\rho_{23})^{-1}\hat b\|_2} \Big) \\
 & &  \hspace{4cm} 1_{\{\|\sigma(\bar\rho_{12}, \bar\rho_{13}, \bar\rho_{23})^{-1}\hat b\|_2 > \delta\}} \Sigma(\bar\rho_{12}, \bar\rho_{13}, \bar\rho_{23})^{-1}\hat b.
 \enqs
 \item [(ii)] If $\hat\kappa^1\hat\kappa^2(\underline \rho_{12}, \underline \rho_{13}, \bar\rho_{23}) <0$, and $\hat\kappa^1\hat\kappa^3( \underline \rho_{12}, \underline \rho_{13}, \bar\rho_{23})<0$, then an optimal portfolio strategy is given by
 \beqs
\alpha_t^* &=& \big[x_0 +\frac{1}{2\lambda} e^{(\|\sigma(\underline \rho_{12}, \underline \rho_{13}, \bar\rho_{23})^{-1}\hat b\|_2 -\delta)^2T}-X_t^*\big]\Big(1 -\frac{\delta}{\|\sigma(\underline \rho_{12}, \underline \rho_{13}, \bar\rho_{23})^{-1}\hat b\|_2}\Big) \\
& & \hspace{4cm} 1_{\{\|\sigma(\underline \rho_{12}, \underline \rho_{13}, \bar\rho_{23})^{-1}\hat b\|_2 > \delta\}}\Sigma(\underline \rho_{12}, \underline \rho_{13}, \bar\rho_{23})^{-1}\hat b.
 \enqs
 \item [(iii)]
 If $\hat\kappa^1\hat\kappa^2(\bar\rho_{12}, \underline \rho_{13}, \underline\rho_{23}) >0$, and $\hat\kappa^1\hat\kappa^3(\bar\rho_{12}, \underline\rho_{13}, \underline \rho_{23})<0$, then an optimal portfolio strategy is given by
 \beqs
 \alpha_t^* &=& \big[x_0 +\frac{1}{2\lambda}e^{(\|\sigma(\bar\rho_{12}, \underline \rho_{13}, \underline \rho_{23})^{-1}\hat b\|_2-\delta)^2T}-X_t^*\big]\Big(1-\frac{\delta}{\|\sigma(\bar\rho_{12}, \underline \rho_{13}, \underline \rho_{23})^{-1}\hat b\|_2} \Big) \\
 & & \hspace{4cm} 1_{\{\|\sigma(\bar\rho_{12}, \underline \rho_{13}, \underline \rho_{23})^{-1}\hat b\|_2 > \delta\}} \Sigma(\bar\rho_{12}, \underline \rho_{13}, \underline \rho_{23})^{-1}\hat b.
 \enqs
 \item [(iv)]If $\hat\kappa^1\hat\kappa^2(\underline \rho_{12}, \bar\rho_{13}, \underline \rho_{23}) <0$, and $\hat\kappa^1\hat\kappa^3(\underline \rho_{12}, \bar\rho_{13}, \underline \rho_{23})>0$, then an optimal portfolio strategy is given by
 \beqs
 \alpha_t^* &=& \big[x_0 +\frac{1}{2\lambda}e^{\|(\sigma(\underline \rho_{12}, \bar\rho_{13}, \underline \rho_{23})^{-1}\hat b\|_2 -\delta)^2 T}-X_t^*\big]\Big(1 -\frac{\delta}{\|\sigma(\underline \rho_{12}, \bar\rho_{13}, \underline \rho_{23})^{-1}\hat b\|_2}\Big) \\
 & & \hspace{4cm}1_{\{\|\sigma(\underline \rho_{12}, \bar\rho_{13}, \underline \rho_{23})^{-1}\hat b\|_2 > \delta\}} \Sigma(\underline \rho_{12}, \bar\rho_{13}, \underline\rho_{23})^{-1}\hat b.
 \enqs
 \end{itemize}
 \end{itemize}
\end{Theorem}

{\bf Proof}. See Section \ref{appenpro3} in Appendix.

\begin{Remark}[Financial interpretation]
{\rm
{{In case {\bf 1} when $\hat\varrho_{12} \in [\underline \rho_{12}, \bar\rho_{12}]$, $\hat\varrho_{13}$ $\in$ $[\underline\rho_{13}, \bar\rho_{13}]$, it is optimal to invest only in the first asset, namely the one with the highest estimated   Sharpe ratio, which is consistent with the anti-diversification result obtained in Theorem \ref{thmThetafull} (see also Remark \ref{remanti}). Case {\bf 1} happens when the range of $[\underline \rho_{1j}, \bar\rho_{1j}]$, $j$ $=$ $2$, $3$, is larger than $\max\{1 + \hat\varrho_{1j}, 1 - \hat\varrho_{1j}\}$}}.

 In case  {\bf 2}, corresponding to a large correlation ambiguity for the third asset, the investor does not invest in the third asset, but only in the first and second assets. Large correlation ambiguity for the third asset is quantified by the fact that the function $(\rho_{13},\rho_{23})$
$\mapsto$ $\hat\kappa(\bar\rho_{12},\rho_{13},\rho_{23})$ evaluated at the lower bounds   $(\underline\rho_{13}, \underline\rho_{23})$ and the
upper bounds $(\bar\rho_{13}, \bar\rho_{23})$ have opposite signs. Moreover, depending on whether the correlation of the first and second assets is small or large compared to the Sharpe ratio proximity ($\bar\rho_{12} < \hat\varrho_{12}$ or $\underline\rho_{12} > \hat\varrho_{12}$),
the investment in the first and second assets follows a directional trading or a spread trading.

We have a similar under-diversification effect in  cases {\bf 3} and {\bf 4}, and notice that it may happen that one does not invest in the first asset even though it has the highest  reference Sharpe ratio. The result in case {\bf 4} is quite  interesting and is {\it a priori} unexpected. Intuitively, an investor should always invest in the asset with the greatest absolute Sharpe ratio. For example, this is the case when anti-diversification occurs and also in cases {\bf 1}, {\bf 2}, {\bf 3}, {\bf 5}. However, the case {\bf 4} means that the asset with the greatest absolute Sharpe ratio (the first asset) may not be traded in the optimal portfolio while the one with the smallest absolute Sharpe ratio (the third asset) may be traded. The idea is that depending on the drift and correlation ambiguity levels, investing in  the two other assets may achieve  higher risk premium than investing in the first asset.  Take case {\bf 4}(i) for example, in this case, the risk premium is $R(\hat b_{-1}, \bar\rho_{23})$ $=$ $\hat\beta_{-1}\trans C(\bar\rho_{23})^{-1}\hat\beta_{-1}$ where $\hat\beta_{-1}$ $:=$ $(\hat\beta_2, \hat\beta_3)\trans$, especially, $R(\hat b_{-1}, \bar\rho_{23})$ $=$ $|\hat\beta_2|^2 + |\hat\beta_3|^2$ when $\bar\rho_{23}$ $=$ $0$. It follows from \reff{Rcase1} in appendix that $R(\hat b_{-1}, \bar\rho_{23})$ $>$ $|\hat\beta_1|^2$. In \cite{LZ2017}, the authors constructed a simple example where such scenario occurs in a single period model in the case where the second and third assets are independent, hence with  no correlation ambiguity. 

Finally, in case {\bf 5}, corresponding to a small correlation ambiguity,  the investor has incentive to well-diversify her portfolio among the three assets. More precisely, Case {\bf 5} involves explicitly  the signs of $\hat\kappa^1\hat\kappa^2$ and $\hat\kappa^1\hat\kappa^3$ at the correlation bounds.
 Assuming that these functions $\hat\kappa^1\hat\kappa^2$ and $\hat\kappa^1\hat\kappa^3$  do not vanish at some point $\rho$ $\in$
 $[\underline \rho_{12}, \bar\rho_{12}]$ $\times$ $[\underline \rho_{13}, \bar\rho_{13}]$ $\times$ $[\underline \rho_{23}, \bar\rho_{23}]$, then by continuity, and provided that the range of these correlation bounds are small enough, we see that one should fall into one of the 4 subcases {\bf 5.}(i), (ii), (iii), (iv), and for which the worst-case correlation is  obtained on the  upper or lower correlation bounds.
}
\epR
\end{Remark}
\section{Portfolio Sharpe ratio}
\setcounter{Theorem}{0} \setcounter{Proposition}{0}
\setcounter{Corollary}{0} \setcounter{Lemma}{0}
\setcounter{Definition}{0} \setcounter{Remark}{0}

In this section, we illustrate through two examples how drift estimation error $\delta(t)$ and correlation estimation interval denoted by $\epsilon(t)$ affect the
portfolio Sharpe ratio of a strategy.

\subsection{The impact of drift estimation error  \texorpdfstring{$\delta(t)$}{Lg} on portfolio Sharpe ratio}

We consider a market with one risky asset, and assume that the true dynamics of the stock price is given by the Black-Scholes model
\beqs
dS_t &=& S_t(b^odt + \sigma^o dW_t),
\enqs
where the true drift $b^o$ $>$ $0$ and the true volatility $\sigma^o$ $>$ $0$ are constants, and $W$ is Brownian motion under some  probability measure $\P$. The portfolio Sharpe ratio of a strategy $\alpha$ $\in$ $\Ac$ over the finite horizon $T$ is defined by
\beqs
SR_T(\alpha) &=& \frac{\E[X_T^\alpha] -x_0}{\sqrt{{\rm Var}(X_T^\alpha)}},
\enqs
that is the excess of the expected return per unit of the standard deviation under the true probability measure $\P$.

\vspace{1mm}

\noindent $\bullet$  Let us first consider an  investor who knows the true drift $b^o$ and true volatility $\sigma^o$. In other words, she knows that the stock price is governed by a Black-Scholes model of parameter $(b^o, \sigma^o)$. Therefore, from our Section \ref{nouncertainty}, the optimal mean-variance portfolio strategy of this  investor with risk-aversion parameter $\lambda$ $>$ $0$, and initial capital $x_0$ is given by
\beqs
\alpha_t^* &=& \big[ x_0 + \frac{e^{|\beta^o|^2 T}}{2\lambda} -  X_t^*\big]\frac{b^o}{|\sigma^o|^2},
\enqs
where $\beta^o$ $=$ $b^o/\sigma^o$, and $X_t^*$ is the wealth process with feedback strategy $\alpha^*$. By noting that the evolution of her wealth process $X_t^*$ under $\P$ is
governed by
\beqs
d X_t^* &=& \alpha_t^* b^o dt + \alpha_t^* \sigma^o dW_t,
\enqs
we get, after straightforward calculation,  that her terminal wealth  is given by
\beqs
X_T^* -x_0 &=& \frac{1}{2\lambda} \Big[e^{|\beta^o|^2T} - e^{-\frac{1}{2}|\beta^o|^2T -\beta^o W_T}\Big],\;\;\;\; \P-a.s.
\enqs
Therefore, its  expectation and variance under $\P$ are explicitly given by
\beqs
\E[X_T^*] -x_0 &=& \frac{1}{2\lambda} \big[e^{{|\beta^o|}^2T} -1\big],\;\;\;  {\rm Var}(X_T^*) \;=\; \frac{1}{4\lambda^2}\big[e^{|\beta^o|^2 T}-1],
\enqs
and thus  the portfolio Sharpe ratio of the first investor following a  portfolio strategy $\alpha^*$ is
\beqs
SR_T^{(1)} \; := \;  SR^{}_T(\alpha^*) &=& \sqrt{e^{|\beta^o|^2T}-1}.
\enqs

\vspace{1mm}

\noindent $\bullet$ Let us  next consider a second investor with risk-aversion parameter $\lambda$, initial capital $x_0$, who knows the true volatility but is uncertain about the drift: she
believes that the drift lies in an ellipsoidal set around $b^o$ with constant $\delta$.
From Proposition \ref{alphaellip}, her robust optimal portfolio strategy denoted by $\tilde\alpha$ is given by
\beqs
\tilde\alpha_t &=& \big[x_0 + \frac{1}{2\lambda} e^{(\beta^o-\delta)^2 T}-\tilde X_t\big](1 -\frac{\delta}{\beta^o})1_{\{\beta^o > \delta\}} \frac{b^o}{|\sigma^o|^2},
\enqs
where $\tilde X_t$ is the wealth process associated to $\tilde\alpha$. By noting that the evolution of $\tilde X$ under the true probability measure $\P$ is
\beqs
d\tilde X_t &=& \tilde \alpha_t b^o dt + \tilde\alpha_t\sigma^odW_t,
\enqs
we get its explicit expression  under true probability measure $\P$
\beqs
\tilde X_T -x_0
&=&
\frac{1}{2\lambda} \Big[e^{(\beta^o-\delta)^2T}- e^{\frac{1}{2}(\delta^2 -|\beta^o|^2)T + (\delta -\beta^o)W_T} \Big]1_{\{\beta^o > \delta\}},\;\;\; \P-a.s.
\enqs
It follows that  the excess expected return and variance under $\P$ are explicitly given by
\beqs
\E[\tilde X_T]-x_0 &=& \frac{1}{2\lambda}  \big[e^{(\beta^o-\delta)^2T} - e^{\delta(\delta-\beta^o)T}\big]1_{\{\beta^o> \delta\}},\\
{\rm Var}(\tilde X_T) &=& \frac{1}{4\lambda^2}[e^{(\beta^o-\delta)(\beta^o-3\delta)T} - e^{2\delta(\delta-\beta^o)T}]1_{\{\beta^o > \delta\}}.
\enqs
Therefore, the portfolio Sharpe ratio of the second investor following a  portfolio strategy $\tilde\alpha$ is
\beqs
SR_T^{(2)} \; := \; SR^{}_T(\tilde\alpha)
&=& \frac{e^{\beta^o(\beta^o-\delta)T}-1}{\sqrt{e^{(\beta^o-\delta)^2T}-1}} 1_{\{\beta^o > \delta\}},
\enqs
with the convention that $SR_T^{(2)}$ $=$ $0$ when $\tilde X_T$ $=$ $x_0$.

\vspace{1mm}

\noindent $\bullet$
Finally, let us consider a third investor with risk-aversion parameter $\lambda$, initial capital $x_0$, who knows the true volatility and has ambiguity about the drift which lies in ellipsoidal set around $b^o$, but compared to the second investor, she learns information about drift  by performing MLE or recursive point estimator with new coming observation, so that her estimation error is  $\delta(t)$ $=$ $\frac{\delta}{\sqrt{1+t}}$, hence decreasing with time. It is consistent with \cite{biechecia17} in which the recursive point estimator of drift converges to the true value with convergence rate $\frac{1}{\sqrt{1+t}}$. It follows from Proposition \ref{alphaellip} that an optimal robust portfolio strategy $\hat\alpha$ is
\beqs
\hat\alpha_t &=& [x_0 + \frac{1}{2\lambda} e^{\int_0^T (\beta^o - \frac{\delta}{\sqrt{1+ s}} )^2 1_{\{\beta^o > \frac{\delta}{\sqrt{1+s}}\}} ds} -\hat X_t \big](1 -\frac{\delta}{\beta^o\sqrt{1+t}})1_{\{\beta^o > \frac{\delta}{\sqrt{1+t}}\}}\frac{b^o}{|\sigma^o|^2}.
\enqs
Similarly as the previous two investors,  we compute the terminal  wealth associated to $\hat\alpha_t$ under the true probability measure $\P$ as
\beqs
\hat X_T -x_0
&=& \frac{1}{2\lambda}\Big[e^{\int_0^T(\beta^o-\frac{\delta}{\sqrt{1+t}})^2 1_{\{\beta^o > \frac{\delta}{\sqrt{1+t}}\}}dt}\\
& & \;\;\;\;\; -\; e^{\int_0^T \frac{1}{2}(\frac{\delta^2}{1+t}-|\beta^o|^2)1_{\{\beta^o > \frac{\delta}{\sqrt{1+t}}\}}dt -\int_0^T(\beta^o-\frac{\delta}{\sqrt{1+t}})1_{\{\beta^o > \frac{\delta}{\sqrt{1+t}}\}}dW_t}\Big]\\
&=& \frac{1}{2\lambda} \Big[e^{\int_{t^o}^T(\beta^o-\frac{\delta}{\sqrt{1+t}})^2dt}- e^{\int_{t^o}^T \frac{1}{2}(\frac{\delta^2}{1+t}-|\beta^o|^2)dt -(\beta^o-\frac{\delta}{\sqrt{1+t}})dW_t}\Big]1_{\{\beta^o > \frac{\delta}{\sqrt{T + 1}}\}},
\enqs
where we set $t^o$ $=$ $\max(\frac{\delta^2}{|\beta^o|^2}-1, 0)$.  Therefore, the expectation and variance of $\hat X_T$ under $\P$ are given by
\beqs
\E[\hat X_T]-x_0
&=& \frac{1}{2\lambda} \Big[e^{\int_{t^o}^{T}(\beta^o - \frac{\delta}{\sqrt{1+ t}})^2dt}- e^{\int_{t^o}^{T}(\frac{\delta^2}{1+t}-\frac{\delta\beta^o}{\sqrt{1+t}})dt}\Big]1_{\{\beta^o >\frac{\delta}{\sqrt{T+1}}\}},\\
{\rm Var}(\hat X_T) &=& \frac{1}{4\lambda^2}\Big[e^{\int_{t^o}^T (\beta^o -\frac{3\delta}{\sqrt{1+t}})(\beta^o -\frac{\delta}{\sqrt{1+t}})dt} -e^{\int_{t^o}^T (\frac{2\delta^2}{1+t} -\frac{2\delta\beta^o}{\sqrt{1+t^o}})dt}\Big]1_{\{\beta^o > \frac{\delta}{\sqrt{T+1}}\}}.
\enqs
It follows that  the portfolio Sharpe ratio of the third investor following the strategy $\hat\alpha$ is
\beqs
SR^{(3)}_T \; := \: SR_T(\hat\alpha)  &=&
\frac{e^{\int_{t^o}^T \beta^o(\beta^o-\frac{\delta}{\sqrt{1+t}})dt}-1}{\sqrt{e^{\int_{t^o}^T (\beta^o-\frac{\delta}{\sqrt{1+t}})^2dt}-1}} 1_{\{\beta^o > \frac{\delta}{\sqrt{1+T}}\}}
\enqs
with the convention that $SR_T^{(3)}$ $=$ $0$ when $\hat X_T$ $=$ $x_0$.

\vspace{2mm}

\begin{figure}[htb]
\centering
\subfigure
{
\includegraphics[height=6cm, width=10.5cm]{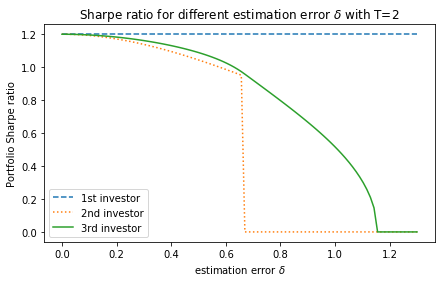}
}
\subfigure
{
\includegraphics[height=6cm, width=10.5cm]{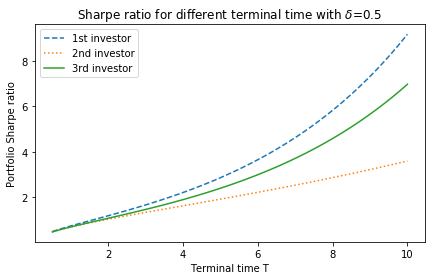}
}

\caption{Portfolio Sharpe ratios for different estimation errors (above) and terminal horizons}
\label{fig diff delta}
\end{figure}

Let us illustrate numerically the effect of the estimation error on the robust optimal portfolio strategy.
We take $b^o$ $=$ $20$$\%$, $\sigma^o$ $=$ $30$$\%$. Figure \ref{fig diff delta} shows the portfolio Sharpe ratio of investors when varying the estimation error $\delta$ at investment horizon $T$ $=$ $2$, and varying terminal time $T$ with estimation error $\delta$ $=$ $0.5$. We see that the Sharpe ratio of the first investor who knows true parameters is always better than the  two other  investors who have ambiguity on drift. Moreover, we notice that the Sharpe ratio decreases as the estimation error increases, and that when investment horizon $T$ is large, the Sharpe ratio of the third investor who learns information with time performs noticeably better than the one of the second investor.
\subsection{The impact of correlation uncertainty level \texorpdfstring{$\epsilon(t)$}{Lg} on portfolio Sharpe ratio}\label{sec5.2}
We consider a market with two risky assets, and assume that the true dynamics of the stock prices $S$ $=$ $(S^1, S^2)$ is governed by
\beqs
dS_t & = &  {\rm diag}(S_t) (b^o dt + \sigma(\rho^o)d W_t)\\
& = &  {\rm diag}(S_t)\left(\left(\begin{matrix}b_1^o \\ b_2^o\end{matrix}\right) dt + \left(\begin{matrix}\sigma_1\sqrt{1 - |\rho^o|^2} & \sigma_1\rho^o \\ 0  & \sigma_2 \end{matrix}\right)\left(\begin{matrix} dW_t^1\\ dW_t^2\end{matrix}\right)
\right),
\enqs
where the drift $b^o$ $=$ $(b^o_1, b^o_2)\trans$, $\sigma_1$ $>$ $0$, $\sigma_2$ $>$ $0$, and the true correlation $\rho^o$ $\in$ $(-1, 1)$ are known constants under some probability measure $\P$. Recall $\beta^o_i$ $=$ $\frac{b_i^o}{\sigma_i}$, $i$ $=$ $1$, $2$, we assume w.l.o.g. $\beta_1^o$ $\geq$ $\beta_2^o$ $>$ $0$, and denote by $\varrho_{12}^o$ $=$ $\frac{\beta_2^o}{\beta_1^o}$ the Sharpe ratio proximity. Note that the true correlation $\rho^o$ is not necessarily equal to $\varrho_{12}^o$.\\

\noindent $\bullet$ Let us consider the first investor who knows the true correlation $\rho^o$. Therefore, from our Section \ref{nouncertainty}, the optimal mean-variance portfolio strategy of this investor with risk-aversion parameter $\lambda$ $>$ $0$, and initial capital $x_0$ is given by
\beqs
\alpha_t^* & = &[x_0 + \frac{1}{2\lambda} e^{R^o T} - X_t^*] \Sigma(\rho^o)^{-1} b^o,
\enqs
where $R^o$ $=$ $(b^o)\trans\Sigma(\rho^o)^{-1} b^o$ and $X_t^*$ is the wealth process with feedback strategy $\alpha^*$. Since her wealth process $X_t^*$ under $\P$ is given by  
\beqs
d X_t^* & = & (\alpha_t^*)\trans( b^o dt +  \sigma(\rho^o) dW_t),
\enqs
 we get, after straightforward calculations, that her  terminal wealth is given by
\beqs
X_T^* - x_0 & =& \frac{1}{2\lambda} \big[ e^{R^o T} - e^{-\frac{1}{2} R^o T - (b^o)\trans\sigma(\rho^o)^{-1} W_T}\big], \;\; \P-a.s.
\enqs
Therefore, its expectation and variance under $\P$ are explicitly given by
\beqs
\E[X_T^*] - x_0 & = & \frac{1}{2\lambda} [e^{R^o T} -1],\;\;\; {\rm Var}(X_T^*) \;=\; \frac{1}{4\lambda^2} [e^{R^o T} -1],
\enqs
hence the portfolio Sharpe ratio of this investor associated to $\alpha^*$ is
\beqs
SR_T^{(1)} : = SR_T(\alpha^*)  = \sqrt{e^{R^o T} -1}.
\enqs
\noindent $\bullet$ Let us consider the second investor who knows the true expected rate of return $b^o$ but is uncertain about correlation.
She believes that the correlation lies in an interval $[\underline\rho, \bar\rho]$ $=$ $[\rho^o -\epsilon, \rho^o + \epsilon]$ $\in$ $(-1, 1)$, with $\epsilon$ a positive constant.
From Theorem \ref{theoopt2}, the robust optimal portfolio strategy denoted by $\tilde\alpha^{(2)}$ of the second investor is given by
\beqs
\tilde\alpha_t^{(2)} &=& \big[x_0 + \frac{1}{2\lambda} e^{R(\rho^{(2),*})T} - \tilde X_t^{(2)}\big]\Sigma(\rho^{(2), *})^{-1} b^o,
\enqs
where $R(\rho)$ $=$ $(b^o)\trans\Sigma(\rho)^{-1}b^o$, $\tilde X_t^{(2)}$ is wealth process associated to $\tilde\alpha^{(2)}$, and
\beq
\label{5rho*2nd}
\rho^{(2), *} &=& \varrho^o_{12}1_{\{\varrho^o_{12}\in [\underline \rho, \bar\rho]\}} + \bar\rho 1_{\{\bar\rho < \varrho^o_{12}\}} + \underline\rho1_{\{\underline \rho > \varrho^o_{12}\}},\\
R(\rho^{(2),*}) &=& |\beta^o_1|^2 1_{\{\varrho^o_{12} \in [\underline\rho, \bar\rho]\}}+ (b^o)\trans\Sigma(\bar\rho)^{-1} b^o 1_{\{\bar\rho < \varrho^o_{12}\}} + (b^o)\trans\Sigma(\underline\rho)^{-1} b^o  1_{\{\underline\rho > \varrho^o_{12}\}} \nonumber.
\enq
By noting that the evolution of $\tilde X^{(2)}$ under probability measure $\P$ is
\beqs
d\tilde X_t^{(2)} &=& (\tilde\alpha_t^{(2)})\trans(b^o dt + \sigma(\rho^o) d W_t),
\enqs
we get its explicit expression under $\P$
\beqs
\tilde X_T^{(2)} - x_0 &=& \frac{1}{2\lambda} \big[e^{R(\rho^{(2),*})T} -e^{-(b^o)\trans\Sigma(\rho^{(2),*})^{-1}\sigma(\rho^o)W_T -\frac{1}{2}(b^o)\trans \Sigma(\rho^{(2),*})^{-1}\Sigma(\rho^o)\Sigma(\rho^{(2),*})^{-1}b^o T}\big],
\enqs
and thus the expectation and variance of $\tilde X_T^{(2)}$ under $\P$ are
\beqs
\E[\tilde X_T^{(2)}] - x_0 &=& \frac{1}{2\lambda} \big[e^{R(\rho^{(2), *})T} -1\big],\\
{\rm Var} (\tilde X_T^{(2)}) &=& \frac{1}{4\lambda^2} (e^{(b^o)\trans \Sigma(\rho^{(2),*})^{-1}\Sigma(\rho^o)\Sigma(\rho^{(2), *})^{-1}b^o T}-1),
\enqs
therefore the portfolio Sharpe ratio of the investor is given by
\beqs
SR_T^{(2)} \;:=\; SR_T(\tilde\alpha^{2}) \;=\; \frac{e^{R(\rho^{(2),*})T} -1}{\sqrt{e^{(b^o)\trans \Sigma(\rho^{(2),*})^{-1}\Sigma(\rho^o)\Sigma(\rho^{(2),*})^{-1}b^o T}-1}}.
\enqs
Substituting $\underline\rho$ $=$ $\rho^o -\epsilon$ and $\bar\rho$ $=$ $\rho^o + \epsilon$ into \reff{5rho*2nd}, we get the explicit form of $\rho^{(2), *}$.
\begin{itemize}
\item [(i)] If $\rho^o$ $=$ $\varrho_{12}^o$, then $\rho^{(2), *}$ $=$ $\varrho_{12}^o$.
\item [(ii)] If $\rho^o$ $<$ $\varrho_{12}^o$, then
\beqs
\rho^{(2), *} &=&
\left\{
\begin{array}{rcl}
\rho^o +\epsilon, & \mbox{ if }\;\; \epsilon < \varrho_{12}^o -\rho^o,\\
\varrho_{12}^o,  & \mbox{ if }\;\; \epsilon \geq \varrho_{12}^o -\rho^o,
\end{array}
\right.\\
\enqs
\item [(iii)] If $\rho^o$ $>$ $\varrho_{12}^o$, then
\beqs
\rho^{(2), *} &=&
\left\{
\begin{array}{rcl}
\rho^o -\epsilon, & \mbox{ if }\;\;\epsilon <\rho^o- \varrho_{12}^o,\\
\varrho_{12}^o,    & \mbox{ if } \;\; \epsilon > \rho^o -\varrho_{12}^o,
\end{array}
\right.
\enqs
\end{itemize}
\noindent $\bullet$  The third investor is more informed than the second investor, and believes that the correlation lies in an interval $[\underline\rho(t), \bar\rho(t)]$ $=$ $[\rho^o-\frac{\epsilon}{\sqrt{1+t}}, \rho^o + \frac{\epsilon}{\sqrt{1+t}}]$ $\subset$ $(-1, 1)$.
From  Theorem \ref{theoopt2}, the robust optimal portfolio strategy denoted by $\tilde\alpha^{(3)}$ of the third investor is given by
\beqs
\tilde\alpha_t^{(3)} &=& \big[x_0 + \frac{1}{2\lambda} e^{\int_0^T R(\rho^{(3),*}(s))ds} - \tilde X_t^{(3)}\big]\Sigma(\rho^{(3), *}(t))^{-1} b^o,
\enqs
where $\tilde X_t^{(3)}$ is wealth process associated to $\tilde\alpha^{(3)}$, and
\beq
\label{5rho*3rd}
\rho^{(3), *}(t) &=& \varrho^o_{12}1_{\{\varrho^o_{12}\in [\underline \rho(t), \bar\rho(t)]\}} + \bar\rho(t) 1_{\{\bar\rho(t) < \varrho^o_{12}\}} + \underline\rho(t)1_{\{\underline \rho(t) > \varrho^o_{12}\}},\\
R(\rho^{(3),*}(t)) &=& |\beta^o_1|^2 1_{\{\varrho^o_{12} \in [\underline\rho(t), \bar\rho(t)]\}}+ (b^o)\trans\Sigma(\bar\rho(t))^{-1} b^o 1_{\{\bar\rho(t) < \varrho^o_{12}\}} + (b^o)\trans\Sigma(\underline\rho(t))^{-1} b^o  1_{\{\underline\rho(t) > \varrho^o_{12}\}}. \nonumber
\enq
The portfolio Sharpe ratio of the third investor is computed in the same way as the second investor, and is given by:
\beqs
SR_T^{(3)} \;:=\; SR_T(\tilde\alpha^{3}) \;=\; \frac{e^{\int_0^T R(\rho^{(3),*}(s))ds} -1}{\sqrt{e^{\int_0^T (b^o)\trans \Sigma(\rho^{(3),*}(s))^{-1}\Sigma(\rho^o)\Sigma(\rho^{(3),*}(s))^{-1}b^o ds}-1}}.
\enqs
Plugging $\underline\rho(t)$ $=$ $\rho^o -\frac{\epsilon}{\sqrt{1+t}}$, and $\bar\rho(t)$ $=$ $\rho^o + \frac{\epsilon}{\sqrt{1+t}}$ into \reff{5rho*3rd}, we get the explicit expression of $\rho^{(3), *}(t)$.
\begin{itemize}
\item [(i)] If $\rho^o$ $=$ $\varrho_{12}^o$, then $\rho^{(3), *}(t)$ $=$ $\varrho_{12}^o$.
\item [(ii)] If $\rho^o$ $<$ $\varrho_{12}^o$, then
\beqs
\rho^{(3), *}(t) &=&
\left\{
\begin{array}{rcl}
\rho^o + \frac{\epsilon}{\sqrt{1+t}}, & &\mbox{ if }\;\; \epsilon < \varrho_{12}^o -\rho^o,\\
\varrho_{12}^o 1_{\{t \in [0, \frac{\epsilon^2}{|\varrho_{12}^o-\rho^o|^2} -1]\}} + (\rho^o + \frac{\epsilon}{\sqrt{1+t}}) 1_{\{t \in [\frac{\epsilon^2}{|\varrho_{12}^o-\rho^o|^2} -1, T]\}}, & &\mbox{ if }\;\; \frac{\epsilon}{\sqrt{1+T}} <  \varrho_{12}^o -\rho^o < \epsilon,\\
\varrho_{12}^o,            & &\mbox{ if }\;\; \frac{\epsilon}{\sqrt{1+T}} \geq  \varrho_{12}^o -\rho^o.
\end{array}
\right.
\enqs
\item [(iii)] If $\rho^o$ $>$ $\varrho_{12}^o$, then
\beqs
\rho^{(3), *}(t) &=&
\left\{
\begin{array}{rcl}
\rho^o -\frac{\epsilon}{\sqrt{1+t}}, & & \mbox{ if }\;\;\epsilon <\rho^o- \varrho_{12}^o,\\
\varrho_{12}^o 1_{\{t \in [0, \frac{\epsilon^2}{|\varrho_{12}^o-\rho^o|^2} -1]\}} + (\rho^o - \frac{\epsilon}{\sqrt{1+t}}) 1_{\{t \in [\frac{\epsilon^2}{|\varrho_{12}^o-\rho^o|^2} -1, T]\}}, & &\mbox{ if } \;\; \frac{\epsilon}{\sqrt{1+T}} <  \varrho^o -\varrho_{12}^o < \epsilon,\\
\varrho_{12}^o,  &  & \mbox{ if }\;\; \frac{\epsilon}{\sqrt{1+T}} > \rho^o -\varrho_{12}^o.
\end{array}
\right.
\enqs
\end{itemize}

\begin{figure}[htb]
\centering
\subfigure
{
\includegraphics[height=6cm, width=10.5cm]{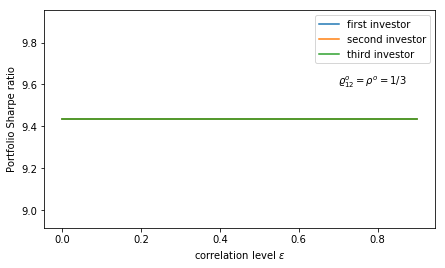}
}
\subfigure
{
\includegraphics[height=6cm, width=10.5cm]{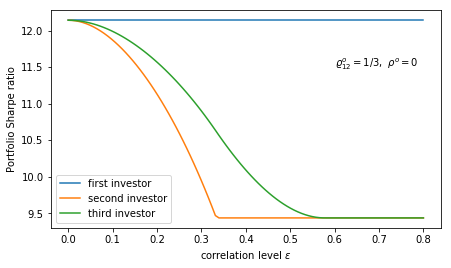}
}
\subfigure
{
\includegraphics[height=6cm, width=10.5cm]{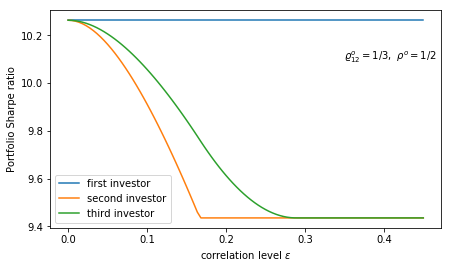}
}
\caption{Portfolio Sharpe ratio for different correlation levels $\epsilon$}
\label{fig diff corrdelta}
\end{figure}

Let us illustrate numerically the effect of the correlation ambiguity on the robust optimal portfolio strategy. We fix investment horizon $T$ $=$ $2$ and take $\beta_1^o$ $=$ 1.5, $\beta_2^o$ $=$ 0.5, hence giving Sharpe ratio proximity $\varrho_{12}^o$ $=$ $\frac{1}{3}$. We select $\rho^o$ $=$ $\varrho_{12}^o$ $=$ $\frac{1}{3}$, or $\rho^o$ $=$ $0$ $<$ $\varrho_{12}^o$, or $\rho^o$ $=$ $\frac{1}{2}$ $>$ $\varrho_{12}^o$. Figure \ref{fig diff corrdelta} shows the effect of the correlation ambiguity level 
on the portfolio Sharpe ratio. We see that portfolio Sharpe ratio decreases with correlation ambiguity level except when  true correlation equals Sharpe ratio proximity.  In this case, when true correlation equals Sharpe ratio proximity under which under-diversification occurs,  whatever the correlation level $\epsilon$ is, portfolio Sharpe ratio is a constant.

\paragraph{Conclusion of Part II.} 
In this part, we have provided a complete picture for two-asset and three-asset setting. The same patterns can hold true for $d-1$-asset and $d$-asset settings with $d>3$. More precisely, we can apply the following induction argument. 
Suppose that we have all the scenarios for $d-1$ assets: anti-diversification(investment in one asset), under-diversification (investment in $2, \ldots, d-2$ assets), and well-diversification (investment in $d-1$ assets),  then we obtain a complete picture of $d$ assets depending on whether $d$-th asset is invested or not,  which corresponds to the evaluation of variance risk ratios $\hat\kappa^j(\rho)$, $\hat\kappa^d(\rho)$ on $\hat\varrho_{jd}$, $\underline\rho_{jd}$, $\bar\rho_{jd}$, $1 \leq j \leq d-1$. Mathematically, we have to compute the minima of risk premium function (convex) on $\prod_{1 \leq i \neq  j \leq d}[\underline\rho_{ij}, \bar\rho_{ij}]  \subset \C_{>+}^d$. Since there are too many sub-cases, we do not provide all the details for $d >3$.

\bigskip
{\bf Acknowledgement:}
We would like to thank the Associate Editor and the two referees for their careful reading of our paper and their valuable comments. We feel that, thanks to all the comments we received, the clarity and the scientific value of the paper have been greatly improved.

\clearpage


\appendix

\setcounter{Theorem}{0} \setcounter{Proposition}{0}
\setcounter{Corollary}{0} \setcounter{Lemma}{0}
\setcounter{Definition}{0} \setcounter{Remark}{0}

\renewcommand\thesection{Appendix}

\section{}

\renewcommand\thesection{\Alph{subsection}}

\renewcommand\thesubsection{\Alph{subsection}}

\subsection{Separation principle for robust portfolio selection}

\setcounter{Theorem}{0} \setcounter{Proposition}{0}
\setcounter{Corollary}{0} \setcounter{Lemma}{0}
\setcounter{Definition}{0} \setcounter{Remark}{0}

We show in this section that the separation principle is actually quite general,  valid not only for mean-variance problem, but also for other classes in decision making problems,
like popular expected utility criteria. It turns out that the proof of the separation principle for standard utility criteria
is simpler than  for mean-variance problems.

We consider a model uncertainty setting as in Section \ref{subsecmodel}.
The investor's preferences are  represented by utility functions $U$ defined from $\R$ to $\R$,  strictly concave and  increasing.
The wealth process $X^\alpha$ is defined as in \reff{Xalphadyna},
and the set $\Ac$  of admissible controls consists of  $\R^d$-valued $\F$-progressively measurable process $\alpha$ such that
the process $U(X^\alpha)$ is uniformly integrable.   The robust portfolio selection problem is then formulated as
\beq \label{robustutility}
V_0 &=& \sup_{\alpha\in\Ac} \inf_{\theta_.\in\Vc_\bTheta} \E_{\theta_.} \big[ U(X_T^\alpha)].
\enq
This is a rather standard min-max stochastic control problem, which is associated by the dynamic programming principle to the Bellman-Isaacs partial differential equation:
\begin{equation}
\left\{
\begin{array}{rcl}
\Dt{v}  + \Sup_{a\in\R^d} \Inf_{\theta\in\Theta(t)} H_t( v_x,  v_{xx},a,\theta) &=& 0, \;\;\;\;\;\;\;  \mbox{ on }   [0,T)\times \R \\
v(T,.) &=& U, \;\;\;
\end{array}
\right.
\end{equation}
(assuming that $v(t,x)$ is smooth and strictly concave in $x$), where  for $t$ $\in$ $[0,T]$, $H_t$ is the Hamiltonian  function defined on $\R\times(-\infty,0)\times\R^d\times\Theta(t)$   by
\beqs
H_t(p,M,a,\theta) &=& p a\trans b + \frac{1}{2} M a\trans \Sigma(\rho) a,  \;\;\;\;\;  p \in \R, M < 0,  a \in \R^d, \theta =(b,\rho) \in \Theta(t).
\enqs

In the no-uncertainty model case, i.e., $\Theta(t)$ is reduced to a singleton $\Theta(t)$ $=$ $\{\theta^o(t)= (b^o(t),\rho^o(t))\}$, $t$ $\in$ $[0,T]$, corresponding to a multi-dimensional Black-Scholes model with deterministic mean return vector $b^o(t)$, covariance matrix $\Sigma^o(t)$ $=$ $\Sigma(\rho^o(t))$, and determi\-nistic risk premium
$R^o(t)$ $=$ $R(\theta^o(t))$ $=$ $(b^o(t))\trans(\Sigma^o(t))^{-1}b^o(t)$,
the Bellman-Isaacs equation reduces to the Bellman equation arising in  classical expected utility maximization, and called Black-Scholes-Merton Bellman PDE:
\begin{equation} \label{HJBstandardMerton}
\left\{
\begin{array}{rcl}
\Dt{v^o}  - \frac{R^o(t)}{2} \frac{(v_x^o)^2}{v_{xx}^o} &=& 0,  \;\;\;\;\;\;\;  \mbox{ on }   [0,T)\times \R \\
v^o(T,.) &=& U.
\end{array}
\right.
\end{equation}
Moreover, when there exists a smooth  function $v^o(t,x)$ to \reff{HJBstandardMerton}, strictly concave in $x$, it is known by classical verification theorem (see e.g. \cite{pha09})  that the optimal portfolio strategy is given by
\beqs
\alpha_t^{o,*} &=&
 \Rc^o(t,X_t^*) (\Sigma^o(t))^{-1} b^o(t), \;\;\; 0 \leq t \leq T,
\enqs
where $\Rc^o(t,x)$ $:=$ $-\frac{v_x^o(t,x)}{v_{xx}^o(t,x)}$ is the so-called risk tolerance function,
$X^*$ is the wealth process associated to $\alpha^*$. For example, when $U$ is a CRRA utility function, i.e., $U(x)$ $=$ $x^\gamma$, $x$ $>$ $0$, with $0<\gamma< 1$,
we obtain the famous Merton solution: $v^o(t,x)$ $=$  $\exp\big(\frac{\int_t^T R^o(s) ds}{2}\frac{\gamma}{1-\gamma} \big)U(x)$, and $\Rc^o(t,x)$ $=$ $x/(1-\gamma)$.
When $U$ is of CARA type, i.e., $U(x)$ $=$ $-e^{-\eta x}$, with $\eta$ $>$ $0$, we have
$v^o(t,x)$ $=$  $\exp\big(\frac{\int_t^T R^o(s) ds}{2} \big)U(x)$, and $\Rc^o(t,x)$ $=$ $1/\eta$.

\vspace{3mm}

In our general model uncertainty setting under {\bf (H$\bTheta$)}, a key lemma is the following saddle point property:

\begin{Lemma} \label{lemsaddle2}
Fix $t$ $\in$ $[0,T]$, and assume that there exists $\theta^*(t)$ $=$ $(b^*(t),\rho^*(t))$ $\in$ ${\rm arg}\Min_{\theta\in\Theta(t)} R(\theta)$.  Let us denote by
\beq \label{defa*}
a^*(t,p,M) & = & - \frac{p}{M} (\Sigma(\rho^*(t))^{-1} b^*(t), \;\;\; p \in \R, \; M < 0.
\enq
Then, for $p$ $\in$ $\R$, $M$ $<$ $0$,  the pair $(a^*(t,p,M),\theta^*(t))$ is a saddle-point of $(a,\theta)$ $\in$ $\R^d\times\Theta(t)$ $\mapsto$ $H_t(p,M,a,\theta)$, i.e.
\beqs
H_t(p,M,a^*(t,p,M),\theta) & \leq &  H_t(p,M,a^*(t,p,M),\theta^*(t)) \; = \;  - \frac{1}{2} \frac{p^2}{M} R(\theta^*(t)) \\
& \leq & H_t(p,M,a,\theta^*(t)), \;\;\; \forall a \in \R^d,\; \theta \in \Theta(t).
\enqs
\end{Lemma}
{\bf Proof.}  For any $n$ $\in$ $\N\setminus\{0\}$,
let us introduce the compact set $\C^d_{n,>+}$  of all elements $\rho$ $=$ $(\rho_{ij})_{1\leq i < j\leq d}$ in the open set $\C^d_{>+}$, such that
$|\rho_{ij}| \leq 1 - \frac{1}{n}, \;  1\leq i < j\leq d$. Given the ambiguity set $\Theta(t)$, let us consider  the sequence of sets:
\beqs
\Theta_n(t) & = & \Big\{ \theta = (b,\rho)  \in \Theta(t):  \rho \in \C^d_{n,>+}  \Big\}, \;\;\; n >0,
\enqs
and notice by {\bf (H$\bTheta$)} that its closure $\overline{\Theta_n(t)}$ is a compact convex set of $\R^d\times\C^d_{>+}$. For fixed  $(t,p,M)$ $\in$
$[0,T]\times\R\times(-\infty,0)$, it is clear that the function $H_t(p,M,.,.)$ is concave in $a$ $\in$ $\R^d$, and  linear (hence convex) in $\theta$ lying in the convex-compact set
$\overline{\Theta_n(t)}$.  By the min-max theorem (see e.g. Theorem 45.8 in \cite{stra85}), we then get the so-called Isaacs relation:
\beq \label{IsaacsHt}
\sup_{a \in \R^d} \inf_{\theta \in \overline{\Theta_n(t)}} H_t(p,M,a,\theta) & = &   \inf_{\theta \in \overline{\Theta_n(t)}}  \sup_{a \in \R^d}   H_t(p,M,a,\theta).
\enq
By square completion, we  can rewrite the function $H_t$  as:
\beq \label{squareH}
H_t(p,M,a,\theta) &=& \frac{M}{2} \big( a - \bar a(p,M,\theta) \big)\trans\Sigma(\rho)(a - \bar a(p,M,\theta)\big)
- \frac{1}{2} \frac{p^2}{M} R(\theta), \\
\mbox{ with } \;\;\; \bar a(p,M,\theta) &:=& -\frac{p}{M}\Sigma^{-1}(\rho)b, \;\;\;\;\; \theta = (b,\rho) \in \Theta(t),  \nonumber
\enq
from which we get
\beq \label{infH}
\sup_{a \in \R^d}  H_t(p,M,a,\theta) &=&  H_t(p,M,\bar a(p,M,\theta),\theta) \; = \;    - \frac{1}{2} \frac{p^2}{M} R(\theta).
\enq
Observe that for $n$ large enough, $n$ $\geq$ $N^*$,  the element $\theta^*(t)$ $=$
$(b^*(t),\rho^*(t))$ $\in$ ${\rm arg}\Min_{\theta\in\Theta(t)} R(\theta)$ lies in $\overline{\Theta_n(t)}$, and thus, using also the continuity of $R(.)$ on $\R^d\times\C^d_{>+}$:
$\Inf_{\theta \in \Theta(t)} R(\theta)$ $=$ $\Inf_{\theta \in \overline{\Theta_n(t)}}  R(\theta)$ $=$ $R(\theta^*(t))$. We deduce with \reff{infH} that for $n$ $\geq$ $N^*$,
\beq
H_t^*(p,M) & := &  \inf_{\theta \in \Theta(t)}  \sup_{a \in \R^d}   H_t(p,M,a,\theta) \; = \; \inf_{\theta \in \overline{\Theta_n(t)}}  \sup_{a \in \R^d}   H_t(p,M,a,\theta) \nonumber \\
&=&  - \frac{1}{2} \frac{p^2}{M} R(\theta^*(t)).   \label{interH*}
\enq

On the other hand, we see that the continuous function $a$ $\in$ $\R^d$ $\mapsto$ $\underline{H}_t(p,M,a)$ $:=$
$\inf_{\theta \in \overline{\Theta_n(t)}} H_t(p,M,a,\theta)$ is concave in $a$, and goes  to $-\infty$, as $|a|$ goes to infinity (recall that $M$ $<$ $0$).  This implies that
$\underline{H}_t(p,M,.)$ attains its supremum at some point $\tilde a(t,p,M)$, and we then have
\beqs
\inf_{\theta \in \overline{\Theta_n(t)}} H_t(p,M,\tilde a(t,p,M),\theta) &=&  \sup_{a \in \R^d} \inf_{\theta \in \overline{\Theta_n(t)}} H_t(p,M,a,\theta)  \\
&=&  \inf_{\theta \in \overline{\Theta_n(t)}}  \sup_{a \in \R^d}   H_t(p,M,a,\theta)  \;= \; H_t^*(p,M) \\
&=&  \sup_{a \in \R^d}  H_t(p,M,a,\theta^*(t)) \; \geq \;  H_t(p,M,a,\theta^*(t)), \;\;\; \forall a \in \R^d,
\enqs
where we used Isaacs condition \reff{IsaacsHt}  in the second equality,  \reff{interH*} in the third equality, and \reff{infH} for $\theta$ $=$ $\theta^*(t)$ in the fourth one.
We then deduce
\beq
H_t(p,M,\tilde a(t,p,M),\theta^*(t)) & \geq &  \inf_{\theta \in \overline{\Theta_n(t)}} H_t(p,M,\tilde a(t,p,M),\theta) \; = \; H_t^*(p,M)  \nonumber \\
& \geq &  H_t(p,M,a,\theta^*(t)), \;\;\; \forall a \in \R^d. \label{saddle1Ht}
\enq
Similarly, we have for any $n$ $\geq$ $N^*$,
\beqs
\sup_{a \in \R^d } H_t(p,M,a,\theta^*(t)) &=&  \inf_{\theta \in \overline{\Theta_n(t)}}  \sup_{a \in \R^d}  H_t(p,M,a,\theta)  \\
&=&    \sup_{a \in \R^d}  \inf_{\theta \in \overline{\Theta_n(t)}}     H_t(p,M,a,\theta) \;= \; H_t^*(p,M) \\
&=&    \inf_{\theta \in \overline{\Theta_n(t)}}  H(p,M,\tilde a(t,p,M),\theta) \; \leq \;  H_t(p,M,\tilde a(t,p,M),\theta), \;\;\; \forall \theta \in \overline{\Theta_n(t)},
\enqs
which implies that
\beq
H_t(p,M,\tilde a(t,p,M),\theta^*(t)) & \leq &  \sup_{a \in\R^d} H_t(p,M,a,\theta^*(t)) \;  = \;  H_t^*(p,M)  \nonumber \\
& \leq &  H_t(p,M,\tilde a(t,p,M),\theta), \;\;\; \forall \theta \in \Theta(t), \label{saddle2Ht}
\enq
since any $\theta$ $\in$ $\Theta(t)$ lies in $\Theta_n(t)$ for  $n$ large enough.
Relations \reff{saddle1Ht}-\reff{saddle2Ht} mean the saddle-point property of the pair $(\tilde a(t,p,M),\theta^*(t))$ for the function
$(a,\theta)$ $\in$ $\R^d\times\Theta(t)$ $\mapsto$ $H_t(p,M,a,\theta)$, and also imply that
$H_t(p,M,\tilde a_t(p,M),\theta^*(t))$ $=$ $H_t^*(p,M)$.  Recalling the expression \reff{squareH} of $H_t$, this is written as
\beqs
& & \frac{M}{2} \big( \tilde a(t,p,M) -  \bar a(p,M,\theta^*(t)) \big)\trans\Sigma(\rho^*(t))(\tilde a(t,p,M) -  \bar a(p,M,\theta^*(t))  \big)  +  H_t^*(p,M) \\
&=& H_t^*(p,M).
\enqs
This proves that $\tilde a(t,p,M)$ $=$ $\bar a(p,M,\theta^*(t))$ $=$ $a^*(t,p,M)$ as defined in \reff{defa*}, and ends the proof.
\ep

\vspace{3mm}

\begin{Proposition}
[Separation Principle for utility criteria]
\label{robustoptimalU}
Suppose that  there exists a pair $\btheta^*= (\theta^*(t))_t = (\bb^*,\brho^*) = (b^*(t),\rho^*(t))_t$ $\in$  $\bTheta$ solution to
$\arg\Min_{\btheta \in \bTheta} \bR(\btheta)$, i.e., $\theta^*(t)$ $\in$ $\arg\Min_{\theta \in \Theta(t)} R(\theta)$, for all $t$ $\in$ $[0,T]$, and a  smooth solution
$v(t,x)$, strictly concave in $x$,  to the Black-Scholes Merton Bellman PDE:
\begin{equation} \label{HJBstandard}
\left\{
\begin{array}{rcl}
\Dt{v}  - \frac{R^*(t)}{2} \frac{(v_x)^2}{v_{xx}} &=& 0,  \;\;\;\;\;\;\;  \mbox{ on }   [0,T)\times \R \\
v(T,.) &=& U,
\end{array}
\right.
\end{equation}
where we set $R^*(t)$ $:=$ $R(\theta^*(t))$, and satisfying the growth condition $v(t,x)$ $\leq$ $C(1+ |U(x)|)$.
Then the robust  utility maximization  problem \reff{robustutility} admits an optimal portfolio strategy given by
\beqs
\alpha_t^{*} &=&
 \Rc^*(t,X_t^*) (\Sigma^*(t))^{-1} b^*(t), \;\;\; 0 \leq t \leq T,  \;\;  \Pc^\bTheta-q.s.,
\enqs
where $\Rc^*(t,x)$ $:=$ $-\frac{v_x(t,x)}{v_{xx}((,x)}$,  $\Sigma^*(t)$ $:=$ $\Sigma(\rho^*(t))$, and $X^*$ is the state process associated to $\alpha_t^*$.  Moreover,
\beq
V_0 \; = \;  v(0,x_0) & = &  \sup_{\alpha\in\Ac} \inf_{\theta_.\in\Vc_\bTheta} \E_{\theta_.} \big[ U(X_T^\alpha)] \; = \;
\E_{\btheta^*} \big[ U(X_T^{\alpha^*})] \nonumber \\
&=& \inf_{\theta_.\in\Vc_\bTheta}   \sup_{\alpha\in\Ac}  \E_{\theta_.} \big[ U(X_T^\alpha)].  \label{valuesaddle}
\enq
\end{Proposition}
{\bf Proof.} For any $\alpha$ $\in$ $\Ac$,  and $\theta_.$ $\in$ $\Vc_\bTheta$, the dynamics of $v(t,X_t^\alpha)$ under $\P^{\theta_.}$ is given by It\^o's formula by
\beqs
d v(t,X_t^\alpha) &=&  D_t^{\alpha,\theta_.} dt + v_x(t,X_t^\alpha) \alpha_t\trans \sigma(\rho_t) dW_t^\theta,
\enqs
where
\beqs
D_t^{\alpha,\theta_.} &=& \Dt{v}(t,X_t^\alpha) +  H_t(v_x(t,X_t^\alpha),v_{xx}(t,X_t^\alpha),\alpha_t,\theta_t), \;\;\; 0 \leq t \leq T.
\enqs
Observe that $\alpha_t^*$ $=$ $a^*(t,v_x(t,X_t^*),v_{xx}(t,X_t^*))$ as defined in Lemma \ref{lemsaddle2}, and the Black-Scholes Merton Bellman PDE for $v$ is written as
\beqs
\Dt{v}(t,x) + H_t(v_x(t,x),v_{xx}(t,x),a^*(t,v_x(t,x),v_{xx}(t,x)),\theta^*(t))) &=& 0, \;\;\; (t,x) \in [0,T) \times\R.
\enqs
From the saddle point property in this Lemma \ref{lemsaddle2},, we then have
\beqs
D_t^{\alpha,\btheta^*} & \leq \;  0  & \leq \;  D_t^{\alpha^*,\theta_.}, \;\;\; 0 \leq t < T, \;\; \forall \alpha\in\Ac, \theta_. \in \Vc_\bTheta.
\enqs
This implies that the process $(v(t,X_t^\alpha))_t$  is a local supermartingale under $\P^{\btheta^*}$, for any $\alpha$ $\in$ $\Ac$, while
$(v(t,X_t^{\alpha^*}))_t$  is a local submartingale under $\P^{\theta_.}$, for any $\theta_.$ $\in$ $\Vc_\bTheta$.
By considering a sequence of localizing stopping times $(\tau_n)_n$ converging to $T$ as $n$ goes to infinity, we then have
\beqs
\E_{\btheta^*}[v(\tau_n,X_{\tau_n}^\alpha)] \; \leq \; v(0,X_0) \; \leq \; \E_{\theta_.}[ v(\tau_n,X_{\tau_n}^{\alpha^*}], \;\;\; \forall \alpha \in \Ac, \; \theta_. \in \Vc_\bTheta.
\enqs
From the growth condition on $v$, and as $U(X^\alpha)$ is uniformly integrable for $\alpha$ $\in$ $\Ac$, we deduce by sending $n$ to infinity, and recalling that $v(T,x)$ $=$ $U(x)$:
\beqs
\E_{\btheta^*}[U(X_{T}^\alpha)] \; \leq \; v(0,x_0) \; \leq \; \E_{\theta_.}[ U(X_{T}^{\alpha^*})], \;\;\; \forall \alpha \in \Ac, \; \theta_. \in \Vc_\bTheta.
\enqs
As the deterministic process $\btheta^*$ lies in particular in $\Vc_\bTheta$, and noting that $\sup_{\alpha\in\Ac} \inf_{\theta_.\in\Vc_\bTheta}$ $\leq$
$\inf_{\theta_.\in\Vc_\bTheta} \sup_{\alpha\in\Ac}$, this above saddle-point relation yields \reff{valuesaddle}.
\ep

\vspace{2mm}

\begin{Remark}
{\rm A similar separation principle holds for robust portfolio selection in a discrete-time setting. We first compute at any time $t$ $=$ $0,\ldots,T-1$,
the parameter $\theta^*(t)$ $=$ $(b^*(t),\rho^*(t))$, which achieves the minimum of the risk premium function $R(\theta)$ $=$ $b\trans\Sigma(\rho)^{-1}b$, over
$\theta$ $=$ $(b,\rho)$ lying in the ambiguity set $\Theta(t)$ at time $t$. The solution to the robust portfolio selection problem is then given by the solution to the
portfolio selection problem in the discrete-time model with mean return  $b^*(t)$ and covariance matrix $\Sigma(\rho^*(t))$ at time $t$.
}
\ep
\end{Remark}

\subsection{Proofs of some Lemmas, Propositions and Theorems}

\setcounter{Theorem}{0} \setcounter{Proposition}{0}
\setcounter{Corollary}{0} \setcounter{Lemma}{0}
\setcounter{Definition}{0} \setcounter{Remark}{0}

\noindent {\bf Notations, differentiation and characterization of convex function}

\vspace{1mm}

\noindent Let us introduce some notations and state some results which will be used frequently in the proof of some Lemmas and Propositions.

\begin{itemize}
\item [\bf 1.] We introduce the so-called variance risk ratios
\beqs
\label{hatkappa}
\hat\kappa_t(\rho) & := & \Sigma(\rho)^{-1}\hat b(t)  \;=\; (\hat\kappa_t^1(\rho), \ldots, \hat\kappa_t^d(\rho))\trans,\\
\label{kappabrho}
\kappa(b, \rho) & := & \Sigma(\rho)^{-1} b  \;=\;(\kappa^1(b, \rho), \ldots, \kappa^d(b, \rho))\trans.
\enqs
\item [\bf 2.]From some matrix calculations (see e.g. corollary 95 and corollary 105 in \cite{Dhr78}), we obtain the explicit expressions of the first partial derivatives of $R(b, \rho)$ $=$
$b\trans\kappa(b,\rho)$ with respect to $b_i$, $\rho_{ij}$ denoted by $\frac{\partial R(b, \rho)}{\partial b_i}$ and $\frac{\partial R(b, \rho)}{\partial \rho_{ij}}$, $1$ $\leq$ $i$ $<$ $j$ $\leq$ $d$,
\beq \label{firstorder}
\frac{\partial R(b, \rho)}{\partial b_i} \; =  \;  2\kappa^i(b, \rho),& &  \frac{\partial R(b, \rho)}{\partial \rho_{ij}} \;=\; -\sigma_i\sigma_j\kappa^i(b, \rho)\kappa^j(b, \rho).
\enq
 We also denote by
$\nabla_b R(b, \rho)$ and $\nabla_\rho R(b, \rho)$ the gradients of $R(b, \rho)$ with respect to $b$ and $\rho$ respectively,
\beq \label{nablabrho}
\left\{
\begin{array}{rcl}
\nabla_b R(b, \rho) &= & (\frac{\partial R(b, \rho)}{\partial b_1}, \ldots, \frac{\partial R(b, \rho)}{\partial b_d})\trans \\
\nabla_\rho R(b, \rho)&=& (\frac{\partial R(b, \rho)}{\partial \rho_{12}}, \ldots, \frac{\partial R(b, \rho)}{\partial \rho_{1d}}, \ldots, \frac{\partial R(b, \rho)}{\partial \rho_{(d-1)d}})\trans
\end{array}
\right.
\enq

\item [\bf 3.]({\it Sufficient and necessary optimality condition}). It is known (see e.g. Lemma 2.2 in \cite{DingChenHuang18}) that $R(b, \rho)$ is jointly convex in $b$ and $\rho$.
Similarly, $R(\hat b(t), \rho)$ is convex in $\rho$. Then   $\rho^*(t)$ is a global minimum of $\rho$ $\mapsto$ $R(\hat b(t), \rho)$ over $\Gamma(t)$ convex set of $\C^d_+$
if and only if, for any $\rho$ $\in$ $\Gamma(t)$ (see e.g. section 4.2.3 in \cite{BoydVand04}),
    \beqs
  (\rho -\rho^*(t))\trans \nabla_\rho R(\hat b(t), \rho^*(t)) &=&  \Sum_{j=1}^d\Sum_{i=1}^{j-1} \frac{\partial R(\hat b(t), \rho^*(t))}{\partial \rho_{ij}}(\rho_{ij} -\rho_{ij}^*) \; \geq \;  0,
    \enqs
    which is written from  \reff{firstorder} as,
    \beq \label{minimumellip}
    \Sum_{j=1}^d \Sum_{i=1}^{j-1}\sigma_i\sigma_j\hat\kappa_t^i\hat\kappa_t^j(\rho^*(t))(\rho_{ij} -\rho_{ij}^*(t)) & \leq &  0.
    \enq

\end{itemize}

\subsubsection{Proof of Lemma \ref{lemH}} \label{appenlemH}

\setcounter{Theorem}{0} \setcounter{Proposition}{0}
\setcounter{Corollary}{0} \setcounter{Lemma}{0}
\setcounter{Definition}{0} \setcounter{Remark}{0}

\noindent  Notice that if there exists $(b^*(t), \rho^*(t))$ $\in$ $\arg\Min_{\theta\in\Theta(t)} R(\theta)$, then the first-order condition implies that for any $(b, \rho)$ lying in the convex set
 $\Theta(t)$:
\beq \label{appenlemHprofprod}
(b-b^*(t))\trans \nabla_b R(\theta^*(t))+(\rho-\rho^*(t))\trans \nabla_\rho R(\theta^*(t))\; \geq\; 0.
\enq
Recalling the expression of $H_t(b, \rho)$ in \reff{defH} and explicit expressions \reff{firstorder}, \reff{nablabrho} of $\nabla_b R(\theta^*(t))$ and $\nabla_\rho R(\theta^*(t))$, we have
\beqs
H_t(b^*(t), \rho)- H_t(b^*(t), \rho^*(t))&=& \Sum_{j=1}^d\Sum_{i=1}^{j-1}\kappa^i(b^*(t), \rho^*(t))\kappa^j(b^*(t), \rho^*(t))\sigma_i\sigma_j(\rho_{ij}-\rho_{ij}^*(t)) \\
&=& (\rho^*(t)-\rho)\trans \nabla_\rho R(\theta^*(t)),  \\
2\big(H_t(b^*(t), \rho^*(t)) - H_t(b, \rho^*(t))\big)&=& 2\Sum_{i=1}^d(b_i^*(t)-b_i)\kappa^i(b^*(t), \rho^*(t))\\
 &=&(b^*(t)-b)\trans \nabla_b R(\theta^*(t)),
\enqs
where by convention, we set:  $\Sum_{i=1}^0 \cdot$ $=$ $0$ .
It follows immediately from the sum of the above two equalities that
\beqs
& & H_t(b^*(t), \rho^*(t)) - 2H_t(b, \rho^*(t)) + H_t(b^*(t), \rho)\\
&=&  (\rho^*(t)-\rho)\trans \nabla_\rho R(\theta^*(t)) + (b^*(t)-b)\trans \nabla_b R(\theta^*(t)) \; \leq \; 0,
\enqs
where we used \reff{appenlemHprofprod} in the last inequality.

\ep

\vspace{3mm}

\subsubsection{Proof of Proposition \ref{alphaellip} } \label{appenlemellipoid}

\setcounter{Theorem}{0} \setcounter{Proposition}{0}
\setcounter{Corollary}{0} \setcounter{Lemma}{0}
\setcounter{Definition}{0} \setcounter{Remark}{0}

Proposition \ref{alphaellip} is an immediate combination of Theorem \ref{robustoptimal} and the following Lemma \ref{lemellipsoid}.
\begin{Lemma} \label{lemellipsoid}
Fix $t$ $\in$ $[0, T]$. Let  $\Theta(t)$ be an ellipsoidal set as in \reff{Thetabeyond} with $p$ $=$ $1$,  and assume that there exists $\rho^*(t)$ $\in$ ${\rm arg}\Min_{\rho\in\Gamma(t)} R(\hat b(t),\rho)$ $=$
${\rm arg}\Min_{\rho\in\Gamma(t)}\big\|\sigma(\rho)^{-1} \hat b(t)\big\|_{_2}$.
Then $\theta^*(t)$ $=$ $(b^*(t),\rho^*(t))$ with
\beq
\label{b*ellip}
b^*(t) &=&  \Big(1 -\frac{\delta(t)}{\|\sigma(\rho^*(t))^{-1}\hat b(t)\|_{_2}} \Big)  1_{\{\|\sigma(\rho^*(t))^{-1}\hat b(t)\|_{_2} > \delta(t)\}} \; \hat b(t), \\
\mbox{ and } \;\;\; R(\theta^*(t)) &=&  \big( \|\sigma(\rho^*(t))^{-1}\hat b(t)\|_{_2} -\delta(t) \big)^21_{\{\|\sigma(\rho^*(t))^{-1} \hat b(t)\|_{_2} > \delta(t)\}}.
\label{Rtheta*}
\enq
\end{Lemma}
{\bf Proof of Theorem \ref{lemellipsoid}} Due to the dependence of $b$ on $\rho$ in the ellipsoidal set $\Theta(t)$ written as $\Theta(t)$ $=$ $\{ (b,\rho) \in \R^d\times\Gamma(t): b \in \Delta_\rho(t)\}$ where
$\Delta_\rho(t)$ $: =$ $\{b \in \R^d: \|\sigma(\rho)^{-1}(b-\hat b(t))\|_2 \leq \delta(t)\}$, we use a Lagrangian approach.
\vspace{1mm}

\noindent For fixed $\rho$ $\in$ $\Gamma(t)$,  let us first focus on the inner minimization
\beq \label{minRellip}
\Min_{b \in \Delta_\rho(t)} R(b, \rho).
\enq
The Lagrangian  with nonnegative multiplier $\mu$ associated to this constrained minimization problem is
\beq \label{L1bmu}
L_t^1(b, \mu) &=& R(b, \rho)-\mu\Big(|\delta(t)|^2 -(b-\hat b(t))\trans \Sigma(\rho)^{-1}(b-\hat b(t)) \Big),
\enq
and the first-order condition gives
\beqs
\frac{\partial L_t^1(b, \mu)}{\partial b} &=& 2 \Sigma(\rho)^{-1} b + 2\mu\Sigma(\rho)^{-1}(b-\hat b(t)) \;=\; 0, \\
\frac{\partial L_t^1(b, \mu)}{\partial \mu} &=& |\delta(t)|^2 -(b-\hat b(t))\trans \Sigma(\rho)^{-1}(b-\hat b(t)) \; = \; 0.
\enqs
Solving these two equations for fixed $\rho$, and recalling that the Lagrange multiplier is nonnegative, yield
\begin{equation} \label{b*rho}
\left\{
\begin{array}{rcl}
\mu_t^*(\rho) & =& (\frac{\|\sigma(\rho)^{-1}\hat b(t)\|_{_2}}{\delta(t)}-1)1_{\{\|\sigma(\rho)^{-1}\hat b(t)\|_{_2} > \delta(t)\}}, \\
b_t^*(\rho) & = & \hat b(t)(1 -\frac{\delta(t)}{\|\sigma(\rho)^{-1}\hat b(t)\|_{_2}}) 1_{\{\|\sigma(\rho)^{-1}\hat b(t)\|_2 > \delta(t)\}}.
\end{array}
\right.
\end{equation}
Substituting these expressions into  the Lagrangian \reff{L1bmu}, we get
\beqs
L_t^1(b_t^*(\rho), \rho) &=& R(b_t^*(\rho), \rho) \; = \;  \big(\|\sigma(\rho)^{-1}\hat b(t)\|_{_2} -\delta(t) \big)^2 1_{\{\|\sigma(\rho)^{-1}\hat b(t)\|_2 > \delta(t)\}},
\enqs
and thus, the original problem $\Inf_{\Theta(t)} R(\theta)$ is reduced to
\beq
\Inf_{\theta=(b,\rho) \in \Theta(t)}R(\theta) &=& \Inf_{\rho \in \Gamma(t)}\Inf_{b \in \Delta_\rho(t)} R(b, \rho) \; = \; \Inf_{\rho \in \Gamma(t)}  R(b_t^*(\rho),\rho) \nonumber \\
&=& \Inf_{\rho \in \Gamma(t)} \Big\{ \big(\|\sigma(\rho)^{-1}\hat b(t)\|_{_2} -\delta(t)\big)^2 1_{\{\|\sigma(\rho)^{-1}\hat b(t)\|_{_2} > \delta(t)\}} \Big\} \nonumber\\
&=&\Big(\Inf_{\rho \in \Gamma(t)}\|\sigma(\rho)^{-1}\hat b(t)\|_{_2} -\delta(t)\Big)^2 1_{\big\{\Inf_{\rho \in \Gamma(t)}\|\sigma(\rho)^{-1}\hat b(t)\|_{_2} > \delta(t)\big\}}. \label{MinRexpl}
\enq
Therefore, whenever  $\rho^*(t)$ $\in$ $\arg\Min_{\Gamma(t)} \|\sigma(\rho)^{-1}\hat b(t)\|_2$ exists, we see from \reff{MinRexpl} that $R$ attains its infimum at
$\theta^*(t)$ $=$ $(b^*(t),\rho^*(t))$ with
$b^*(t)$ $=$ $b_t^*(\rho^*(t))$ as in \reff{b*rho} with $\rho$ $=$ $\rho^*(t)$, which leads to the expressions of $(b^*(t), \rho^*(t))$ and $R(\theta^*(t))$ as described in \reff{b*ellip} and \reff{Rtheta*}.
\ep

\subsubsection{Proof of Theorem \ref{thmThetafull} } \label{appenproThetafull}

From the formula \reff{robustalpha} of the optimal portfolio strategy in Theorem \ref{robustoptimal}, we only need to calculate compute $\int_0^T R(b^*(s), \rho^*(s))ds$ and vector $\Sigma(\rho^*(t))^{-1}b^*(t)$. The rest of this subsection is to calculate $R(b^*(t), \rho^*(t))$ and $\Sigma(\rho^*(t))^{-1} b^*(t)$.

\noindent For any $t$ $\in$ $[0, T]$, let us introduce for $1$ $\leq$ $l$ $\leq$ $p$,
\beq \label{Thetal}
\Theta_l(t) &=& \{(b, \rho) \in \R^d \times \Gamma(t): \|\sigma_{J_l}(\rho)^{-1}(b_{J_l} -\hat b_{J_l}(t))\|_2 \leq \delta_l\},
\enq
\begin{Lemma}\label{lemThetalfull} Let $\Theta_l(t)$ be an ellipsoidal set as in \reff{Thetal} with $\Gamma(t)$ $=$ $\C_{>+}^d$, and assume there exists $m_l$ $=$ $m_l(t)$ $\in$ $J_l$ s.t. $|\hat\beta_{m_l}(t)|$ $>$ $\Max_{j \in J_l, j \neq m_l}|\hat\beta_j(t)|$, then we have $\arg\Min_{(b, \rho) \in \Theta_l(t)} \|\sigma_{J_l}(\rho)^{-1}b_{J_l}\|_2$ $\neq$ $\emptyset$ attained at $(b^*(t), \rho^*(t))$ with
\beqs
\rho_{m_lj}^*(t) &=& \hat\varrho_{m_lj}(t), \; j \in J_l, \; j \neq m_l,\\
b_{J_l}^*(t) &=& \hat b_{J_l}(t) (1 -\frac{\delta_l(t)}{|\hat\beta_{m_l}(t)|})1_{\{|\hat\beta_{m_l}(t)| > \delta_l(t)\}}.
\enqs
Moreover,
\beqs
\Min_{\Theta_l(t)}b_{J_l}\trans\Sigma_{J_l}(\rho)^{-1} b_{J_1} \;=\; (|\hat\beta_{m_l}(t)| -\delta_l(t))^2 1_{\{|\hat\beta_{m_l}(t)| > \delta_l(t)\}}.
\enqs
\end{Lemma}
{\bf Proof.}
W assume w.l.o.g. that $l$ $=$ $1$ and $m_l$ $=$ $1$, i.e., $|\hat\beta_1(t)|$ $>$ $\Max_{j \in J_1, j\neq 1}|\hat\beta_j(t)|$, otherwise we rearrange the assets. By noting that $b_{J_1}\trans\Sigma_{J_1}(\rho)^{-1}b_{J_1}$ is actually (squared) risk premium associated to the assets in the subclass $J_1$, and that Lemma \ref{lemellipsoid} is valid for any number of assets. we deduce from Lemma \ref{lemellipsoid} that
\beqs
& & \Inf_{(b, \rho) \in \Theta_1(t)} b_{J_1}\trans \Sigma_{J_1}(\rho)^{-1} b_{J_1}\\
&=& \Big(\Inf_{\rho \in \C_{>+}^d}\|\sigma_{J_1}(\rho)^{-1}\hat b_{J_1}(t)\|_{_2} -\delta_1(t)\Big)^2 1_{\big\{\Inf_{\rho \in \C_{>+}^d}\|\sigma_{J_1}(\rho)^{-1}\hat b_{J_1}(t)\|_{_2} > \delta_1(t)\big\}}.
\enqs
Let us show that if $|\hat\beta_1(t)|$ $>$ $\Max_{j \in J_1, j\neq 1}|\hat\beta_j(t)|$, then $\Inf_{\rho \in \C_{>+}^d}\|\sigma_{J_1}(\rho)^{-1}\hat b_{J_1}(t)\|_2 $ is attained over $\C_{>+}^d$. The key point is that $(\hat b_{J_1}(t))\trans \Sigma_{J_1}(\rho)^{-1} \hat b_{J_1}(t)$ is written as the sum of two nonnegative parts by matrix transformations. The procedures are as follows: we express $\Sigma_{J_1}(\rho)$ as the form of block matrix
\beqs
\Sigma_{J_1}(\rho) &=&\left(\begin{array}{c|c}
   \sigma_1^2                   &             \Sigma_{J_1-1, 1}(\rho)\trans \\
    \hline
    \Sigma_{J_1-1, 1}(\rho)         &             \Sigma_{J_1-1}(\rho)
\end{array}\right),
\enqs
by transforming $\Sigma_{J_1}(\rho)$ to block diagonal matrix, and taking the inverse, we then obtain
\beq
\Sigma_{J_1}(\rho)^{-1} &=&
\left(\begin{array}{cc}
1           &        -\frac{\Sigma_{J_1 -1, 1}(\rho)\trans}{\sigma_1^2}\\
0_{J_1-1,1} &     I_{J_1 -1}\end{array}\right)
\left(\begin{array}{cc}
\frac{1}{\sigma_1^2}   &    0_{J_1-1, 1}\trans\\
0_{J_1-1, 1}           &    A_{J_1 -1}(\rho)^{-1}
\end{array}\right)\left(\begin{array}{cc}
1           &        -\frac{\Sigma_{J_1 -1, 1}(\rho)\trans}{\sigma_1^2}\\
0_{J_1-1,1} &     I_{J_1 -1}\end{array}\right)\trans
\label{appenSigmaJ1-1}
\enq
where $I_{J_1-1}$ is $(|J_1|-1)$ $\times$ $(|J_1|-1)$ unit matrix and $A_{J_1 -1}(\rho)$ $:=$ $\Sigma_{J_1-1}(\rho)-\frac{\Sigma_{J_1-1,1}(\rho)\Sigma_{J_1-1,1}(\rho)\trans}{\sigma_1^2}$ lies in $\S_{>+}^d$.\\
From the expression of $\Sigma_{J_1}(\rho)^{-1}$ in \reff{appenSigmaJ1-1} and by writing $\hat b_{J_1}(t)$ in corresponding block form $(\hat b_1(t), (\hat b_{J_1 -1}(t))\trans)\trans$, we get
\beqs
& & (\hat b_{J_1}(t))\trans\Sigma_{J_1}(\rho)^{-1} \hat b_{J_1}(t)\\
&=& |\hat\beta_1(t)|^2 + \big(\hat b_{J_1-1}(t) -\frac{\hat b_1(t)}{\sigma_1^2} \Sigma_{J_1-1,1}(\rho)\big)\trans A_{J_1 -1}(\rho)^{-1}
\big(\hat b_{J_1-1}(t) -\frac{\hat b_1(t)}{\sigma_1^2} \Sigma_{J_1-1,1}(\rho)\big)\\
& \geq & |\hat\beta_1(t)|^2,
\enqs
and thus $\|\sigma_{J_1}(\rho^*(t))^{-1}\hat b_{J_1}(t)\|_2$ $=$ $|\hat\beta_1(t))|$ only when the second term is zero
\beq \label{appenfulliff3}
\hat b_{J_1-1}(t) -\frac{\hat b_1(t)}{\sigma_1^2} \Sigma_{J_1-1,1}(\rho^*(t)) = 0,
\enq
which has the explicit solution
\beq \label{appenlemmfullrho*}
\rho_{1j}^*(t) &=& \hat\varrho_{1j}(t) \;\in\; (-1, 1) ,\;\;\; j \in J_1, \; j \neq 1.
\enq
Therefore, we deduce from \reff{appenlemmfullrho*} and \reff{b*ellip} that
\beq \label{appenlemmfullrho*b*}
\rho_{1j}^*(t) &=& \hat\varrho_{1j}(t), j \in J_1, \; j \neq 1, \;\;\; b^*_{J_1}(t) \;=\; \hat b_{J_1}(t) (1 -\frac{\delta_1(t)}{|\hat\beta_1(t)|}) 1_{\{|\hat\beta_1(t)| > \delta_1(t)\}}.
\enq
Once $(\rho_{1j}^*(t))_{j \in J_1, j \neq 1}$ is given in the above equality, we can complete the other values of $\rho_{ij}^*(t)$ such that $\rho^*(t)$ $\in$ $\C_{>+}^d$. For instance, $\rho_{1j}^*(t)$ $=$ $\hat\varrho_{1j}(t)$, $1$ $<$ $j$ $\leq$ $d$, $\rho_{ij}^*(t)$ $=$ $\hat\varrho_{1i}(t)\hat\varrho_{1j}(t)$, $1$ $<$ $i$ $\neq$ $j$ $\leq$ $d$. It is easy to check that such a construction of $\rho^*(t)$ $\in$ $\C_{>+}^d$.
Moreover in this case, it follows from \reff{appenSigmaJ1-1} and \reff{appenfulliff3} that
\beq \label{appenvarratio}
\Sigma_{J_1}(\rho^*(t))^{-1} \hat b_{J_1}(t) &=& (\frac{\hat b_1(t)}{\sigma_1^2}, 0, \ldots, 0)\trans.
\enq
\ep

\vspace{2mm}

Let us  now prove Theorem \ref{thmThetafull}: if $|\hat\beta_{m_l}(t)|$ $>$ $\Max_{ j \in J_l, j \neq m_l}|\hat\beta_j(t)|$ and $|\hat\beta_{m_{k}}(t)|$ $-$ $\delta_{k}(t)$
$>$ $\Max_{1 \leq l \leq p, l \neq k}(|\hat\beta_{m_l}(t)| -\delta_{l}(t))$ , then $\Inf_{\Theta(t)} R(b, \rho)$ exists and $(b^*(t), \rho^*(t))$ $\in$ $\arg\Min_{(b, \rho) \in \Theta(t)} R(b, \rho)$ is computed explicitly. Let us consider w.l.o.g. the case of $p$ $=$ $2$ subsets, and reorder the assets s.t. $J_1$ $=$ $\{1, \ldots, k^o-1\}$, $J_2$ $=$ $\{k^o, \ldots, d\}$ and $m_1$ $=$ $1$, $m_2$ $=$ $k^o$ for some $1$ $\leq$ $k^o$ $\leq$ $d$, i.e.
\beq\label{propbiggest}
& & |\hat\beta_1(t)|\;>\; \Max_{j \in J_1, j \neq 1}|\hat\beta_j(t)|,\; |\hat\beta_{k^o}(t)|\;>\;
\Max_{j \in J_2, j \neq k^o}|\hat\beta_j(t)|, \;|\hat\beta_1(t)| -\delta_1(t)\;>\;|\hat\beta_{k^o}(t)|-\delta_2(t).
\enq
Notice that $R(b, \rho)$ can be expressed as the sum of two nonnegative parts in the same way as $(\hat b_{J_1}(t))\trans\Sigma_{J_1}(\rho)^{-1}\hat b_{J_1}(t)$ in Lemma \ref{lemThetalfull}.  $\Sigma(\rho)$ is written in a form of blocks as follows
\beqs
\Sigma(\rho)&=& \left(\begin{array}{c|c} \Sigma_{J_1}(\rho) & \Sigma_{J_{21}}(\rho)\trans\\
\hline
\Sigma_{J_{21}}(\rho) & \Sigma_{J_2}(\rho)\end{array}\right),
\enqs
and its inverse $\Sigma(\rho)$ is in the form
\beq
\Sigma(\rho)^{-1} &=& \left(\begin{array}{cc}I_{J_1} & -\Sigma_{J_1}(\rho)^{-1}\Sigma_{J_{21}}(\rho)\trans\\ 0_{J_{21}} & I_{J_2}\end{array}\right)
\left(\begin{array}{cc}\Sigma_{J_1}(\rho)^{-1} & 0_{J_{12}}\\ 0_{J_{21}} & A_{J_2}(\rho)^{-1}\end{array}\right) \nonumber\\
& & \;\;\;\;\; \left(\begin{array}{cc}I_{J_1} & -\Sigma_{J_1}(\rho)^{-1}\Sigma_{J_{21}}(\rho)\trans\\ 0_{J_{21}} & I_{J_2}\end{array}\right)\trans,
\label{appenSigma-1}
\enq
where $I_{J_l}$, $l$ $=$ $1$, $2$, is $|J_l|$ $\times$ $|J_l|$ unit matrix, and $A_{J_2}(\rho)$ $:=$ $\Sigma_{J_2}(\rho)$ $-$ $\Sigma_{J_{21}}(\rho)\Sigma_{J_1}(\rho)^{-1}\Sigma_{J_{21}}(\rho)\trans$ lies in $\S_{>+}^d$.\\
Recalling that $R(b, \rho)$ $=$ $b\trans \Sigma(\rho)^{-1} b$ and rewriting vector $b$ in corresponding block matrix form $(b_{J_1}\trans| b_{J_2}\trans)\trans$, together with \reff{appenSigma-1},
we express $R(b, \rho)$ as two nonnegative terms
\beqs
& & R(b, \rho)\\
&=& b_{J_1}\trans \Sigma_{J_1}(\rho)^{-1} b_{J_1} +\big(b_{J_2}-\Sigma_{J_{21}}(\rho)\Sigma_{J_1}(\rho)^{-1} b_{J_1}\big)\trans A_{J_2}(\rho)^{-1}\big(b_{J_2}-\Sigma_{J_{21}}(\rho)\Sigma_{J_1}(\rho)^{-1} b_{J_1}\big)\\
& \geq & b_{J_1}\trans \Sigma_{J_1}(\rho)^{-1} b_{J_1}
\;\geq \;  (|\hat\beta_1(t)| -\delta_1(t))^21_{\{|\hat\beta_1(t)| > \delta_1(t)\}},
\enqs
where we used  $A_{J_2}(\rho)$ $\in$ $\S_{>+}^d$ in the first inequality, and the second inequality is from $\Theta(t)$ $=$ $\Theta_1(t)$ $\cap$ $\Theta_2(t)$, hence $\inf_{\Theta(t)} b_{J_1}\trans\Sigma_{J_1}(\rho)^{-1} b_{J_1}$ $\geq$ $\inf_{\Theta_1(t)} b_{J_1}\trans\Sigma_{J_1}(\rho)^{-1} b_{J_1}$, and Lemma \ref{lemThetalfull}.

Therefore,

$R(b^*(t), \rho^*(t))$ $=$ $(|\hat\beta_1(t)| -\delta_1(t))^21_{\{|\hat\beta_1(t)| > \delta_1(t)\}}$ is minimum if and only if $(b^*(t), \rho^*(t))$ $\in$ $\Theta(t)$ satisfies \reff{appenlemmfullrho*b*} and
\beq
b_{J_2}^*(t) -\Sigma_{J_{21}}(\rho^*(t))\Sigma_{J_1}(\rho^*(t))^{-1}b_{J_1}^*(t) &=& 0,
\label{appenpropfulliff}
\enq
which yields, together with \reff{appenvarratio}, the explicit form,
\beq \label{appenfullb*J2}
\beta_j^*(t) &=& \beta_1^*(t)\rho_{1j}^*(t) \;=\; \hat\beta_1(t)(1-\frac{\delta_1(t)}{|\hat\beta_1(t)|})1_{\{|\hat\beta_1(t)| > \delta_1(t)\}}\rho_{1j}^*(t), \;\; j \in J_2.
\enq

In fact when \reff{propbiggest} holds, there exists an element $\theta^*(t)$ $=$ $(b^*(t), \rho^*(t))$ $\in$ $\Theta(t)$ attaining this infimum. For instance, we construct $(b^*(t), \rho^*(t))$ satisfying \reff{appenlemmfullrho*b*}-\reff{appenfullb*J2} in the form
\beq
& & \left\{
\begin{array}{rcl}
\beta_{J_1}^*(t) &=& (1 -\frac{\delta_1(t)}{|\hat\beta_1(t)|})1_{\{|\hat\beta_1(t)| > \delta_1(t)\}}\hat \beta_{J_1}(t),\\
\beta_{J_2}^*(t) &=& (1 -\frac{\delta_2(t)}{|\hat\beta_{k^o}(t)|})1_{\{|\hat\beta_{k^o}(t)| > \delta_2(t)\}}\hat \beta_{J_2}(t),
\end{array}
\right.
 \label{appenfullbeta*}\\
 \mbox{ and }  & &  \left\{
\begin{array}{rcl}
\rho_{1j}^*(t) &=& \beta_j^*(t)/\beta_1^*(t)\;=\; \hat\varrho_{1j}(t), \;\;\;\; j \in J_1, j \neq 1,\\
\rho_{k^oj}^*(t)&=& \beta_j^*(t)/\beta_{k^o}^*(t)\;=\; \hat\varrho_{k^oj}(t),\;\;\;\; j \in J_2, \; j \neq k^o,\\
\rho_{ij}^*(t) &=& \hat\varrho_{1i}(t)\hat\varrho_{1j}(t),\;\;\;\; 1< i \neq j \in J_1,\\
\rho_{ij}^*(t) &=&  \hat\varrho_{k^oi}(t)\hat\varrho_{k^oj}(t),\;\;\;\;  k^o < i \neq j \in J_2,\\
\rho_{1j}^*(t) &=& \beta_j^*(t)/\beta_1^*(t) \;=\; \frac{(1-\frac{\delta_2(t)}{|\hat\beta_{k^o}(t)|})\hat\beta_j(t)}{(1-\frac{\delta_1(t)}{|\hat\beta_1(t)|})\hat\beta_1(t)} \;\in\; (-1, 1),\;\;\;\; j \in J_2,\\
\rho_{ij}^*(t) &=& \rho_{1i}^*(t)\rho_{1j}^*(t),\;\;\;\; \mbox { otherwise },
\end{array}
\right.
\label{appenfullrho*}
\enq

The rest of the proof is to check that $(b^*(t), \rho^*(t))$ given in \reff{appenfullbeta*}-\reff{appenfullrho*} belongs to $\Theta(t)$, i.e., $\rho^*(t)$ $\in$ $\C_{>+}^d$ and $\|\sigma_{J_l}(\rho^*(t))^{-1}(b_{J_l}^*(t)-\hat b_{J_l}(t))\|_2$ $\leq$ $\delta_l(t)$, $l$ $=$ $1$, $2$. \\
\noindent $\bullet$  Let us first check that $\rho^*(t)$ in \reff{appenfullrho*} belongs to $\C_{>+}^d$.
In this case, $C(\rho^*(t))$ is written in the form (in what follows we often omit the dependence in $t$ of $\beta^*$ $=$ $\beta^*(t)$ and $\rho^*$ $=$ $\rho^*(t)$),
\beqs
C(\rho^*) = \left(\begin{array}{cccc|cccc}
1 & \hat\varrho_{12} & \ldots & \hat\varrho_{1k^o-1} & \frac{\beta_{k^o}^*}{\beta_1^*} & \frac{\beta_{k^o+1}^*}{\beta_1^*} &\ldots & \frac{\beta_d^*}{\beta_1^*}
\\
\hat\varrho_{12} & 1 & \ldots & \hat\varrho_{12}\hat\varrho_{1k^o-1} & \frac{\beta_2^*\beta_{k^o}^*}{|\beta_1^*|^2} & \frac{\beta_2^*\beta_{k^o+1}^*}{|\beta_1^*|^2} &\ldots & \frac{\beta_2^*\beta_d^*}{|\beta_1^*|^2}
\\
\vdots & \vdots & \ddots & \vdots  & \vdots & \vdots & \ldots & \vdots
\\
\hat\varrho_{1k^o-1} & \hat\varrho_{12}\hat\varrho_{1k^o-1} & \ldots & 1 & \frac{\beta_{k^o-1}^*\beta_{k^o}^*}{|\beta_1^*|^2} & \frac{\beta_{k^o-1}^*\beta_{k^o+1}^*}{|\beta_1^*|^2} & \ldots & \frac{\beta_{k^o-1}^*\beta_d^*}{|\beta_1^*|^2}
\\
\hline
\frac{\beta_{k^o}^*}{\beta_1^*} & \frac{\beta_2^*\beta_{k^o}^*}{|\beta_1^*|^2} & \ldots & \frac{\beta_{k^o-1}^*\beta_{k^o}^*}{|\beta_1^*|^2} & 1 & \hat\varrho_{k^ok^o+1} & \ldots & \hat\varrho_{k^od}
\\
\frac{\beta_{k^o+1}^*}{\beta_1^*} & \frac{\beta_2^*\beta_{k^o+1}^*}{|\beta_1^*|^2} & \ldots & \frac{\beta_{k^o-1}^*\beta_{k^o+1}^*}{|\beta_1^*|^2} & \hat\varrho_{k^ok^o+1} & 1 & \ldots & \hat\varrho_{k^ok^o+1}\hat\varrho_{k^od}
\\
\vdots & \vdots & \ldots & \vdots & \vdots & \vdots & \ddots & \vdots
\\
\frac{\beta_d^*}{\beta_1^*} & \frac{\beta_2^*\beta_d^*}{|\beta_1^*|^2} & \ldots & \frac{\beta_{k^o-1}^*\beta_{k^o}^*}{|\beta_1^*|^2} & \hat\varrho_{k^od} &\hat\varrho_{k^ok^o+1}\hat\varrho_{k^od} &  \ldots & 1
\end{array}\right),
\enqs
observe that
\beqs
 \left(\begin{array}{cccc}1 & 0 & \ldots & 0\\
                         -\frac{\beta_2^*}{\beta_1^*} & 1 & \ldots & 0\\
                         \vdots & \vdots & \ddots & \vdots\\
                         -\frac{\beta_d^*}{\beta_1^*} & 0 & \ldots & 1
\end{array}\right)C(\rho^*) \left(\begin{array}{cccc}1 & 0 & \ldots & 0\\
                         -\frac{\beta_2^*}{\beta_1^*} & 1 & \ldots & 0\\
                         \vdots & \vdots & \ddots & \vdots\\
                         -\frac{\beta_d^*}{\beta_1^*} & 0 & \ldots & 1
\end{array}\right)\trans &=& \left(\begin{array}{c|c} \hat C_{J_1}(\rho^*) & 0_{J_{21}}\trans \\ \hline 0_{J_{21}} & \hat C_{J_2}(\rho^*) \end{array}\right)
\enqs
where $\hat C_{J_1}(\rho^*)$ is a diagonal matrix in the form ${\rm diag}(1, 1-|\hat\varrho_{12}|^2, \ldots, 1-|\hat\varrho_{1k^o-1}|^2)$,
and
\beqs
\hat C_{J_2}(\rho^*) &=& \left(\begin{array}{cccc}
1-|\frac{\beta_{k^o}^*}{\beta_1^*}|^2 & \frac{\beta_{k^o+1}^*}{\beta_{k^o}^*}(1 -|\frac{\beta_{k^o}^*}{\beta_1^*}|^2) & \ldots & \frac{\beta_d^*}{\beta_{k^o}^*}(1-|\frac{\beta_{k^o}^*}{\beta_1^*}|^2)\\
\frac{\beta_{k^o+1}^*}{\beta_{k^o}^*}(1-|\frac{\beta_{k^o}^*}{\beta_1^*}|^2) & 1 -|\frac{\beta_{k^o+1}^*}{\beta_1^*}|^2 & \ldots & \frac{\beta_{k^o+1}^*\beta_d^*}{|\beta_{k^o}^*|^2}-\frac{\beta_{k^o+1}^*\beta_d^*}{|\beta_1^*|^2}\\
\vdots & \vdots & \ddots & \vdots\\
\frac{\beta_d^*}{\beta_{k^o}^*}(1-|\frac{\beta_{k^o}^*}{\beta_1^*}|^2) & \frac{\beta_{k^o+1}^*\beta_d^*}{|\beta_{k^o}^*|^2} -\frac{\beta_{k^o+1}^*\beta_d^*}{|\beta_1^*|^2} & \ldots & 1-|\frac{\beta_d^*}{\beta_1^*}|^2
\end{array}\right),
\enqs
then notice that
\beqs
& & \left(\begin{array}{cccc} 1 & 0 & \ldots & 0 \\
                          -\frac{\beta_{k^o+1}^*}{\beta_{k^o}^*} & 1 & \ldots & 0\\
                          \vdots & \vdots & \ddots & \vdots\\
                          -\frac{\beta_d^*}{\beta_{k^o}^*} & 0 & \ldots & 1
\end{array}\right) \hat C_{J_2}(\rho^*)
\left(\begin{array}{cccc} 1 & 0 & \ldots & 0 \\
                          -\frac{\beta_{k^o+1}^*}{\beta_{k^o}^*} & 1 & \ldots & 0\\
                          \vdots & \vdots & \ddots & \vdots\\
                          -\frac{\beta_d^*}{\beta_{k^o}^*} & 0 & \ldots & 1
\end{array}\right)\trans \\
&=& {\rm diag}(1-|\frac{\beta_{k^o}^*}{\beta_1^*}|^2, 1-|\hat\varrho_{k^ok^o+1}|^2, \ldots, 1-|\hat\varrho_{k^od}|^2).
\enqs
Since $\hat C_{J_1}(\rho^*)$ and $\hat C_{J_2}(\rho^*)$ are both symmetric positive definite matrices, then we have $\rho^*(t)$ as in \reff{appenfullrho*} belongs to $\C_{>+}^d$.

\vspace{2mm}

\noindent $\bullet$ Then it is  easily checked that $\|\sigma_{J_l}(\rho^*(t))^{-1}(b_{J_l}^*(t) -\hat b_{J_l}(t))\|_2$ $\leq$ $\delta_l$, $l$ $=$ $1$, $2$, with $(b^*(t), \rho^*(t))$ given by \reff{appenfullbeta*}-\reff{appenfullrho*} because we have
\beqs
\|\sigma_{J_l}(\rho^*(t))^{-1}(b_{J_l}^*(t) -\hat b_{J_l}(t))\|_2 &=& \frac{\delta_l(t)}{|\hat\beta_{m_l}(t)|}\|\sigma_{J_l}(\rho^*(t))^{-1}\hat b_{J_l}(t)\|_2 1_{\{|\hat\beta_{m_l}(t)| > \delta_l(t)\}}\\
&=& \delta_l(t)1_{\{|\hat\beta_{m_l}(t)| > \delta_l(t)\}} \; \leq \; \delta_l(t),
\enqs
where we used $\|\sigma_{J_l}(\rho^*(t))^{-1}\hat b_{J_l}(t)\|_2$ $=$ $|\hat\beta_{m_l}(t)|$, $l$ $=$ $1$, $2$ in Lemma \ref{lemThetalfull}.\\
Consequently, we deduce that $(b^*(t), \rho^*(t))$ $\in$ $\Theta(t)$ given by \reff{appenfullbeta*}-\reff{appenfullrho*} achieves the minimal risk premium
\beq \label{appenRtheta*}
R(b^*(t), \rho^*(t)) &=& (|\hat\beta_1(t)| -\delta_1(t))^21_{\{|\hat\beta_1(t)| > \delta_1(t)\}},
\enq
and with \reff{appenSigma-1}, \reff{appenpropfulliff} and \reff{appenvarratio} that
\beq
\Sigma(\rho^*(t))^{-1} b^*(t) &=&
(\frac{\hat b_1(t)}{\sigma_1^2}, 0, \ldots, 0)\trans (1 -\frac{\delta_1(t)}{|\hat\beta_1(t)|}) 1_{\{|\hat\beta_1(t)| > \delta_1(t)\}},
\label{Rbrho*full}
\enq
which completes the proof.
\ep

\subsubsection{Proof of  Theorem \ref{theoopt2}} \label{appenpro2}

In light of formula \reff{optimalalpha*ellip} of the optimal portfolio strategy in Proposition \ref{alphaellip}, we only need to compute $\int_0^T(\|\sigma(\rho^*(s))^{-1}\hat b(s)\|_2 -\delta(s))^2 1_{\{\|\sigma(\rho^*(s))^{-1}\hat b(s)\|_2 > \delta(s)\}} ds$
 and vector $\hat\kappa_t(\rho^*(t))$ $=$ $\Sigma(\rho^*(t))^{-1}\hat b(t)$.

 As $\Gamma(t)$ $=$ $[\underline \rho(t), \bar \rho(t)]$ is compact for fixed $t$ $\in$ $[0, T]$, we know that $\rho^*(t)$ $=$ ${\rm arg}\Min_{\rho\in\Gamma(t)} R(\hat b(t),\rho)$ exists, and from Lemma \ref{lemellipsoid}, we only need to compute the minimum of the function $\rho$ $\mapsto$ $R(\hat b(t),\rho)$ over $\Gamma(t)$.
From \reff{minimumellip} with $d$ $=$ $2$, we obtain the sufficient and necessary condition of
$\rho^*(t)$ for being global minimum of $R(\hat b(t), \rho)$ over $\Gamma(t)$:
\beq \label{rhonecsufglobal}
\sigma_1\sigma_2\hat \kappa^1_t(\rho^*(t))\hat \kappa^2_t(\rho^*(t))(\rho -\rho^*(t)) \leq 0, \;\;\; \text{for all}\; \rho \in [\underline \rho(t), \bar\rho(t)],
\enq
where $\hat\kappa_t(\rho)$ is explicitly written  as
\beq \label{hatkappa-2}
\hat\kappa_t(\rho)\;=\;\frac{1}{1-\rho^2}\left(\begin{matrix}\frac{\hat b_1(t)}{\sigma_1^2} -\frac{\hat b_2(t)}{\sigma_1\sigma_2}\rho\\ \frac{\hat b_2(t)}{\sigma_2^2} -\frac{\hat b_1(t)}{\sigma_1\sigma_2}\rho\end{matrix}\right)  \; = \;
\frac{1}{1-\rho^2}\left(\begin{matrix}\frac{\hat \beta_1(t) - \hat\beta_2(t) \rho}{\sigma_1}\\ \frac{\hat \beta_2(t) - \hat\beta_1(t)\rho}{\sigma_2}\end{matrix}\right).
\enq
From \reff{rhonecsufglobal}, we have following possible cases for fixed $t$:
\begin{itemize}
\item  $\hat\kappa^1_t(\rho^*(t)) \hat\kappa^2_t(\rho^*(t))$ $=$ $0$.  From the explicit expression \reff{hatkappa-2} of $\hat\kappa(\rho^*(t))$ and definition of $\hat\varrho_{12}(t)$ in \reff{hatvarrhoij}, we obtain $(\hat\varrho_{12}(t) - \rho^*(t))(1-\hat\varrho_{12}(t)\rho^*(t))$ $=$ $0$, and as $\rho^*(t)$ has to belong to
$[\underline \rho(t), \bar \rho(t)]$ $\subset$ $(-1,1)$, we obtain $\rho^*(t)$ $=$ $\hat\varrho_{12}(t)$,
and so $R(\hat b(t), \rho^*(t))$ $=$ $(\max(|\hat \beta_1(t)|, |\hat\beta_2(t)|))^2$.
\item $\hat\kappa^1_t(\rho^*(t))\hat\kappa^2(\rho^*(t))$ $>$ $0$. Then \reff{rhonecsufglobal} is satisfied iff $\rho^*(t)$ $=$ $\bar\rho(t)$. Moreover, from the above explicit expression of $\hat\kappa_t(\rho^*(t))$, we obtain $\bar\rho(t)$ $<$ $\hat\varrho_{12}(t)$.
\item $\hat\kappa^1_t(\rho^*(t))\hat\kappa^2_t(\rho^*(t))$ $<$ $0$. Then \reff{rhonecsufglobal} is satisfied iff  $\rho^*(t)$ $=$ $\underline \rho(t)$. Moreover, from the explicit expression of
$\hat\kappa_t(\rho^*(t))$, we obtain $\underline \rho(t)$ $>$ $\hat\varrho_{12}(t)$.
\end{itemize}
We obtain $\alpha^*(t)$ described as in Theorem \ref{theoopt2}.
\ep

\subsubsection{Proof of Theorem \ref{theoopt3}} \label{appenpro3}


In view of formula \reff{optimalalpha*ellip} of the optimal portfolio strategy in Proposition \ref{alphaellip}, we only need to compute
 $\hat\kappa(\rho^*)$ $=$ $\Sigma(\rho^*)^{-1}\hat b$, and $\|\sigma(\rho^*)^{-1}\hat b\|_2$, i.e., $R(\hat b, \rho^*)$.

As $\Gamma$ $=$ $\Prod_{j=1}^3\Prod_{i=1}^{j-1}$ $[\underline \rho_{ij}, \bar\rho_{ij}]$ is compact, we already know that $\rho^*$ $=$ $\arg\Min_{\rho \in \Gamma}R(\hat b, \rho)$ exists. From Lemma \ref{lemellipsoid}, we only need to compute the minimum of the function $\rho$ $\mapsto$ $R(\hat b, \rho)$ over $\Gamma$ by applying the optimality principle \reff{minimumellip} when  $d$ $=$ $3$,

\beq \label{rhoij*ellip}
\Sum_{j=1}^3 \Sum_{i=1}^{j-1}\sigma_i\sigma_j\hat\kappa^i(\rho^*)\hat\kappa^j(\rho^*)(\rho_{ij} -\rho_{ij}^*) \leq 0 \;\;\;\; \mbox {for any} \;\rho \in \Gamma.
\enq

Observe from \reff{rhoij*ellip} that, each $\rho_{ij}^*$, $1$ $\leq$ $i$ $<$ $j$ $\leq$ $3$ may be lower bound $\underline \rho_{ij}$, upper bound $\bar\rho_{ij}$, or an interior point in $(\underline \rho_{ij}, \bar\rho_{ij})$, which corresponds to $\kappa^i\kappa^j(\rho^*)$ $>$ $0$, $\kappa^i\kappa^j(\rho^*)$ $<$ $0$, or $\kappa^i\kappa^j(\rho^*)$ $=$ $0$ respectively. Therefore, let us consider the following possible exclusive  cases depending on the number of zero components in $\hat\kappa(\rho^*)$:
\begin{itemize}
\item [{\bf 1.}]
$\hat\kappa^1\hat\kappa^2(\rho^*)$ $=$ $0$, $\hat\kappa^1\hat\kappa^3(\rho^*)$ $=$ $0$, $\hat\kappa^2\hat\kappa^3(\rho^*)$ $=$ $0$.\\
In this case, \reff{rhoij*ellip} is immediately satisfied. As we assume that $\hat b$ $\neq$ $0$, $\hat\kappa(\rho^*)$ is not zero, i.e., at least one component of $\hat\kappa(\rho^*)$ is nonzero. Then, two components of $\hat\kappa(\rho^*)$ are zero.
Under the assumption that $|\hat\beta_1|$ $\geq$ $|\hat\beta_2|$ $\geq$ $|\hat\beta_3|$, \reff{Rbrho*full}, \reff{appenRtheta*} in Section \ref{appenproThetafull} yield the explicit expressions of $\rho^*$, $\hat\kappa(\rho^*)$ and $R(\hat b, \rho^*)$
    \beqs
   \rho_{12}^*  \;= \; \hat\varrho_{12} \in [\underline \rho_{12}, \bar\rho_{12}],\; \rho_{13}^* \; =\;\hat\varrho_{13} \in [\underline \rho_{13}, \bar\rho_{13}],\; \mbox{any}\; \rho_{23}^* \in [\underline \rho_{23}, \bar\rho_{23}]
    \enqs
    and
    \beq \label{kappaR1}
     \hat\kappa(\rho^*) \;= \; (\frac{\hat b_1}{\sigma_1^2}, 0, 0 )\trans,\;\;\;\; R(\hat b, \rho^*) \;=\; |\hat\beta_1|^2.
    \enq
 Let us show that $|\hat\beta_1|^2$ in \reff{kappaR1} is strict minimum value in the sense that $R(\hat b, \rho^*)$ $=$ $|\hat\beta_1|^2$ if and only if $\rho_{12}^* =\hat\varrho_{12}$ $\in$
$[\underline \rho_{12}, \bar\rho_{12}]$, $\rho_{13}^*$ $=$ $\hat\varrho_{13}$ $\in$ $[\underline \rho_{13}, \bar\rho_{13}]$ and any $\rho_{23}^*$ $\in$ $[\underline \rho_{23}, \bar\rho_{23}]$. We express $\Sigma(\rho)$ as the following block matrix
\beqs
\Sigma(\rho) =\left(\begin{matrix}\sigma_1^2 & C_1\trans \\ C_1 & \Sigma_{-1}(\rho_{23})\end{matrix}\right),
\enqs
where the vector $C_1=(\sigma_1\sigma_2\rho_{12}, \sigma_1\sigma_3\rho_{13})\trans$.

Noting that
\beq \label{Sigma1appen}
\left(\begin{matrix}1 & {\bf 0}_{1 \times 2} \\ -\frac{C_1}{\sigma_1^2} & I_{2 \times 2} \end{matrix}\right)\left(\begin{matrix}\sigma_1^2 & C_1\trans \\ C_1& \Sigma_{-1}(\rho_{23})\end{matrix}\right)\left(\begin{matrix}1 & -\frac{C_1\trans}{\sigma_1^2} \\{\bf 0}_{2 \times 1} & I_{2 \times 2}\end{matrix}\right)  &=& \left(\begin{matrix}\sigma_1^2 & {\bf 0}_{1 \times 2} \\ {\bf 0}_{2 \times 1} & A\end{matrix}\right),
\enq
where $I_{2 \times 2}$ denotes 2 $\times$ 2 identity matrix and $A$ $=$ $\Sigma_{-1}(\rho_{23}) - \frac{C_1 C_1\trans}{\sigma_1^4}$ is $2$ $\times$ $2$ positive definite matrix, and inverting on both sides of \reff{Sigma1appen}, we get
\beq \label{Sigma1-1appen}
\Sigma^{-1}(\rho) &=& \left(\begin{matrix}1 & -\frac{C_1\trans}{\sigma_1^2} \\{\bf 0}_{2 \times 1} & I_{2 \times 2}\end{matrix}\right) \left(\begin{matrix}\sigma_1^{-2} & {\bf 0}_{1 \times 2} \\ {\bf 0}_{2 \times 1} & A^{-1}\end{matrix}\right) \left(\begin{matrix}1 & {\bf 0}_{1 \times 2} \\ -\frac{C_1}{\sigma_1^2} & I_{2 \times 2} \end{matrix}\right).
\enq
We express $\hat b$ as $(\hat b_1, \hat b_{-1}\trans)\trans$ and then write $R(\hat b, \rho)$ as two nonnegative decompositions from \reff{Sigma1-1appen},
\beq \label{Rcase1}
R(\hat b, \rho)
&=& |\hat\beta_1|^2 + (\hat b_{-1} -\frac{\hat b_1}{\sigma_1^2} C_1)\trans A^{-1}(\hat b_{-1} -\frac{\hat b_1}{\sigma_1^2} C_1)\\
& \geq & |\hat\beta_1|^2, \nonumber
\enq
where in the last inequality,  `=' holds if and only if $\hat b_{-1} -\frac{\hat b_1}{\sigma_1^2} C_1$ $=$ $0$, i.e., $\rho_{12}^*$ $=$ $\hat\varrho_{12}$, $\rho_{13}^*$ $=$ $\hat\varrho_{13}$. This corresponds to case {\bf 1.}  of Theorem \ref{theoopt3}.

\item [{\bf 2.}] $\hat\kappa^1\hat\kappa^2(\rho^*)$ $\neq$ $0$, $\hat\kappa^1\hat\kappa^3(\rho^*)$ $=$ $0$, $\hat\kappa^2\hat\kappa^3(\rho^*)$ $=$ $0$.\\
     In this case,
we express $\Sigma(\rho)$ as the following block-matrix form for convenience,
\beqs
\Sigma(\rho) &=& \left(\begin{matrix}\Sigma_{-3}(\rho_{12}) & C_3 \\ C_3\trans & \sigma_3^2 \end{matrix}\right),
\enqs
where the vector $C_3$ $=$ $(\sigma_1\sigma_3\rho_{13}, \sigma_2\sigma_3\rho_{23})\trans$.\\
By first transforming $\Sigma(\rho)$ to block diagonal matrix as \reff{Sigma1-1appen} and then taking inverse, we obtain
\beq \label{Sigmarho-3}
\Sigma(\rho)^{-1} &=& \left(\begin{matrix}I_{2 \times 2} & -\Sigma_{-3}(\rho_{12})^{-1}C_3\\ {\bf 0}_{1 \times 2} & 1\end{matrix}\right)\left(\begin{matrix}\Sigma_{-3}(\rho_{12})^{-1} & {\bf 0}_{2 \times 1} \\{\bf 0}_{1 \times 2} & a(\rho)^{-1}\end{matrix}\right) \nonumber\\
 & & \;\;\; \left(\begin{matrix} I_{2 \times 2} & {\bf 0}_{2 \times 1} \\ -C_3\trans\Sigma_{-3}(\rho_{12})^{-1} & 1\end{matrix}\right),
\enq
where
$a(\rho)$ $=$ $\sigma_3^2 -C_3\trans\Sigma_{-3}(\rho_{12})^{-1} C_3$ is positive.

Recalling the definition of $\kappa(\hat b, \rho)$ and $R(\hat b, \rho)$ , we obtain from \reff{Sigmarho-3}
\beq \label{2hatkappaappen}
\left\{
\begin{array}{rcl}
\left(\begin{matrix}\hat\kappa^1(\rho) \\ \hat\kappa^2(\rho)\end{matrix}\right) &=& \Sigma_{-3}(\rho_{12})^{-1}\hat b_{-3} -\hat\kappa^3(\rho)\Sigma_{-3}(\rho_{12})^{-1} C_{3}\\
\hat\kappa^3(\rho)
 &=& \frac{1}{a(\rho)}(\hat b_3 -C_3\trans \Sigma_{-3}(\rho_{12})^{-1}\hat b_{-3}),
\end{array}
\right.
\enq
and
\beq \label{2Rappen}
R(\hat b, \rho) &=& \hat b_{-3}\trans\Sigma_{-3}(\rho_{12})^{-1} \hat b_{-3} + a(\rho)|\hat\kappa^3(\rho)|^2.
\enq
In the following, we write $\hat b_{-3}\trans \Sigma_{-3}(\rho_{12})^{-1}\hat b_{-3}$ as $R(\hat b_{-3}, \rho_{12})$.\\
As $\hat\kappa^3(\rho^*)$ $=$ $0$, we obtain from \reff{rhoij*ellip} that
\beq
\sigma_1\sigma_2\hat\kappa^1\hat\kappa^2(\rho_{12}^*)(\rho_{12}-\rho_{12}^*) & \leq & 0 \;\;\;\; \mbox{for all}\; \rho_{12} \in [\underline \rho_{12}, \bar\rho_{12}]
\enq
and  from \reff{2hatkappaappen} and \reff{2Rappen} that
\beq \label{hatkappa2g}
\left\{
\begin{array}{rcl}
\left(\begin{matrix}
\hat\kappa^1(\rho_{12}^*) \\ \hat\kappa^2(\rho_{12}^*)\end{matrix}\right) &=& \Sigma_{-3}(\rho_{12}^*)^{-1}\hat b_{-3} \\
R(\hat b, \rho^*) &=& R(\hat b_{-3}, \rho_{12}^*).
\end{array}
\right.
\enq
This is the case of ambiguous correlation in the two risky assets: the first and second assets with ambiguous correlation $\rho_{12}$ in $[\underline \rho_{12}, \bar\rho_{12}]$. In this case, $\hat\kappa^1(\rho^*)$ and $\hat\kappa^2(\rho^*)$ are not zero, therefore we have that from Theorem \ref{theoopt2}
\beq \label{rho12*appen}
\rho_{12}^* &=& \bar\rho_{12} 1_{\{\bar\rho_{12} < \hat\varrho_{12}\}} \; + \; \underline \rho_{12} 1_{\{\underline \rho_{12} > \hat\varrho_{12}\}}.
\enq
By setting $g(\rho_{12}^*, \rho_{13}, \rho_{23})$ $:=$ $ a(\rho_{12}^*, \rho_{13}, \rho_{23})\kappa^3(\rho_{12}^*, \rho_{13}, \rho_{23})$ for fixed $\rho_{12}^*$ in \reff{rho12*appen}, we deduce from \reff{2hatkappaappen} that the function
\beqs
(\rho_{13}, \rho_{23}) \mapsto g(\rho_{12}^*, \rho_{13}, \rho_{23}) & = &\hat b_3 -\sigma_1\sigma_3\hat\kappa^1(\rho_{12}^*)\rho_{13}  - \sigma_2\sigma_3\hat\kappa^2(\rho_{12}^*)\rho_{23},
\enqs
 is linear in $(\rho_{13}, \rho_{23})$ $\in$ $[\underline \rho_{13}, \bar\rho_{13}]$ $\times$ $[\underline \rho_{23}, \bar\rho_{23}]$, and has the same sign as $\hat\kappa^3(\rho_{12}^*, \rho_{13}, \rho_{23})$ due to the positiveness of $a(\rho_{12}^*, \rho_{13}, \rho_{23})$. 
 To study the condition of $\kappa^3(\rho^*)$ $=$ $0$ on $\Gamma$, we discuss it in the following two cases:
\begin{itemize}
\item [(i)]
if $\bar\rho_{12}$ $<$ $\hat\varrho_{12}$, then $\hat\kappa^1\hat\kappa^2(\bar\rho_{12})$ $>$ $0$, the function $(\rho_{13}, \rho_{23})$ $\mapsto$ $g(\bar\rho_{12}, \rho_{13}, \rho_{23})$ has the same monotonicity with respect to $\rho_{13}$, $\rho_{23}$. Therefore, to ensure that the function $g(\bar\rho_{12}, \rho_{13}, \rho_{23})$ has a root in $[\underline \rho_{13}, \bar\rho_{13}]$ $\times$ $[\underline \rho_{23}, \bar\rho_{23}]$, we need $g(\bar\rho_{12}, \underline \rho_{13}, \underline \rho_{23})$$g(\bar\rho_{12}, \bar\rho_{13}, \bar\rho_{23}) \leq 0$, or equivalently $\hat\kappa^3(\bar\rho_{12}, \underline\rho_{13}, \underline \rho_{23})$ $\hat\kappa^3(\bar\rho_{12}, \bar \rho_{13}, \bar \rho_{23}) \leq0$.

\item [(ii)] if $\underline \rho_{12}$ $>$ $\hat\varrho_{12}$, then $\hat\kappa^1\hat\kappa^2(\underline \rho_{12})$ $<$ $0$, the function $(\rho_{13}, \rho_{23})$ $\mapsto$ $g(\underline \rho_{12}, \rho_{13}, \rho_{23})$ has the opposite monotonicity with respect to $\rho_{13}$, $\rho_{23}$. Therefore, when
$g(\underline \rho_{12}, \bar\rho_{13}, \underline \rho_{23})$ $g(\underline \rho_{12}, \underline \rho_{13}, \bar\rho_{23})$ $\leq $ $0$, or equivalently
$\hat\kappa^3(\underline \rho_{12}, \bar\rho_{13}, \underline \rho_{23})\hat\kappa^3(\underline \rho_{12}, \underline \rho_{13}, \bar\rho_{13})\leq 0$, the function $g(\underline \rho_{12}, \rho_{13}, \rho_{23})$ has a root in $[\underline \rho_{13}, \bar\rho_{13}]$ $\times$ $[\underline \rho_{23}, \bar\rho_{23}]$.
\end{itemize}
Therefore, we deduce that $R(\hat b, \rho)$ $\geq$ $R(\hat b_{-3}, \rho_{12})$ $\geq$ $R(\hat b_{-3}, \bar\rho_{12}1_{\{\bar\rho_{12} < \hat\varrho_{12}\}} + \underline \rho_{12} 1_{\{\underline \rho_{12} > \hat\varrho_{12}\}})$ and that `=' holds
if and only if $\rho_{12}^*$ $=$ $\bar\rho_{12} 1_{\{\bar\rho_{12} < \hat\varrho_{12}\}}$ $+$ $\underline \rho_{12} 1_{\{\underline \rho_{12} > \hat\varrho_{12}\}}$ and $\rho_{13}^*$, $\rho_{23}^*$ satisfies $\hat\kappa^3(\rho_{12}^*, \rho_{13}^*, \rho_{23}^*)$ $=$ $0$. This corresponds to subcases {\bf 2.}(i) and {\bf 2.}(ii) of  Theorem \ref{theoopt3}.

\item[\bf 3.] $\hat\kappa^1\hat\kappa^2(\rho^*)$ $=$ $0$, $\hat\kappa^1\hat\kappa^3(\rho^*)$ $\neq$ $0$, $\hat\kappa^2(\rho^*)\hat\kappa^3(\rho^*)$ $=$ $0$.\\
In this case, we make permutations as follows,
\beq \label{R-2}
\left(\begin{matrix} \hat\kappa^{-2}(\rho)\\ \hat\kappa^2(\rho)\end{matrix}\right)&=&\left(\begin{matrix}\Sigma_{-2}(\rho_{13}) & C_2 \\ C_2\trans & \sigma_2^2\end{matrix}\right)^{-1}\left(\begin{matrix}\hat b_{-2} \\ \hat b_2 \end{matrix}\right),
\enq
where $\hat\kappa^{-2}(\rho)$ $:=$ $(\hat\kappa^1(\rho), \hat\kappa^3(\rho))\trans$ and $C_2$ $:=$ $(\sigma_1\sigma_2\rho_{12}, \sigma_2\sigma_3\rho_{23})\trans$. Using \reff{R-2} and proceeding with the same arguments as in  the case {\bf 2.}, we obtain the result of $\hat\kappa^2(\rho^*)$ $=$ $0$, $\hat\kappa^1(\rho^*)$$\hat\kappa^3(\rho^*)$ $\neq$ $0$ as described in the subcases {\bf 3.}(i) and {\bf 3.}(ii) of Theorem \ref{theoopt3}.
\item [\bf 4.]
$\hat\kappa^1\hat\kappa^2(\rho^*)$ $=$ $0$, $\hat\kappa^1\hat\kappa^3(\rho^*)$ $=$ $0$, $\hat\kappa^2(\rho^*)\hat\kappa^3(\rho^*)$ $\neq$ $0$.\\
Notice that
\beq \label{R-1}
\left(\begin{matrix}\hat\kappa^{-1}(\rho)\\ \hat\kappa^1(\rho)\end{matrix}\right)&=& \left(\begin{matrix}\Sigma_{-1}(\rho_{23}) & C_1\\ C_1\trans & \sigma_1^2\end{matrix}\right)^{-1} \left(\begin{matrix}\hat b_{-1} \\ \hat b_1\end{matrix}\right),
\enq
where $\hat\kappa^{-1}(\rho)$ $:=$ $(\hat\kappa^2(\rho), \hat\kappa^3(\rho))\trans$ and $C_1$ $:=$ $(\sigma_1\sigma_2\rho_{12}, \sigma_1\sigma_3\rho_{13})\trans$. Using \reff{R-1} and proceeding with the same arguments as in the case {\bf 2.}, we obtain the result of $\hat\kappa^1(\rho^*)$ $=$ $0$, $\hat\kappa^2(\rho^*)$$\hat\kappa^3(\rho^*)$ $\neq$ $0$ as described in subcases {\bf 4.}(i) and {\bf 4.}(ii) of Theorem \ref{theoopt3}.
\item [\bf 5.] 
$\hat\kappa^1\hat\kappa^2(\rho^*)$ $\neq$ $0$, $\hat\kappa^1\hat\kappa^3(\rho^*)$ $\neq$ $0$, $\hat\kappa^2(\rho^*)\hat\kappa^3(\rho^*)$ $\neq$ $0$.\\
In this case, we see from \reff{rhoij*ellip} that each $\rho_{ij}^*$ takes value in $\{\underline \rho_{ij}, \bar\rho_{ij}\}$ relying on the sign of $\hat\kappa^i\hat\kappa^j(\rho^*)$. Notice that once the signs of $\hat\kappa^1\hat\kappa^2(\rho^*)$ and $\hat\kappa^1\hat\kappa^3(\rho^*)$ are known, the sign of $\hat\kappa^2(\rho^*)\hat\kappa^3(\rho^*)$ is determined. Therefore, by combination, there are 4 possible sub-cases as described in the case {\bf 5.} of Theorem \ref{theoopt3}.\\
As $\hat\kappa^i(\rho^*)\hat\kappa^j(\rho^*)$ $\neq$ $0$ in each subcase, left right hand of \reff{rhoij*ellip} is strictly negative for any $\rho$ $\in$ $\Gamma$ $\setminus$ $\{\rho^*\}$.

From the first-order characterization for convexity of $R(\hat b, \rho)$ (see e.g. Section 3.1.3 in \cite{BoydVand04}) and \reff{firstorder}, 
we obtain for any $\rho$ $\in$ $\Gamma$ $\setminus$ $\{\rho^*\}$,
\beqs
R(\hat b, \rho) & \geq & R(\hat b, \rho^*) + (\rho -\rho^*)\trans \nabla_\rho R(\hat b, \rho^*)\\
&=& R(\hat b, \rho^*) - \Sum_{j=1}^3\Sum_{i=1}^{j-1}\sigma_i\sigma_j \hat\kappa^i(\rho^*)\hat\kappa^j(\rho^*)(\rho_{ij} -\rho_{ij}^*)\\
& > & R(\hat b, \rho^*),
\enqs
which indicates that $\rho^*$ in each sub-case of case {\bf 5.} in Theorem \ref{theoopt3} is a strict minimum of $R(\hat b, \rho)$.
\end{itemize}
 As $R(\hat b, \rho^*)$ in each subcase is strict minimum value, we conclude that each subcase in Theorem \ref{theoopt3} is exclusive.
 By combining this with Lemma \ref{lemellipsoid}, we obtain $b^*$ described as in Theorem \ref{theoopt3}.
\ep

\small
\nocite{*}
\bibliographystyle{apacite}  
\bibliography{MFrefs.bib}

\end{document}